\documentclass[a4paper,10pt]{article} 

\usepackage{amsthm,amsmath,amssymb,amsfonts}
\usepackage{amscd}
\usepackage[bookmarksnumbered=true]{hyperref}

\usepackage{color}

\usepackage{titlesec}
\titleformat*{\section}{\Large \normalfont \bfseries}
\titleformat*{\subsection}{\large \normalfont \bfseries}
\usepackage{setspace}
\usepackage{relsize}
\numberwithin{equation}{section}

\usepackage[left=2cm,right=2cm,top=2cm,bottom=2cm ,bindingoffset=0cm]{geometry}

\usepackage{color}
\usepackage{tikz}
\usepackage{pgffor}
\usepackage{upgreek}

\usetikzlibrary{arrows, positioning, cd, calc}
\usetikzlibrary{decorations.pathreplacing, decorations.markings,math, shapes}

\tikzset{
	on each segment/.style={
		decorate,
		decoration={
			show path construction,
			moveto code={},
			lineto code={
				\path [#1]
				(\tikzinputsegmentfirst) -- (\tikzinputsegmentlast);
			},
			curveto code={
				\path [#1] (\tikzinputsegmentfirst)
				.. controls
				(\tikzinputsegmentsupporta) and (\tikzinputsegmentsupportb)
				..
				(\tikzinputsegmentlast);
			},
			closepath code={
				\path [#1]
				(\tikzinputsegmentfirst) -- (\tikzinputsegmentlast);
			},
		},
	},
	mid arrow/.style={postaction={decorate,decoration={
				markings,
				mark=at position .5 with {\arrow[#1]{s\colDth}}
	}}},
}

\def\sgn{\mathrm{sgn}}
\def\deg{\mathrm{deg}}
\def\End{\mathrm{End}}

\def\X{\mathcal{X}}
\def\P{\mathrm{P}}

\def\A{\mathcal{A}}
\def\F{\mathrm{F}}
\def\V{\mathrm{V}}
\def\C2{\mathbb{C}^2}
\def\LBS{\mathrm{L}_{\mathrm{BS}}}
\def\LCL{\mathrm{L}_{\mathrm{CL}}}
\def\LCLm{\overline{\mathrm{L}}_{\mathrm{cl}}}
\def\Rmut{\mu_\mathcal{R}}
\def\Rtet{\mathcal{R}}

\def\z{\zeta}
\def\Zig{\mathrm{Z}}
\def\Z{\mathcal{Z}}
\def\zig{z}

\def\XG{\mathrm{X}_G}

\def\GLb{\widehat{\mathrm{PGL}}}
\def\T2{\mathbb{T}^2}
\def\H1{\mathrm{H}_1}
\def\Hc1{\mathrm{H}^1}
\def\ha{h_A}
\def\hb{h_B}

\def\Newt{\Delta}

\def\L{\mathrm{L}}

\def\xa{x_a}
\def\ya{y_a}
\def\ka{k_a}
\def\la{\lambda_a}
\def\ma{\mu_a}
\def\spA{\alpha_1} 
\def\spB{\alpha_2} 
\def\spC{\alpha_3} 
\def\xb{x_b}
\def\yb{y_b}
\def\kb{k_b}
\def\lb{\lambda_b}
\def\mb{\mu_b}
\def\xc{x_c}
\def\yc{y_c}
\def\kc{k_c}
\def\lc{\lambda_c}
\def\mc{\mu_c}

\def\Lat{L}

\def\x{\mathbf{x}}

\def\colP{blue}
\def\colZ{teal}
\def\colMut{red}

\def\colA{\colP}
\def\colB{\colP}
\def\colC{\colP}
\def\colD{\colP}

\title{\textbf{Solution of tetrahedron equation and cluster algebras}}

\author{P. Gavrylenko$^{1,2}$\thanks{pasha.145@gmail.com}, M. Semenyakin$^{1,2}$\thanks{semenyakinms@gmail.com}, Y. Zenkevich$^{3-8}$\thanks{yegor.zenkevich@gmail.com}}
\date{
{\small\textit{$^1$ Center for Advanced Studies, Skoltech, Moscow, Russia}}\\
{\small\textit{$^2$ Faculty of Mathematics, HSE University, Moscow, Russia}} \\
{\small\textit{$^3$ SISSA, via
      Bonomea 265, 34136 Trieste, Italy}}\\
  {\small\textit{$^4$ INFN, Sezione di Trieste}}\\
  {\small\textit{$^5$ IGAP, via Beirut 2/1, 34151 Trieste, Italy}}\\
  {\small\textit{$^6$ ITEP, Bolshaya Cheremushkinskaya street 25, 117218
      Moscow, Russia}}\\
  {\small\textit{$^7$ ITMP MSU, Leninskie gory 1, 119991 Moscow, Russia}}\\
  {\small\textit{$^8$ MIPT, Institutski per. 9, 141701 Dolgoprudny, Russia}}}

\begin{document}
\tikzset{
	styleDiagonals/.style={color=violet!50,very thin},
	styleFill/.style={fill,gray,opacity=0.2},
	styleArrow/.style={
		postaction={decorate},
		decoration={markings,mark=at position 0.75 with {\arrow{Stealth[scale=1]}}},
	},
		styleArrowPointer/.style={
		postaction={decorate},
		decoration={markings,mark=at position 1 with {\arrow{Stealth[scale=1]}}},
	},
	styleArrowDimer/.style={
		postaction={decorate},
		decoration={markings,mark=at position 0.8 with {\arrow{Stealth[scale=0.8]}}},
		blue,
		line width=0.5mm
	},
		styleArrowShort/.style={
		postaction={decorate},
		decoration={markings,
		mark=at position 0.5 with {\arrow{Stealth[scale=1]}}
		}
	},
	styleArrowShifted/.style={
		postaction={decorate},
		decoration={markings,
		mark=at position 0.35 with {\arrow{Stealth[scale=1]}}
		}
	},
	styleTextEdges/.style={
		scale = 0.8
	},
	styleQuiverEdge/.style={
		line width=0.7pt, 
		postaction={decorate},
		decoration={markings,
		mark=at position 0.6 with {\arrow{Stealth[scale=1]}}
		}
	},
	styleDirectedEdge/.style={ 
		postaction={decorate},
		decoration={markings,
		mark=at position 0.6 with {\arrow{Stealth[scale=0.8]}}
		}
	},
	quiverVertex/.style={fill=black,thick,radius=0.07},
	blackCircle/.style={fill=black,thick,radius=0.1,inner sep=0},
	greyCircle/.style={fill=gray,thick,radius=0.1,inner sep=0},
	whiteCircle/.style={fill=white,thick,radius=0.1,inner sep=0},
	cross/.style={cross out, draw=black, fill=none, minimum size=2*(#1-\pgflinewidth), inner sep=0pt, outer sep=0pt}, 		cross/.default={4pt}
}

\tikzset{
toda_one_text/.pic=
{
\tikzset{
	scale=1
}

\draw[black,styleArrow, thick] (1,1)--(0,0);
\draw[black,styleArrow, thick] (1,1)--(2,2);
\draw[black,styleArrowShifted, thick] (1,1)--(1,3);
\draw[black,styleArrow, thick] (1,1)--(1,0);
\draw[thick] (1,2) -- (1,4);
\draw[thick] (0,2)--(2,4);

\draw[whiteCircle] (1,3) circle;
\draw[blackCircle] (1,1) circle;

}
}

\tikzset{
toda_two_text/.pic=
{
\tikzset{
	scale=1
}

\draw[black,styleArrow, thick] (3,1)--(1,3);
\draw[black,styleArrow, thick] (3,1)--(1,1);
\draw[black,styleArrow, thick] (3,1)--(4,2);
\draw[black,styleArrow, thick] (3,1)--(2,0);
\draw[black,styleArrow, thick] (3,3)--(4,4);
\draw[black,styleArrow, thick] (3,3)--(1,3);
\draw[thick] (1,1) -- (0,0);
\draw[thick] (1,3)--(0,2);
\draw[thick] (2,4)--(1,3);

\draw[whiteCircle] (1,1) circle;
\draw[whiteCircle] (1,3) circle;

\draw[blackCircle] (3,1) circle;
\draw[blackCircle] (3,3) circle;
}
}

\tikzset{
toda_casi_hor/.pic=
{
\tikzset{
	scale=1
}

\draw[blue,styleArrow, thick] (1,1)--(0,0);
\draw[orange,styleArrow, thick] (1,1)--(2,2);
\draw[orange,styleArrowShifted, thick] (1,3)--(1,1);
\draw[blue,styleArrow, thick] (1,0)--(1,1);
\draw[orange, thick] (1,2) -- (1,3);
\draw[blue, thick] (1,3) -- (1,4);
\draw[orange, thick] (0,2)--(1,3);
\draw[blue, thick] (2,4)--(1,3);

\draw[whiteCircle] (1,3) circle;
\draw[blackCircle] (1,1) circle;

}
}

\tikzset{
toda_casi_hor_twist/.pic=
{
\tikzset{
	scale=1
}

\draw[green,styleArrow, thick] (0,0)--(1,1);
\draw[green,styleArrowShifted, thick] (1,1)--(1,3);
\draw[green, thick] (2,4)--(1,3);

\draw[red,styleArrow, thick] (2,2)--(1,1);
\draw[red,styleArrow, thick] (1,1)--(1,0);
\draw[red, thick] (1,3) -- (1,4);
\draw[red, thick] (0,2)--(1,3);

\draw[whiteCircle] (1,3) circle;
\draw[blackCircle] (1,1) circle;

}
}

\tikzset{
quiver_cross/.pic=
{
\tikzset{
	scale=1, font = \small 
}

\draw[styleQuiverEdge] (0,1) -- (1,1);
\draw[styleQuiverEdge] (1,1) -- (1,0);
\draw[styleQuiverEdge] (1,1) -- (1,2);
\draw[styleQuiverEdge] (2,1) -- (1,1);

\draw[quiverVertex] (0,1) circle;
\draw[quiverVertex] (1,0) circle;
\draw[quiverVertex] (1,2) circle;
\draw[quiverVertex] (2,1) circle;
\draw[quiverVertex] (1,1) circle;

}
}

\tikzset{
bipart_text/.pic=
{
\tikzset{
	scale=0.5, font = \small 
}

\draw[black, thick] (0,2)--(2,4);
\draw[black, thick] (0,2)--(2,0);
\draw[black, thick] (4,2)--(2,4);
\draw[black, thick] (4,2)--(2,0);

\draw[whiteCircle] (2,4) circle;
\draw[whiteCircle] (2,0) circle;
\draw[blackCircle] (0,2) circle;
\draw[blackCircle] (4,2) circle;

}
}

\tikzset{
bipart_text_expanded/.pic=
{
\tikzset{
	scale=0.33, font = \small 
}

\draw[black, thick] (0,2)--(2,4);
\draw[black, thick] (0,2)--(2,0);
\draw[black, thick] (4,2)--(2,4);
\draw[black, thick] (4,2)--(2,0);

\draw[black, thick] (4,2)--(5,2);
\draw[black, thick] (6,2)--(5,2);
\draw[whiteCircle] (5,2) circle;
  
\draw[black, thick] (2,5)--(2,4);
\draw[blackCircle] (2,5) circle;
\draw[black, thick] (2,5)--(2,6);  

\draw[whiteCircle] (2,4) circle;
\draw[whiteCircle] (2,0) circle;
\draw[blackCircle] (0,2) circle;
\draw[blackCircle] (4,2) circle;

}
}

\tikzset{
bipart_text_expanded_blank/.pic=
{
\tikzset{
	scale=0.5, font = \small 
}

\draw[black, thick] (0,2)--(2,4);
\draw[black, thick] (0,2)--(2,0);
\draw[black, thick] (4,2)--(2,4);
\draw[black, thick] (4,2)--(2,0);

\draw[black, thick] (4,2)--(6,2);
\draw[black, thick] (2,0)--(2,-2);

}
}

\tikzset{
bipart_text_arrows/.pic=
{
\tikzset{
	scale=0.5, font = \small 
}

\draw[black,styleArrow, thick] (0,2)--(2,4);
\draw[black,styleArrow, thick] (0,2)--(2,0);
\draw[black,styleArrow, thick] (4,2)--(2,4);
\draw[black,styleArrow, thick] (4,2)--(2,0);

\draw[whiteCircle] (2,4) circle;
\draw[whiteCircle] (2,0) circle;
\draw[blackCircle] (0,2) circle;
\draw[blackCircle] (4,2) circle;

}
}

\tikzset{
bipart_text_expanded_arrows/.pic=
{
\tikzset{
	scale=0.33, font = \small 
}

\draw[black, styleArrow, thick] (0,2)--(2,4);
\draw[black, styleArrow, thick] (0,2)--(2,0);
\draw[black, styleArrow, thick] (4,2)--(2,4);
\draw[black, styleArrow, thick] (4,2)--(2,0);

\draw[black, styleArrow, thick] (4,2)--(5,2);
\draw[black, styleArrow, thick] (6,2)--(5,2);
\draw[whiteCircle] (5,2) circle;
  
\draw[black, styleArrow, thick] (2,5)--(2,4);
\draw[blackCircle] (2,5) circle;
\draw[black, styleArrow, thick] (2,5)--(2,6);  

\draw[whiteCircle] (2,4) circle;
\draw[whiteCircle] (2,0) circle;
\draw[blackCircle] (0,2) circle;
\draw[blackCircle] (4,2) circle;

}
}

\tikzset{
bipart_casi_hor/.pic=
{
\tikzset{
	scale=0.75, font = \small 
}

\draw[orange, styleArrow, thick] (0,2)--(2,4);
\draw[orange, styleArrow, thick] (2,4)--(4,2);
\draw[blue, styleArrow, thick] (2,0)--(0,2);
\draw[blue, styleArrow, thick] (4,2)--(2,0);

\draw[whiteCircle] (2,4) circle;
\draw[whiteCircle] (2,0) circle;
\draw[blackCircle] (0,2) circle;
\draw[blackCircle] (4,2) circle;

}
}

\tikzset{
bipart_casi_vert/.pic=
{
\tikzset{
	scale=0.75, font = \small 
}

\draw[red, styleArrow, thick] (2,4)--(0,2);
\draw[red, styleArrow, thick] (0,2)--(2,0);
\draw[green, styleArrow, thick] (2,0)--(4,2);
\draw[green, styleArrow, thick] (4,2)--(2,4);

\draw[whiteCircle] (2,4) circle;
\draw[whiteCircle] (2,0) circle;
\draw[blackCircle] (0,2) circle;
\draw[blackCircle] (4,2) circle;

}
}

\tikzset{
Wt_triv_gr/.pic=
{
	\fill[color=gray!10] (0,0)--(1,0)--(1,1)--(0,1)--cycle;
	\draw (0,0)--(1,0);
	\draw (0,1)--(1,1);
}
}
\tikzset{
Wt_triv_gr_no_top/.pic=
{
	\fill[color=gray!10] (0,0)--(1,0)--(1,1)--(0,1)--cycle;
	\draw (0,0)--(1,0);
}
}
\tikzset{
Wt_triv_gr_no_bot/.pic=
{
	\fill[color=gray!10] (0,0)--(1,0)--(1,1)--(0,1)--cycle;
	\draw (0,1)--(1,1);
}
}
\tikzset{
Wt_triv_gr_no_top_bot/.pic=
{
	\fill[color=gray!10] (0,0)--(1,0)--(1,1)--(0,1)--cycle;
}
}

\tikzset{
Wt_triv_wh/.pic=
{
	\fill[color=white!10] (0,0)--(1,0)--(1,1)--(0,1)--cycle;
	\draw (0,0)--(1,0);
	\draw (0,1)--(1,1);
}
}
\tikzset{
Wt_triv_wh_no_top/.pic=
{
	\fill[color=white!10] (0,0)--(1,0)--(1,1)--(0,1)--cycle;
	\draw (0,0)--(1,0);
}
}
\tikzset{
Wt_triv_wh_no_bot/.pic=
{
	\fill[color=white!10] (0,0)--(1,0)--(1,1)--(0,1)--cycle;
	\draw (0,1)--(1,1);
}
}
\tikzset{
Wt_triv_wh_no_top_bot/.pic=
{
	\fill[color=white!10] (0,0)--(1,0)--(1,1)--(0,1)--cycle;
}
}

\tikzset{
Wt_lower_left_gr/.pic=
{
	\fill[color=gray!10] (0,0)..controls(0.5,0)..(1,1)--(0,1)--cycle;
	\draw (0,0)..controls(0.5,0)..(1,1)--(0,1);
}
}

\tikzset{
Wt_lower_left_wh/.pic=
{
	\fill[color=gray!10] (0,0)..controls(0.5,0)..(1,1)--(1,0)--cycle;
	\draw (0,0)..controls(0.5,0)..(1,1)--(0,1);
}
}

\tikzset{
Wt_lower_right_gr/.pic=
{
	\fill[color=gray!10] (1,0)..controls(0.5,0)..(0,1)--(1,1)--cycle;
	\draw (1,0)..controls(0.5,0)..(0,1)--(1,1);
}
}

\tikzset{
Wt_lower_right_wh/.pic=
{
	\fill[color=gray!10] (1,0)..controls(0.5,0)..(0,1)--(0,0)--cycle;
	\draw (1,0)..controls(0.5,0)..(0,1)--(1,1);
}
}

\tikzset{
Wt_upper_left_gr/.pic=
{
	\fill[color=gray!10] (0,1)..controls(0.5,1)..(1,0)--(0,0)--cycle;
	\draw (0,1)..controls(0.5,1)..(1,0)--(0,0);
}
}

\tikzset{
Wt_upper_left_wh/.pic=
{
	\fill[color=gray!10] (0,1)..controls(0.5,1)..(1,0)--(1,1)--cycle;
	\draw (0,1)..controls(0.5,1)..(1,0)--(0,0);
}
}

\tikzset{
Wt_upper_right_gr/.pic=
{
	\fill[color=gray!10] (1,1)..controls(0.5,1)..(0,0)--(1,0)--cycle;
	\draw (1,1)..controls(0.5,1)..(0,0)--(1,0);
}
}

\tikzset{
Wt_upper_right_wh/.pic=
{
	\fill[color=gray!10] (1,1)..controls(0.5,1)..(0,0)--(0,1)--cycle;
	\draw (1,1)..controls(0.5,1)..(0,0)--(1,0);
}
}

\tikzset{
Wt_LambdaL_gr/.pic=
{
	\fill[color=gray!10] (0,0)..controls(0.5,0)..(1,1)--(0,1)--cycle;
	\draw (0,0)..controls(0.5,0)..(1,1);
}
}

\tikzset{
Wt_LambdaL_wh/.pic=
{
	\fill[color=gray!10] (0,0)..controls(0.5,0)..(1,1)--(1,0)--cycle;
	\draw (0,0)..controls(0.5,0)..(1,1);
}
}

\tikzset{
Wt_LambdaU_gr/.pic=
{
	\fill[color=gray!10] (0,0)..controls(0.5,1)..(1,1)--(0,1)--cycle;
	\draw (0,0)..controls(0.5,1)..(1,1);
}
}

\tikzset{
Wt_LambdaU_wh/.pic=
{
	\fill[color=gray!10] (0,0)..controls(0.5,1)..(1,1)--(1,0)--cycle;
	\draw (0,0)..controls(0.5,1)..(1,1);
}
}

\tikzset{
thurston_def/.pic=
{
\tikzset{
	scale=0.5, font = \small 
}

\draw[black,styleArrow] (0,3)--(3,6);
\draw[black,styleArrow] (0,3)--(3,0);
\draw[black,styleArrow] (6,3)--(3,6);
\draw[black,styleArrow] (6,3)--(3,0);

\draw[whiteCircle] (3,6) circle;
\draw[whiteCircle] (3,0) circle;
\draw[blackCircle] (0,3) circle;
\draw[blackCircle] (6,3) circle;

}
}

\tikzset{
thurston_left/.pic=
{
\tikzset{
	scale=0.5, font = \small 
}

\draw[black,styleArrowShort] (6,3)--(4.5,4.5);
\draw[black,styleArrowShort] (6,3)--(4.5,1.5);

}
}

\tikzset{
thurston_top/.pic=
{
\tikzset{
	scale=0.5, font = \small 
}


\draw[black,styleArrowShort] (1.5,1.5)--(3,0);
\draw[black,styleArrowShort] (4.5,1.5)--(3,0);

}
}

\tikzset{
thurston_right/.pic=
{
\tikzset{
	scale=0.5, font = \small 
}


\draw[black,styleArrow] (0,3)--(3,6);
\draw[black,styleArrow] (0,3)--(3,0);
\draw[black,styleArrowShort] (4.5,4.5)--(3,6);
\draw[black,styleArrowShort] (4.5,1.5)--(3,0);

\draw[whiteCircle] (3,6) circle;
\draw[whiteCircle] (3,0) circle;
\draw[blackCircle] (0,3) circle;

}
}

\tikzset{
thurston_bottom/.pic=
{
\tikzset{
	scale=0.5, font = \small 
}

\draw[black,styleArrow] (0,3)--(3,6);
\draw[black,styleArrowShort] (0,3)--(1.5,1.5);
\draw[black,styleArrow] (6,3)--(3,6);
\draw[black,styleArrowShort] (6,3)--(4.5,1.5);

\draw[whiteCircle] (3,6) circle;
\draw[blackCircle] (0,3) circle;
\draw[blackCircle] (6,3) circle;

}
}

\tikzset{
thurston_right_bottom/.pic=
{
\tikzset{
	scale=0.5, font = \small 
}

\draw[black,styleArrow] (0,3)--(3,6);
\draw[black,styleArrow] (0,3)--(1.5,1.5);
\draw[black,styleArrowShort] (4.5,4.5)--(3,6);

\draw[whiteCircle] (3,6) circle;
\draw[blackCircle] (0,3) circle;

}
}

\maketitle
\vspace{-65ex}
\begin{flushright}
  ITEP-TH-22/20
\end{flushright}
\vspace{55ex}

\begin{abstract}\vspace*{2pt}
\noindent
We notice a remarkable connection between the Bazhanov-Sergeev solution of Zamolodchikov tetrahedron equation and certain well-known cluster algebra expression. The tetrahedron transformation is then identified with a sequence of four mutations. As an application of the new formalism, we show how to construct an integrable system with the spectral curve with arbitrary symmetric Newton polygon. Finally, we embed this integrable system into the double Bruhat cell of a Poisson-Lie group, show how triangular decomposition can be used to extend our approach to the general non-symmetric Newton polygons, and prove the Lemma which classifies conjugacy classes in double affine Weyl groups of $A$-type by decorated Newton polygons.
\end{abstract}

\tableofcontents

\newpage

\section{Introduction}

In the theory of integrable systems one usually starts with the $\mathrm{RLL}$ equation
\begin{equation}
\label{eq:RLL}
R_{12} \mathcal{L}_{1,a} \mathcal{L}_{2,a}
=
\mathcal{L}_{2,a} \mathcal{L}_{1,a} R_{12}
\end{equation}
which defines the relation between the $R$-matrix $R: \V \otimes \V \to \V \otimes \V$ intertwining a pair of ``auxiliary spaces'' $\V$, and the Lax operator $\mathcal{L}: \V \otimes \F \to \V \otimes \F$, acting on the tensor product of the auxiliary space and the ``quantum'' space $\F$ of the integrable system. The $\mathrm{RLL}$ equation implies $[\mathrm{tr}_1\, \mathcal{L}_{1,a}, \mathrm{tr}_2\, \mathcal{L}_{2,a}]=0$, i.e.\ that the integrals of motion of the system commute. The renowned Yang-Baxter equation
\begin{equation}
\label{eq:YB}
R_{12} R_{13} R_{23}
=
R_{23} R_{13} R_{12}
\end{equation}
appears in this approach as the associativity condition for the braiding relations~\eqref{eq:RLL}. This condition can be formulated as an equality between two different ways of permuting the product of three Lax operators:
\begin{equation}
\mathcal{L}_{3,a}\mathcal{L}_{2,a} \mathcal{L}_{1,a}
=
\mathcal{J}_{\pm}(\mathcal{L}_{1,a}\mathcal{L}_{2,a} \mathcal{L}_{3,a}),
~~~
\mathcal{J}_{+} = \mathrm{Ad}_{R_{12}} \mathrm{Ad}_{R_{13}} \mathrm{Ad}_{R_{23}},
~~~
\mathcal{J}_{-} = \mathrm{Ad}_{R_{23}} \mathrm{Ad}_{R_{13}} \mathrm{Ad}_{R_{12}}
\end{equation}
A solution of the Yang-Baxter equation allows one to construct an integrable system, e.g.\ a spin chain. Equivalently, in a more abstract language one can use the solution to define a quasitriangular Hopf algebra, e.g.\ a quantum group.

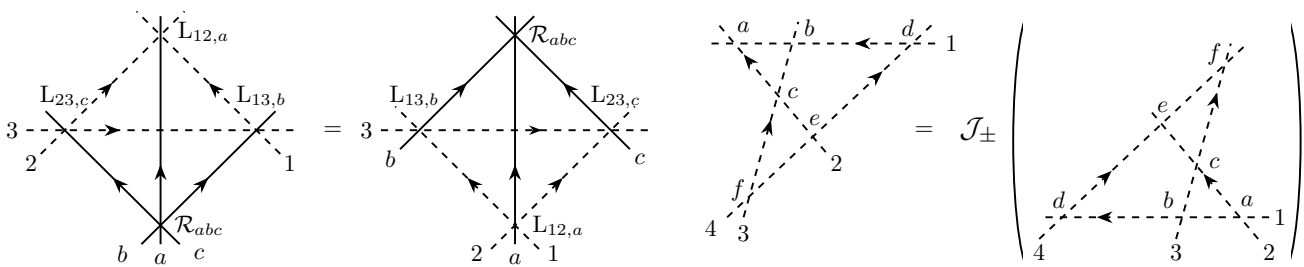
\begin{figure}[h]

\begin{center}
\scalebox{0.9}{
\begin{tikzpicture}

\tikzmath{\xs=0;\ys=0;\d=1.4;};

\draw[thick, dashed, styleArrowShifted] (\xs-1.4*\d,\ys) -- (\xs+1.4*\d,\ys);
\draw[thick, dashed, styleArrowShort] (\xs+1.25*\d,\ys-0.25*\d) -- (\xs-0.25*\d,\ys+1.25*\d);
\draw[thick, dashed, styleArrowShort] (\xs-1.25*\d,\ys-0.25*\d) -- (\xs+0.25*\d,\ys+1.25*\d);
\node at (\xs+1.35*\d,\ys-0.35*\d) {$1$};
\node at (\xs-1.35*\d,\ys-0.35*\d) {$2$};
\node at (\xs-1.55*\d,\ys) {$3$};
	
\draw[thick, styleArrowShifted] (\xs,\ys-1.2*\d) -- (\xs,\ys+1.2*\d);
\draw[thick, styleArrowShort] (\xs-0.2*\d,\ys-1.2*\d) -- (\xs+1.2*\d,\ys+0.2*\d);
\draw[thick, styleArrowShort] (\xs+0.2*\d,\ys-1.2*\d) -- (\xs-1.2*\d,\ys+0.2*\d);
\node at (\xs,\ys-1.35*\d) {$a$};
\node at (\xs-0.4*\d,\ys-1.3*\d) {$b$};
\node at (\xs+0.4*\d,\ys-1.3*\d) {$c$};	

\node at (\xs+0.4*\d,\ys-1*\d) {$\Rtet_{abc}$};
\node at (\xs+0.45*\d,\ys+1.02*\d) {$\L_{12,a}$};
\node at (\xs+1.05*\d,\ys+0.35*\d) {$\L_{13,b}$};
\node at (\xs-1*\d,\ys+0.35*\d) {$\L_{23,c}$};

\node at (1.8*\d, 0) {$=$};

\tikzmath{\xs=3.7*\d;\ys=0;};

\draw[thick, dashed, styleDirectedEdge] (\xs-1.4*\d,\ys) -- (\xs+1.4*\d,\ys);
\draw[thick, dashed, styleArrowShort] (\xs+0.25*\d,\ys-1.25*\d) -- (\xs-1.25*\d,\ys+0.25*\d);
\draw[thick, dashed, styleArrowShort] (\xs-0.25*\d,\ys-1.25*\d) -- (\xs+1.25*\d,\ys+0.25*\d);
\node at (\xs-1.55*\d,\ys) {$3$};
\node at (\xs-0.4*\d,\ys-1.3*\d) {$2$};	
\node at (\xs+0.4*\d,\ys-1.3*\d) {$1$};	
	
\draw[thick, styleArrowShifted] (\xs,\ys-1.2*\d) -- (\xs,\ys+1.2*\d);
\draw[thick, styleArrowShort] (\xs-1.2*\d,\ys-0.2*\d) -- (\xs+0.2*\d,\ys+1.2*\d);

\draw[thick, styleArrowShort] (\xs+1.2*\d,\ys-0.2*\d) -- (\xs-0.2*\d,\ys+1.2*\d);
\node at (\xs,\ys-1.35*\d) {$a$};
\node at (\xs-1.3*\d,\ys-0.3*\d) {$b$};
\node at (\xs+1.3*\d,\ys-0.3*\d) {$c$};

\node at (\xs+0.4*\d,\ys+1*\d) {$\Rtet_{abc}$};
\node at (\xs+0.45*\d,\ys-1.02*\d) {$\L_{12,a}$};
\node at (\xs-1.05*\d,\ys+0.35*\d) {$\L_{13,b}$};
\node at (\xs+1.05*\d,\ys+0.35*\d) {$\L_{23,c}$};

\end{tikzpicture}
\begin{tikzpicture}
\tikzmath{\xs=0;\ys=0;\d=1.6;};

\draw[white] (\xs-1*\d, \ys) -- (\xs-1*\d, \ys-1.2*\d);

\draw[thick, dashed, styleArrowShifted] (\xs+1.5*\d, \ys+0.6*\d) -- (\xs-0.6*\d, \ys+0.6*\d);
\draw[thick, dashed, styleArrow] (\xs+0.5*\d, \ys-0.4*\d) -- (\xs-0.5*\d, \ys+0.8*\d);
\draw[thick, dashed, styleArrowShort] (\xs-0.25*\d, \ys-\d) -- (\xs+0.25*\d, \ys+0.8*\d);
\draw[thick, dashed, styleArrow] (\xs-0.4*\d, \ys-1*\d) -- (\xs+1.5*\d, \ys+0.8*\d);

\node at (\xs+1.65*\d, \ys+0.6*\d) {$1$};
\node at (\xs+0.6*\d, \ys-0.5*\d) {$2$};
\node at (\xs-0.25*\d, \ys-1.15*\d) {$3$};
\node at (\xs-0.55*\d, \ys-1.1*\d) {$4$};

\node at (\xs-0.25*\d, \ys+0.75*\d) {$a$};
\node at (\xs+0.35*\d, \ys+0.75*\d) {$b$};
\node at (\xs+0.2*\d, \ys+0.15*\d) {$c$};
\node at (\xs+1.25*\d, \ys+0.75*\d) {$d$};
\node at (\xs+0.4*\d, \ys-0.1*\d) {$e$};
\node at (\xs-0.3*\d, \ys-0.75*\d) {$f$};


\node at (1.4*\d, \ys-0.2*\d) {$=$};

\tikzmath{\xs=3*\d;\ys=-0.4*\d;};

\node[scale=1.2] at (\xs - 1.1*\d, \ys+0.2*\d) {$\mathcal{J_{\pm}}$};

\begin{scope}
	\clip (\xs-\d,\ys+0.2*\d-1.3*\d) rectangle (\xs-0.7*\d, \ys+0.2*\d+1.3*\d);
	\draw[thick] (\xs-0.6*\d, \ys+0.1*\d) ellipse (0.3 and 2);
\end{scope}

\begin{scope}
	\clip (\xs+1.75*\d,\ys+0.2*\d-1.3*\d) rectangle (\xs+2.5*\d, \ys+0.2*\d+1.3*\d);
	\draw[thick] (\xs+1.65*\d, \ys+0.1*\d) ellipse (0.3 and 2);
\end{scope}

\tikzmath{\xs=3.95*\d;\ys=-0.4*\d;};

\draw[thick, dashed, styleArrow] (\xs+0.6*\d, \ys-0.6*\d) -- (\xs-1.5*\d, \ys-0.6*\d);
\draw[thick, dashed, styleArrowShort] (\xs+0.5*\d, \ys-0.8*\d) -- (\xs-0.5*\d, \ys+0.4*\d);
\draw[thick, dashed, styleArrow] (\xs-0.25*\d, \ys-0.8*\d) -- (\xs+0.25*\d, \ys+\d);
\draw[thick, dashed, styleArrowShifted] (\xs-1.5*\d, \ys-0.8*\d) -- (\xs+0.4*\d, \ys+1*\d);

\node at (\xs+0.7*\d, \ys-0.6*\d) {$1$};
\node at (\xs+0.6*\d, \ys-0.92*\d) {$2$};
\node at (\xs-0.25*\d, \ys-0.92*\d) {$3$};
\node at (\xs-1.5 *\d, \ys-0.92*\d) {$4$};

\node at (\xs+0.4*\d, \ys-0.45*\d) {$a$};
\node at (\xs-0.32*\d, \ys-0.44*\d) {$b$};
\node at (\xs+0.1*\d, \ys-0.1*\d) {$c$};
\node at (\xs-1.32*\d, \ys-0.44*\d) {$d$};
\node at (\xs-0.37*\d, \ys+0.42*\d) {$e$};
\node at (\xs+0.1*\d, \ys+0.9*\d) {$f$};

\end{tikzpicture}
}
\end{center}

\caption{Left. The tetrahedron equation. To view it as a modification of the Yang-Baxter equation one has to look at the transformation of the dashed triangle. Right. The functional tetrahedron equation. The quantum spaces are located in the direction transverse to the plane of the figure.}
\label{fig:tetraFig}
\end{figure}

Zamolodchikov tetrahedron equation \cite{Za1, Za2} is a natural generalization of the Yang-Baxter equation to three dimensions. While the Yang-Baxter equation is an equation on operators corresponding to crossings of lines in a plane, the tetrahedron equation describes triple crossings of planes in a $3d$ space. An analog of the $\mathrm{RLL}$ equation 
\begin{equation}
\label{eq:LLLR}
\L_{12,a} \L_{13,b} \L_{23,c} \Rtet_{abc}
=
\Rtet_{abc} \L_{23,c} \L_{13,b} \L_{12,a},
\end{equation}
drawn in Fig.~\ref{fig:tetraFig}, left, involves two kinds of spaces, $\F$ and $\V$, and two kinds of operators
\begin{equation}
\L: \V \otimes \V \otimes \F \to \V \otimes \V \otimes \F,
~~~
\Rtet: \F\otimes \F \otimes \F \to \F \otimes \F \otimes \F.
\end{equation}
The tetrahedron equation should lead to the structures which are no less profound and much more beautiful, compared to the Yang-Baxter equation. For example, in \cite{KV}, the tetrahedral structure was related to higher algebra and category theory. Its interpretation as a ``higher'' analogue of the Yang-Baxter equation becomes clear, if one assumes invertibility of $\Rtet_{abc}$ and multiplies both sides of the equation by $\Rtet_{abc}^{-1}$ on the right\footnote{We always assume that \(\mathcal{R}\) is invertible.}. This gives a version of Yang-Baxter equation ``up to'' conjugation, i.e.\ it is no longer an \emph{equality} between two ways to permute the $\L$-operators, but their \emph{equivalence.} Considering $\L_{ij,x}$ as a matrix acting in $\V_i\otimes \V_j$, with coefficients in the algebra $\A_x = \End (\F_x)$, we can rewrite Eq.~\eqref{eq:LLLR} as
\begin{equation}
\label{eq:BSTetra}
\L_{12}(\{v_a\})\L_{13}(\{v_b\})\L_{23}(\{v_c\})
=
\L_{23}(\{v'_c\})\L_{13}(\{v'_b\})\L_{12}(\{v'_a\}),
\end{equation}
where by $\{v_x\}$ we denote the set of generators of $\A_x$,
\begin{equation}
\label{eq:3}
\mathrm{L}_{ij,x}=\mathrm{L}_{ij}(\{v_x\}),
\end{equation} and 
\begin{equation}
\label{eq:vprime}
v'_x = \Rtet_{abc} \, v_x \, \Rtet_{abc}^{-1}
\end{equation} is the set of generators conjugated by $\Rtet_{abc}\in \A_a \otimes \A_b \otimes \A_c$. Since conjugation is an inner automorphism of the algebra, generators $v'$ satisfy the same relations as $v$, and all central functions remain unchanged.

We can apply four transformations \eqref{eq:BSTetra} to rearrange six \(\mathrm{L}\)-operators in a different way.
Moreover, there are two different ways to perform this rearrangement (denoted by \(\mathcal{J}_+\) and \(\mathcal{J}_-\)):
\begin{equation}
\begin{array}{cc}
\L_{12,a} \L_{13,b} \L_{23,c} \L_{14,d} \L_{24,e} \L_{34,f}
=
\mathcal{J}_{\pm}(\L_{34,f} \L_{24,e} \L_{14,d} \L_{23,c} \L_{13,b} \L_{12,a})
\\
\mathcal{J}_{+} = \mathrm{Ad}_{\Rtet_{abc}} \mathrm{Ad}_{\Rtet_{ade}} \mathrm{Ad}_{\Rtet_{bdf}} \mathrm{Ad}_{\Rtet_{cef}},
~~~
\mathcal{J}_{-} = \mathrm{Ad}_{\Rtet_{cef}} \mathrm{Ad}_{\Rtet_{bdf}} \mathrm{Ad}_{\Rtet_{ade}} \mathrm{Ad}_{\Rtet_{abc}}.
\end{array}
\end{equation}
See the pictorial representation in Fig.~\ref{fig:tetraFig}, right.
Statement that these two ways are equivalent gives under certain assumptions the functional tetrahedron equation \cite{S97, KKS}
\begin{equation}
\label{eq:tetraFunc}
\Rtet_{cef} \Rtet_{bdf} \Rtet_{ade} \Rtet_{abc}
=
\Rtet_{abc} \Rtet_{ade} \Rtet_{bdf} \Rtet_{cef}.
\end{equation}
The first assumption is that \eqref{eq:BSTetra} fixes \(\mathcal{R}\) uniquely up to constant, or in other words, that centralizer of \(\L_{12,a}\L_{13,b}\L_{23,c}\) in \(\mathcal{A}_a\otimes\mathcal{A}_b\otimes\mathcal{A}_c\) is trivial.
The second assumption is that centralizer of \(\L_{12,a} \L_{13,b} \L_{23,c} \L_{14,d} \L_{24,e} \L_{34,f}\) in \(\mathcal{A}_a\otimes\mathcal{A}_b\otimes\mathcal{A}_c\otimes\mathcal{A}_d\otimes\mathcal{A}_e\otimes\mathcal{A}_f\) is trivial as well.
It will become clear later that (classical limits of) these assumptions are actually satisfied for the \(\mathrm{L}\)-operators that we consider in the present paper, once we identify these 3-fold and 6-fold products with elements in the largest double Bruhat cells in \(PGL(3)\) and \(PGL(4)\), respectively.
These two assumptions are sufficient to derive \eqref{eq:tetraFunc} still up to some extra constant factor.
To prove that this factor is identity one needs either to check some matrix element, or find some extra property (for example, that traces of l.h.s. and r.h.s. are defined and non-zero).
We are not going into such details and refer to \cite{BMS08,BMS09} and references therein.


Forgetting about the space $\F$, one can consider Eq.~(\ref{eq:BSTetra}) as an equation on the $\mathrm{L}$-matrix valued in some algebra $\A$, together with an automorphism of $\A_a \otimes \A_b \otimes \A_c$. Suppose that $\A$ has classical limit to commutative Poisson algebra. Then conjugation with $\Rtet_{abc}$ has to be replaced by some canonical transformation of $\mathbb{C}[\A_a,\A_b,\A_c]$.\\

A solution of tetrahedron equation with $V=\mathbb{C}^2$ and the Lax operator valued in the $q$-oscillator algebra was found in~\cite{S, BS} and further studied in~\cite{KS}, \cite{MBS}, \cite{KOS}, \cite{BMS08}, \cite{S09}, \cite{BMS09}, \cite{BM}. We do not give the quantum solution here as we will not need it here. The classical limit of the solution is an operator $\LBS:\mathbb{C}^2 \otimes \mathbb{C}^2 ~ \to ~ \mathbb{C}^2 \otimes \mathbb{C}^2$ acting as a matrix\footnote{Note that compared with \cite{BS} we use different notation for $4\times 4$ matrices representing operators acting on $\mathbb{C}^2 \otimes \mathbb{C}^2$: indices of the first $\mathbb{C}^2$ encode the position of the $2\times 2$ block while index of the second $\mathbb{C}^2$ encodes matrix elements inside the block.}
\begin{equation}
\label{eq:LaxBS}
\LBS(x,y,\lambda,\mu)=
\left(
\begin{matrix}
1 &  &  &  \\
 & \mu k & -\lambda \mu x & \\
 & y & \lambda k & \\
 &  &  & -\lambda \mu
\end{matrix}
\right),
\end{equation}
where $k^2 = 1-xy$ and the Poisson brackets are
\begin{equation}
\label{eq:BSPoisson}
\{x,y \} = k^2, ~~~ \{x,\lambda \} = \{x,\mu \} = \{y,\lambda \} = \{y,\mu \} = 0.
\end{equation}
The Lax operator~\eqref{eq:LaxBS} satisfies the tetrahedron equation~(\ref{eq:BSTetra}) with the transformed variables being
\begin{equation}
\label{eq:BSvarTrans}
\begin{array}{c}
\begin{array}{ll}
\xa' = \kb'^{-1} \dfrac{\lb}{\lc}\left(\kc \xa - \dfrac{1}{\la \mc}\ka\xb\yc \right),
&
\ya' = \kb'^{-1} \dfrac{\lc}{\lb}\left(\kc \ya - \la \mc \ka\yb\xc \right),
\\
\xb' = \xa \xc + \dfrac{1}{\la \mc} \ka \kc \xb,
&
\yb' = \ya \yc + \la \mc \ka \kc \yb,
\\
\xa' = \kb'^{-1} \dfrac{\mb}{\ma}\left(\ka \xc - \dfrac{1}{\la \mc}\kc\ya\xb \right),
&
\yc' = \kb'^{-1} \dfrac{\ma}{\mb}\left(\ka \yc - \la \mc \kc\xa\yb \right),
\\
\ka' = \ka \dfrac{\kb}{\kb'},
&
\kc' = \kc \dfrac{\kb}{\kb'},
\end{array}
\\
\kb'^{2} = \ka^2 \kb^2 \kc^2 - 2 \ka^2 \kc^2 + \ka^2 + \kc^2 - \dfrac{\ka \kc \ya \xb \yc}{\la \mc} - \la \mc \ka \kc \xa \yb \xc, \\
\end{array}
\end{equation}
The new variables~\eqref{eq:BSvarTrans} satisfy the same Poisson brackets, so the transformation is indeed canonical.
Variables with different labels $a,b,c$ Poisson commute, and $\lambda$'s and $\mu$'s do not change under the transformation (the reason for this is that \(\lambda\) and \(\mu\) are central functions, so after quantization they will not be changed by \eqref{eq:vprime}, and so we demand that they are do not change in the classical limit as well).

By contracting $N$ Lax operators along one space, and taking the trace
\begin{equation}
\label{eq:LaxXXZ}
\mathcal{L}_{2^N} = \mathrm{Tr}_{0} \left( \L_{01,a_1} \L_{02,a_2} ... \L_{0N,a_N} \right)
\end{equation}
one gets the Lax operator with auxiliary space $(\mathbb{C}^2)^{N}$. This solution is called the ``quantum group-like'' solution, as the Lax operator is block-diagonal and preserves the decomposition $\mathcal{L}_{\mathbb{C}^{2^N}
} = \bigoplus_{k=1}^{N} \mathcal{L}_{\Lambda^k \mathbb{C}^N}$, where $\mathcal{L}_{\Lambda^k \mathbb{C}^N}$ is the Lax operator whose auxiliary space is $k$-th fundamental representation of $U_q(\mathfrak{gl}_k)$. In particular, the first non-trivial operator $\mathcal{L}_{\mathbb{C}^{N}}$ in the classical limit satisfies the $r$-matrix Poisson bracket
\begin{equation}
\label{eq:rLLintro}
\{ \mathcal{L}_{\mathbb{C}^{N}}(\lambda) , \mathcal{L}_{\mathbb{C}^{N}}(\mu) \} = [r(\lambda/\mu), \mathcal{L}_{\mathbb{C}^{N}}(\lambda) \otimes \mathcal{L}_{\mathbb{C}^{N}} (\mu)]
\end{equation}
with $r$ being the classical trigonometric $r$-matrix. The quantum version of the Lax operator satisfies the $\mathrm{RLL}$ relation with the quantum trigonometric $R$-matrix \footnote{We do not give here explicit expression for the classical $r$-matrix. Interested reader can find it e.g.\ in~\cite{GSV:2009-1}, \cite{M} or \cite{MS}.}. This implies that by multiplying such Lax operators one obtains monodromy matrix of some integrable system. This system can be identified with the $\mathrm{XXZ}$ spin chain in the $q$-oscillatory representation, or its classical limit.\\

Cluster algebras originally appeared in the theory of Lie groups and algebras (see e.g.~\cite{FZ}) and are now known to provide convenient language in the theory of integrable systems \cite{GSV:2009, GK:2011, EFS, M, GSTV:2014, FM:2014, BGM}. In the present paper we try to make a small step towards fully integrating the tetrahedron equations into the general mathematical physics context, showing how Bazhanov-Sergeev solution naturally appears in the theory of cluster integrable systems. Namely, we show that the Lax operator~(\ref{eq:LaxBS}) can be identified with the transfer matrix of paths on a four-gonal bi-coloured graph shown in Fig.~\ref{fig:fourgon}. The tetrahedron equation for such Lax operators is then translated into the equality between the transfer matrix of a graph composed from three four-gonal blocks and the result of the action of four ``spider moves'' on it (see Fig.~\ref{fig:tetrahedron}). This correspondence allows us to generalize the construction for the spectral curve of the $\mathrm{XXZ}$ chain given in~\cite{BS} to systems with arbitrary symmetric Newton polygon. We shall note here that this block and the sequence of mutations leading to tetrahedron equation already appeared in the related contexts \cite{K95, S, BS, Y16, AGPR}, however the full identification was missing.\\

We start our exposition in Section \ref{s:netw} where we give a brief recapitulation of planar networks, Poisson structure on the variables associated with paths on these networks and the transformations of the networks preserving the Poisson structure and partition function of paths. We give three-parametric family of mappings of Poisson variables corresponding to ``corner'' paths, shown in Fig.~\ref{fig:poisson}, all leading to the usual formulas for the transformations of the face variables.

Then in Section \ref{s:tetra} we show that the Lax operator~(\ref{eq:LaxBS}) coincides with the transfer matrix~(\ref{eq:LaxCL}) of non-intersecting paths on the planar network from Fig.~\ref{fig:fourgon}. We interperet the auxiliary space $\mathbb{C}^2$ in the Lax operator as a space on which the transfer matrix of paths acts. We realize the tetrahedron transformation (\ref{eq:BSTetra}) as a sequence of four spider-moves (and several two-moves) of the planar network shown in Fig.~\ref{fig:tetrahedron} and Fig.~ \ref{fig:tetraMovesDetailed}. Surprisingly, this sequence of transformations appears to be well known in cluster-algebraic literature~\cite{KP, Y16}, however it was not identified before with the Bazhanov-Sergeev solution of the tetrahedron equation. We also show that the transformation of the ``corner'' variables (\ref{eq:tetraClustGamma}) derived from the transformations of the network is consistent with those given by Eq.~(\ref{eq:BSvarTrans}).

In Section~\ref{s:clustint} we extend the construction of the Lax operator~(\ref{eq:LaxXXZ}) for the $\mathrm{XXZ}$ spin chain (which has rectangular Newton polygon) made by contraction of the ``tetrahedron'' Lax operators (\ref{eq:LaxBS}), to integrable systems with arbitrary centrally symmetric Newton polygon. Finally, we discuss this construction from the point of view of the embedding of cluster integrable system into affine group $\GLb(N)$ and extend it to non-symmetric Newton polygons. We also prove a Lemma which shows the converse: it classifies conjugacy classes in double affine Weyl group of $A$-type by Newton polygons.

\section{Perfect networks and flows on them}
\label{s:netw}

In this introductory section we recall notions of perfect networks and flows on them, construct Poisson structure on paths and discuss discrete transformations of networks preserving this structure. This will allow us to construct solution of the tetrahedron equation in Section \ref{s:tetra} and Hamiltonians of cluster integrable system with arbitrary Newton polygon in Section \ref{s:clustint}. The way of exposition, which we follow here, is a mixture of approaches from \cite{GSV:2008}, \cite{Talaska} and \cite{GK:2011}.

\subsection{Flows on perfect networks}

The main actor in considered approach to cluster integrable systems is a (planar) perfect network $N = (G,w)$ --- (planar) perfectly oriented graph in disk, with edges weight function $w$. Orientation is called perfect if all vertices of a graph can be coloured in three colours: all boundary vertices are grey (\tikz{\node[cross, very thick, gray] at (0,0) {};}), all internal vertices are either white (\tikz{\draw[whiteCircle] (0,0) circle;}) (and have exactly one outgoing edge) or black (\tikz{\draw[blackCircle] (0,0) circle;}) (and have exactly one incoming edge). We do not assume graph to be connected, however we assume that there are no 1-loops (edges going form the vertex to itself) and leaves (internal 1-valent vertices). All boundary vertices are assumed to be 1-valent. Boundary vertex with edges oriented away from it is called source. Vertex with edges oriented toward it is sink. We denote the set of sources by $I$, and the set of sinks --- by $J$. It will be useful to put additional grey vertices in the middles of internal edges, and refer to edges connecting black and white vertices with grey vertices as half-edges. We say that the vertex $v$ with the all adjacent half-edges is the fan of vertex $v$, number of half-edges in the fan is degree of the vertex and is denoted by $\deg(v)$. To each half-edge $e$ oriented from black or white to grey vertex we assign weight $w_e\in \mathbb{C}^*$, to half-edge with opposite orientation $-e$ we assign weight $w_{-e}=w_e^{-1}$. Weight of any set of oriented edges $P$ is the product of weights of all half-edges in it $w_P = \prod_{e\in P} w_e$.\\

Flow $p$ on the perfect network $N$ is the set of such non-intersecting and non-self-intersecting paths (\tikz[baseline=-0.5ex]{\draw[ultra thick,blue,opacity=1](0,0)--(0.5,0);}) oriented by $G$ that $\partial p = B - A$ with $A\subset I$, $B \subset J$, i.e. with all begin and end points belonging to the boundary. The set of all flows with starting points $A$ and end points $B$ is called $\mathcal{F}^{B}_{A}$. For example, the set of all closed non-intersecting oriented cycles on graph is $\mathcal{F}_\varnothing^\varnothing$. The sum of weights over all flows from $A$ to $B$
\begin{equation}
\Z_N(A\to B) = \sum_{p\in \mathcal{F}_A^B} w_{p}
\end{equation}
is called partition function of flows from $A$ to $B$. One can find examples of perfect networks and sets of all flows on them in Fig.~ \ref{fig:fourgon}.

The partition function is naturally multiplicative with respect to the gluing of disks: take pair of planar networks $N' = (G_1,w_1)$ and $N_2 = (G_2,w_2)$ on disks $D_1$ and $D_2$. Take intervals $\ell_1\subset \partial D_1$ containing $A_1\subset I_1, B_1\subset J_1$ at boundary of $D_1$, and $\ell_2\subset \partial D_2$ containing $A_2 \subset I_2, B_2\subset J_2$ at boundary of $D_2$. We say that perfect network $N$ in disk $D$ is the result of gluing of $N_1$ over $\ell_1$ to $N_2$ over $\ell_2$ if disks are glued $D = (D_1\sqcup D_2)/{_{\ell_1\sim\ell_2}}$ in such a way that the grey vertices from $A_1$ are glued to $B_2$, and from $B_1$ --- to $A_2$. Set of sources of $N$ is $I = (I_1\backslash A_1)\cup (I_2\backslash A_2)$ and set of sinks is $J = (J_1\backslash B_1)\cup (J_2 \backslash B_2)$. It is easy to see that partition function of flows from $A$ to $B$ on glued network $N$ is given by
\begin{equation}
\label{eq:gluedFlows}
\Z_N(A\to B) =
\mathlarger{\sum}\limits_{C \subset A_1 , E \subset B_1} 
\Z_{N_1}\left(C\cup (A\cap I_1) \to E \cup (B\cap J_1) \right)
\Z_{N_2}\left(E\cup (A\cap I_2) \to C \cup (B\cap J_2) \right),
\end{equation}
where the sum goes over all subsets of $A_1=B_2$ and $B_1=A_2$.

\begin{figure}[h!]
\begin{center}
\begin{tikzpicture}

\tikzmath{\d=2.5;};

\draw[thick, styleDirectedEdge](\d,\d)--(\d,\d+0.5*\d);
\draw[thick, styleDirectedEdge](3*\d,\d)--(3*\d,\d+0.5*\d);
\draw[thick, styleDirectedEdge](2*\d,\d+0.5*\d)--(2*\d,\d);

\tikzmath{\xshift=0;\yshift=0;};

\draw[thick, styleDirectedEdge] (\xshift+0.5*\d,\yshift)--(\xshift+1*\d,\yshift);
\draw[thick, styleDirectedEdge] (\xshift+1*\d,\yshift)--(\xshift+2*\d,\yshift);
\draw[thick, styleDirectedEdge] (\xshift+2*\d,\yshift)--(\xshift+3.0*\d,\yshift);

\draw[thick, styleDirectedEdge] (\xshift+0.5*\d,\yshift+\d)--(\xshift+1*\d,\yshift+\d);
\draw[thick, styleDirectedEdge] (\xshift+1*\d,\yshift+\d)--(\xshift+2*\d,\yshift+\d);
\draw[thick, styleDirectedEdge] (\xshift+2*\d,\yshift+\d)--(\xshift+3.0*\d,\yshift+\d);

\draw[thick, styleDirectedEdge] (\xshift+\d,\yshift+\d)--(\xshift+\d,\yshift+0.5*\d);
\draw[thick, styleDirectedEdge] (\xshift+\d,\yshift+0.5*\d)--(\xshift+\d,\yshift);
\node[cross, very thick, gray] at (\xshift+\d,\yshift+0.5*\d) {};
\draw[thick, styleDirectedEdge] (\xshift+2*\d,\yshift+0.5*\d)--(\xshift+2*\d,\yshift+\d);
\node[cross, very thick, gray] at (\xshift+2*\d,\yshift+0.5*\d) {};
\draw[thick, styleDirectedEdge] (\xshift+2*\d,\yshift)--(\xshift+2*\d,\yshift+0.5*\d);

\draw[blackCircle] (\xshift+\d,\yshift+\d) circle;
\draw[whiteCircle] (\xshift+\d,\yshift+0) circle;
\draw[blackCircle] (\xshift+2*\d,\yshift+0) circle;
\draw[whiteCircle] (\xshift+2*\d,\yshift+\d) circle;







\tikzmath{\xshift=\d;\yshift=-1*\d;};

\draw[thick, styleDirectedEdge] (\xshift-0.5*\d,\yshift)--(\xshift+1*\d,\yshift);
\draw[thick, styleDirectedEdge] (\xshift+1*\d,\yshift)--(\xshift+2*\d,\yshift);
\draw[thick, styleDirectedEdge] (\xshift+2*\d,\yshift)--(\xshift+3.5*\d,\yshift);


\draw[thick, styleDirectedEdge] (\xshift+\d,\yshift+\d)--(\xshift+\d,\yshift);
\draw[thick, styleDirectedEdge] (\xshift+2*\d,\yshift)--(\xshift+2*\d,\yshift+\d);

\draw[whiteCircle] (\xshift+\d,\yshift+0) circle;
\draw[blackCircle] (\xshift+2*\d,\yshift+0) circle;

\tikzmath{\xshift=2*\d;\yshift=0;\op=0.8;};

\draw[thick, styleDirectedEdge] (\xshift+1*\d,\yshift)--(\xshift+2*\d,\yshift);
\draw[thick, styleDirectedEdge] (\xshift+2*\d,\yshift)--(\xshift+2.5*\d,\yshift);

\draw[thick, styleDirectedEdge] (\xshift+1*\d,\yshift+\d)--(\xshift+2*\d,\yshift+\d);
\draw[thick, styleDirectedEdge] (\xshift+2*\d,\yshift+\d)--(\xshift+2.5*\d,\yshift+\d);

\draw[thick, styleDirectedEdge] (\xshift+\d,\yshift+\d)--(\xshift+\d,\yshift+0.5*\d);
\draw[thick, styleDirectedEdge] (\xshift+\d,\yshift+0.5*\d)--(\xshift+\d,\yshift);
\node[cross, very thick, gray] at (\xshift+\d,\yshift+0.5*\d) {};
\draw[thick, styleDirectedEdge] (\xshift+2*\d,\yshift+0.5*\d)--(\xshift+2*\d,\yshift+\d);
\node[cross, very thick, gray] at (\xshift+2*\d,\yshift+0.5*\d) {};
\draw[thick, styleDirectedEdge] (\xshift+2*\d,\yshift)--(\xshift+2*\d,\yshift+0.5*\d);

\draw[blackCircle] (\xshift+\d,\yshift+\d) circle;
\draw[whiteCircle] (\xshift+\d,\yshift+0) circle;
\draw[blackCircle] (\xshift+2*\d,\yshift+0) circle;
\draw[whiteCircle] (\xshift+2*\d,\yshift+\d) circle;

\node[cross, very thick, gray] at (0.5*\d,0*\d) {};
\node[cross, very thick, gray] at (0.5*\d,\d) {};
\node[cross, very thick, gray] at (0.5*\d,-\d) {};

\node[cross, very thick, gray] at (4.5*\d,0*\d) {};
\node[cross, very thick, gray] at (4.5*\d,\d) {};
\node[cross, very thick, gray] at (4.5*\d,-\d) {};

\node[cross, very thick, gray] at (\d,1.5*\d) {};
\node[cross, very thick, gray] at (2*\d,1.5*\d) {};
\node[cross, very thick, gray] at (3*\d,1.5*\d) {};

\node at(0.5*\d-0.3,1*\d){\(1\)};
\node at(1*\d,1.5*\d+0.4){\(2\)};
\node at(2*\d,1.5*\d+0.4){\(3\)};
\node at(3*\d,1.5*\d+0.4){\(4\)};
\node at(4.5*\d+0.3,1*\d){\(5\)};
\node at(4.5*\d+0.3,0*\d){\(6\)};
\node at(4.5*\d+0.3,-\d){\(7\)};
\node at(0.5*\d-0.3,-\d){\(8\)};
\node at(0.5*\d-0.3,0*\d){\(9\)};
\node at(\d+0.3,0.5*\d+0.3){\(10\)};
\node at(2*\d+0.3,0.5*\d+0.3){\(11\)};
\node at(3*\d+0.3,0.5*\d+0.3){\(12\)};
\node at(4*\d+0.3,0.5*\d+0.3){\(13\)};

\draw[dashed](0.5*\d,0.5*\d)--(4.5*\d,0.5*\d);
\draw[dashed,rounded corners=10mm](0.5*\d,0.5*\d)--++(0,\d)--++(4*\d,0)--+(0,-\d);
\draw[dashed,rounded corners=10mm](0.5*\d,0.5*\d)--++(0,-2*\d)--++(4*\d,0)--+(0,2*\d);

\draw[ultra thick,blue,opacity=\op](0.5*\d,\d)--++(0.5*\d,0)--++(0,-\d)--++(\d,0)--++(0,-\d)--++(2.5*\d,0);

\draw[ultra thick,blue,opacity=\op](2*\d,1.5*\d)--++(0,-0.5*\d)--++(\d,0)--++(0,-\d)--++(\d,0)--++(0,\d)--++(0.5*\d,0);

\node[draw,circle,dashed] at (0.4*\d,1.4*\d){\(D_1\)};
\node[draw,circle,dashed] at (0.4*\d,-1.4*\d){\(D_2\)};
\node at(0.5*\d-0.6,0.5*\d){\(\ell_1\sim\ell_2\)};

\end{tikzpicture}
\end{center}
\caption{\label{fig:gluing}Gluing of two planar networks.}
\end{figure}
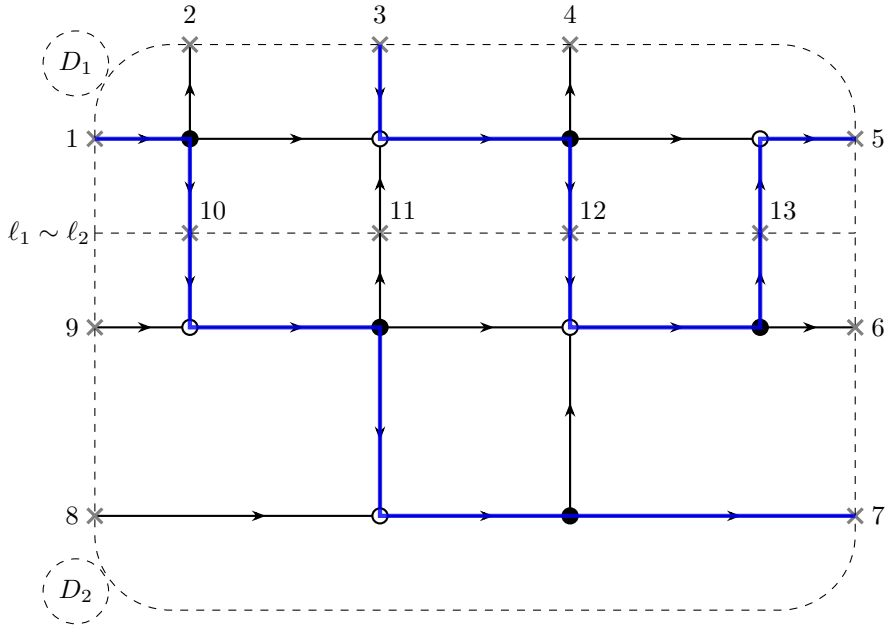

Consider corresponding subsets in example of two planar networks glued together in Figure \ref{fig:gluing}. Sets that depend on planar networks only are
\(I=\{1,3,8\}\) (all sources in \(\partial D\)),
\(J=\{2,4,5,6,7\}\) (all sinks in \(\partial D\)),
\(I_1=\{1,3,13,11\}\) (all sources in \(\partial D_1\)),
\(J_1=\{2,4,5,12,10\}\) (all sinks in \(\partial D_1\)),
\(I_2=\{8,9,10,12\}\) (all sources in \(\partial D_2\)),
\(J_2=\{11,13,6,7\}\) (all sinks in \(\partial D_2\)),
\(A_1=B_2=\{11,13\}\) (passages from \(D_2\) to \(D_1\)),
\(A_2=B_1=\{10,12\}\) (passages from \(D_1\) to \(D_2\)). The particular flow (denoted by \tikz[baseline=-0.5ex]{\draw[ultra thick,blue,opacity=1](0,0)--(0.5,0);}) determines sets
\(A=\{1,3\}\) (starting points of the flow),
\(B=\{5,7\}\) (endpoints of the flow),
\(C=\{13\}\) (passages where flow is allowed to go from \(D_2\) to \(D_1\)),
\(E=\{10,12\}\) (passages where flow is allowed to go form \(D_1\) to \(D_2\)).
This is a single term in summation which goes over all possible subsets \(C\subset A_1=B_2\) and \(E\subset B_1=A_2\), as we sum over all possible flows in disk \(D\).

Formula \eqref{eq:gluedFlows} can be conveniently encoded using transfer matrix of flows $T_N$.
This is an operator $T_N:\, (\mathbb{C}^2)^{\otimes |I|} \to (\mathbb{C}^2)^{\otimes |J|}$ given by
\begin{equation}
T_N =
\mathlarger{\sum}_{A \subset I , B \subset J}
\Z_N(A\to B)  \cdot
\bigotimes\limits_{j\in J} e_{s(j,B),\,j}
\otimes
\bigotimes\limits_{i\in I} e^*_{s(i,A),\,i},
\end{equation}
where $s(k,X) = + \text{ if } k \in X$ and $s(k,X) = - \text{ if } k \notin X$. The vectors $e_{\pm,\,j}$ are basis vectors in $j$-th component of $(\mathbb{C}^2)^{\otimes |J|}$, vectors $e^*_{\pm,\,i}$ are basis co-vectors in $i$-th component of $(\mathbb{C}^2)^{\otimes |I|}$. Using this operator (\ref{eq:gluedFlows}) becomes simply
\begin{equation}
T_N = T_{N_1}\circ T_{N_2},
\end{equation}
where spaces with labels $A_1$ contract with corresponding spaces in $B_1$, and the same for $A_2$ and $B_2$. Transfer matrices for perfect networks drawn in Fig.~\ref{fig:fourgon} are written in (\ref{eq:LaxCL}).

\paragraph{Remark.}{For the reader, familiar with combinatorics of dimers, we note that there is a bijection between bipartite graphs without two-valent vertices with selected perfect matching $\mathsf{D}_0$, and perfect networks without neighbouring vertices of the same colour. The bijection can be established by choosing orientation on the bipartite graph from black to white for the edges not in $\mathsf{D}_0$, and from white to black for those in $\mathsf{D}_0$. There is also similar bijection between perfect matchings on bipartite graphs and flows on perfect networks.}

\subsection{Poisson structure on paths and $\mathcal{X}$-cluster variety}
\label{ss:poissonXclust}
\begin{figure}[!ht]
\begin{center}
\begin{tikzpicture}

\tikzmath{\xs=0;};
\draw[thick] (\xs,0)--(\xs-0.87,0.5);
\draw[blackCircle] (\xs+0,0) circle;
\draw[<-, very thick, blue] (\xs,-1)--(\xs,0)--(\xs+0.87,0.5);

\node at (\xs,-1.5) {$\gamma^*_1$};

\tikzmath{\xs=2.5;};
\draw[thick] (\xs,0)--(\xs+0.87,0.5);
\draw[blackCircle] (\xs+0,0) circle;
\draw[->, very thick, blue] (\xs,-1)--(\xs,0)--(\xs-0.87,0.5);
\node at (\xs,-1.5) {$\gamma^*_2$};

\tikzmath{\xs=5;};
\draw[thick] (\xs,-1)--(\xs,-0.1);
\draw[blackCircle] (\xs+0,0) circle;
\draw[<-, very thick, blue] (\xs+0.87,0.5)--(\xs,0)--(\xs-0.87,0.5);
\node at (\xs,-1.5) {$\gamma^*_3$};

\node at (8,0.7) {$\delta_v({\gamma^*_1},{\gamma^*_2})=\dfrac{1}{2}$};
\node at (8,-0.3) {$\delta_v({\gamma_2^*},{\gamma_3^*})=\dfrac{1}{2}$};
\node at (8,-1.3) {$\delta_v({\gamma^*_3},{\gamma^*_1})=\dfrac{1}{2}$};
\end{tikzpicture}
\end{center}
\caption{Definition of the local pairing on paths at the three-valent vertex, $\gamma_1^*+\gamma_2^* = -\gamma_3^*$. Simple corners are shown by blue.}
\label{fig:poisson}
\end{figure}
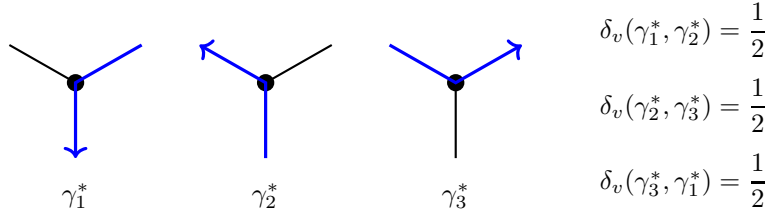

There is a two-parametric family of Poisson brackets on the weights of half-edges, see \cite{GSV:2008}. Here we will use, however, 1-parametric specialization of it restricted to the paths connecting middles of the edges\footnote{The latter can be obtained from the former using procedure of the gauge symmetry reduction in black and white vertices.} considered in \cite{GK:2011}. Any path $p$ on perfect network $N = (G,w)$ which begins and ends at the grey vertices (in the middles of edges or at boundary points) can be decomposed into sum of contributions associated with the fans of internal vertices
\begin{equation}
\label{eq:lattice}
p = \sum\limits_{v\in C_0(G)} n_{i,v} \gamma^*_{v,i}
\end{equation}
where $C_0(G)$ is the set of internal vertices of $G$. Generators $\gamma^*_{v,i}$ are called simple corners and are associated with paths which go through $v$ and connect middles of adjusted edges in the clockwise order, see Fig.~\ref{fig:poisson} for example. They satisfy relation $\sum_{i=1}^{\mathrm{deg}\, v} \gamma^*_{v,i} =0$ which means that by traversing all simple corners associated with one vertex we get trivial path. The logarithmically constant Poisson bracket on weights of paths is
\begin{equation}
\label{eq:poissonClust}
\{w_{p_1}, w_{p_2}\} = \varepsilon(p_1,p_2)w_{p_1}w_{p_2},
\end{equation}
where the skew-linear form $\varepsilon$ is defined as sum of local contributions of each fan
\begin{equation}
\begin{array}{l}
\varepsilon(p_1,p_2) = \mathlarger{\sum}\limits_{v\in C_0(G)} \sgn(v) \delta_v(p_1,p_2), \\
\delta_u(\gamma_{v,i}^*,\gamma_{w,j}^*)=
\left\{\begin{array}{ll}
\pm \dfrac{1}{2}\delta_{u,v}\delta_{u,w} & \text{ if } j=i\pm 1 \\
0 & \text{ otherwise}
\end{array}\right. ,
\end{array}
\end{equation}
where $\sgn(v)=1$ for the black vertices and $\sgn(v)=-1$ for the white. Example of pairing at three-valent black vertex is shown in Fig.~\ref{fig:poisson}. In fact, bracket can be defined by extending it from the bracket on weights of simple corners $\gamma_{v,i} = w_{\gamma^*_{v,i}}$. Thanks to the local structure of the bracket, the gluing of perfect networks is Poisson map, as it was shown in \cite{GSV:2008}. There is an opposite operation of splitting network $N = (G, w)$ on $D$ to $N_1 = (G_1,w_1)$ and $N_2 = (G_2,w_2)$ by cutting $D$ into $D_1$ and $D_2$ along some simple curve, which intersects $G$ only at middles of edges and divide grey vertices into pairs of vertices belonging to different networks. It is not uniquely defined, if only weights of paths connecting boundary vertices of $D$ are known, because of the gauge redundancy under transformations at internal grey vertices, which multiply weights of all paths ending at internal grey vertex $v$ by $x_v\in\mathbb{C}^*$, and all paths starting at $v$ by $x_v^{-1}$. We will face this problem again in Section \ref{ss:tetra}.

The weight of any flow on planar network can be expressed using only the weights of oriented boundaries of faces $\x_i = w_{(\partial \bar{f}_i \cap G)}$. Faces are defined from decomposition $D \backslash G = \bigcup_i f_i$. Note that for unbounded disks (adjacent to $\partial D$) we take only parts belonging to $G$. The face variables $\x_i$ satisfy single relation $\prod_{i} \x_i = 1$, as each edge of $G$ belongs to the boundaries of exactly two faces with the opposite orientations.

Space of face weights admits structure of the toric chart in the $\mathcal{X}$-cluster variety. This means that it is algebraic torus with coordinate functions $\x_i$ satisfying log-constant Poisson bracket
\begin{equation}
\{ \x_i, \x_j \} = \varepsilon_{ij}\, \x_i \x_j
\end{equation}
with some skew-symmetric matrix $\varepsilon$ called exchange matrix. We say that $\x_i$ are $\mathcal{X}$-cluster variables, and those $\x_i$ which come from faces adjacent to $\partial{D}$ are frozen variables. Exchange matrix $\varepsilon$ for perfect networks follows from (\ref{eq:poissonClust}). It is convenient to represent $\varepsilon$ as oriented graph (quiver) with edges with multiplicities, whose oriented adjacency matrix is $\varepsilon$ and vertices correspond to $\x_i$, see examples in Fig.~\ref{fig:fourumove} and \ref{fig:tetrahedron}. Toric charts are glued by transformations of mutations in directions of non-frozen variables $\x_i$. Mutation \(\mu_i\) in direction of variable $\x_i$ is defined by the action
\begin{equation}
\label{eq:mutX}
\mu_{i}(\x_j) =
\left\{
\begin{array}{ll}
\x_j^{-1}, &
i = j
\\
\x_j (1+\x_i^{\sgn\, \varepsilon_{ji}})^{\varepsilon_{ji}}, &
i \neq j
\end{array}
\right.
~~~
\mu_{i}(\varepsilon_{kl}) =
\left\{
\begin{array}{ll}
-\varepsilon_{kl}, &
\text{if} ~ i=k ~ \text{or} ~ {i=l}
\\
\varepsilon_{kl} + \dfrac{|\varepsilon_{ki}|\varepsilon_{il} + \varepsilon_{ki}|\varepsilon_{il}|}{2}, &
\text{otherwise}
\end{array}
\right.
\end{equation}
on cluster variables and exchange matrix. We will call $\mathcal{X}$-cluster variety of face variables of graph $G$ by $\XG$. Realization of mutations as transformations of perfect network will be discussed in the next subsection.

Operation of gluing of the disks results in the product of $\mathcal{X}$-cluster varieties with amalgamation, for details see \cite{FG:2005}. In simple words one has to replace pair of frozen variables corresponding to two unbounded faces, which are glued to one bounded, by the new unfrozen variable (which equals to the product of initial variables), and obtain new exchange matrix from the glued graph. From the point of view of quivers, product with amalgamation is gluing of quivers by vertices corresponding to frozen variables. 

\subsection{Plabic graph transformations}
\label{ss:transform}
\begin{figure}[h!]
\begin{center}
\scalebox{0.9}{
\begin{tikzpicture}

\tikzmath{\xshift=0;\yshift=0;};

\draw[thick] (\xshift-1,\yshift+2)--(\xshift,\yshift+1)--(\xshift+1,\yshift+2);
\draw[thick] (\xshift,\yshift+1)--(\xshift,\yshift-1);
\draw[thick] (\xshift-1,\yshift-2)--(\xshift,\yshift-1)--(\xshift+1,\yshift-2);

\draw[blackCircle] (\xshift,\yshift+1) circle;
\draw[blackCircle] (\xshift,\yshift-1) circle;


\draw[very thick, \colA, ->] (\xshift-0.5,\yshift+1.8) -- (\xshift,\yshift+1.3) -- (\xshift+0.5,\yshift+1.8);
\draw[very thick, \colB, ->] (\xshift+0.8,\yshift+1.5) -- (\xshift+0.2,\yshift+0.9) -- (\xshift+0.2,\yshift+0.3);
\draw[very thick, \colC, ->] (\xshift-0.2,\yshift+0.3) -- (\xshift-0.2,\yshift+0.9) -- (\xshift-0.8,\yshift+1.5);

\node at (\xshift,\yshift+2) {$t_1$};
\node at (\xshift+0.7,\yshift+0.8) {$t_2$};
\node at (\xshift-0.7,\yshift+0.8) {$t_3$};

\draw[very thick, \colA, ->] (\xshift+0.5,\yshift-1.8) -- (\xshift,\yshift-1.3) -- (\xshift-0.5,\yshift-1.8);
\draw[very thick, \colB, ->] (\xshift+0.2,\yshift-0.3) -- (\xshift+0.2,\yshift-0.9) -- (\xshift+0.8,\yshift-1.5);
\draw[very thick, \colC, ->] (\xshift-0.8,\yshift-1.5) -- (\xshift-0.2,\yshift-0.9) -- (\xshift-0.2,\yshift-0.3);

\node at (\xshift,\yshift-2) {$b_1$};
\node at (\xshift+0.7,\yshift-0.8) {$b_3$};
\node at (\xshift-0.7,\yshift-0.8) {$b_2$};

\draw[very thick, ->] (1.3,0) -- (2.3,0);

\tikzmath{\yshift=5;\xshift=0;};

\draw[thick] (\yshift+2,\xshift-1)--(\yshift+1, \xshift)--(\yshift+2, \xshift+1);
\draw[thick] (\yshift+1, \xshift)--(\yshift-1, \xshift);
\draw[thick] (\yshift-2, \xshift-1)--(\yshift-1, \xshift)--(\yshift-2, \xshift+1);

\draw[blackCircle] (\yshift+1, \xshift) circle;
\draw[blackCircle] (\yshift-1, \xshift) circle;


\draw[very thick, \colA, ->] (\yshift+1.8, \xshift+0.5) -- (\yshift+1.3, \xshift) -- (\yshift+1.8, \xshift-0.5);
\draw[very thick, \colB, ->] (\yshift+0.3, \xshift+0.2) -- (\yshift+0.9, \xshift+0.2) -- (\yshift+1.5, \xshift+0.8);
\draw[very thick, \colC, ->] (\yshift+1.5, \xshift-0.8) -- (\yshift+0.9, \xshift-0.2) -- (\yshift+0.3, \xshift-0.2);

\node at (\yshift+2, \xshift) {$r'_1$};
\node at (\yshift+0.8, \xshift+0.7) {$r'_3$};
\node at (\yshift+0.8, \xshift-0.7) {$r'_2$};

\draw[very thick, \colA, ->] (\yshift-1.8, \xshift-0.5) -- (\yshift-1.3, \xshift) -- (\yshift-1.8, \xshift+0.5);
\draw[very thick, \colB, ->] (\yshift-1.5, \xshift+0.8) -- (\yshift-0.9, \xshift+0.2) -- (\yshift-0.3, \xshift+0.2);
\draw[very thick, \colC, ->] (\yshift-0.3, \xshift-0.2) -- (\yshift-0.9, \xshift-0.2) -- (\yshift-1.5, \xshift-0.8);

\node at (\yshift-2, \xshift) {$l'_1$};
\node at (\yshift-0.8, \xshift+0.7) {$l'_2$};
\node at (\yshift-0.8, \xshift-0.7) {$l'_3$};
\end{tikzpicture}
\begin{tikzpicture}
\draw[white] (-0.5,0)--(0.5,0);
\end{tikzpicture}
\begin{tikzpicture}

\tikzmath{\xshift=0;\yshift=0;};

\draw[thick] (\xshift-1,\yshift+2)--(\xshift,\yshift+1)--(\xshift+1,\yshift+2);
\draw[thick] (\xshift,\yshift+1)--(\xshift,\yshift-1);
\draw[thick] (\xshift-1,\yshift-2)--(\xshift,\yshift-1)--(\xshift+1,\yshift-2);

\draw[whiteCircle] (\xshift,\yshift+1) circle;
\draw[whiteCircle] (\xshift,\yshift-1) circle;


\draw[very thick, \colA, ->] (\xshift-0.5,\yshift+1.8) -- (\xshift,\yshift+1.3) -- (\xshift+0.5,\yshift+1.8);
\draw[very thick, \colB, ->] (\xshift+0.8,\yshift+1.5) -- (\xshift+0.2,\yshift+0.9) -- (\xshift+0.2,\yshift+0.3);
\draw[very thick, \colC, ->] (\xshift-0.2,\yshift+0.3) -- (\xshift-0.2,\yshift+0.9) -- (\xshift-0.8,\yshift+1.5);

\node at (\xshift,\yshift+2) {$t_1$};
\node at (\xshift+0.7,\yshift+0.8) {$t_3$};
\node at (\xshift-0.7,\yshift+0.8) {$t_2$};

\draw[very thick, \colA, ->] (\xshift+0.5,\yshift-1.8) -- (\xshift,\yshift-1.3) -- (\xshift-0.5,\yshift-1.8);
\draw[very thick, \colB, ->] (\xshift+0.2,\yshift-0.3) -- (\xshift+0.2,\yshift-0.9) -- (\xshift+0.8,\yshift-1.5);
\draw[very thick, \colC, ->] (\xshift-0.8,\yshift-1.5) -- (\xshift-0.2,\yshift-0.9) -- (\xshift-0.2,\yshift-0.3);

\node at (\xshift,\yshift-2) {$b_1$};
\node at (\xshift+0.7,\yshift-0.8) {$b_2$};
\node at (\xshift-0.7,\yshift-0.8) {$b_3$};

\draw[very thick, ->] (1.3,0) -- (2.3,0);

\tikzmath{\yshift=5;\xshift=0;};

\draw[thick] (\yshift+2,\xshift-1)--(\yshift+1, \xshift)--(\yshift+2, \xshift+1);
\draw[thick] (\yshift+1, \xshift)--(\yshift-1, \xshift);
\draw[thick] (\yshift-2, \xshift-1)--(\yshift-1, \xshift)--(\yshift-2, \xshift+1);

\draw[whiteCircle] (\yshift+1, \xshift) circle;
\draw[whiteCircle] (\yshift-1, \xshift) circle;


\draw[very thick, \colA, ->] (\yshift+1.8, \xshift+0.5) -- (\yshift+1.3, \xshift) -- (\yshift+1.8, \xshift-0.5);
\draw[very thick, \colB, ->] (\yshift+0.3, \xshift+0.2) -- (\yshift+0.9, \xshift+0.2) -- (\yshift+1.5, \xshift+0.8);
\draw[very thick, \colC, ->] (\yshift+1.5, \xshift-0.8) -- (\yshift+0.9, \xshift-0.2) -- (\yshift+0.3, \xshift-0.2);

\node at (\yshift+2, \xshift) {$r'_1$};
\node at (\yshift+0.8, \xshift+0.7) {$r'_2$};
\node at (\yshift+0.8, \xshift-0.7) {$r'_3$};

\draw[very thick, \colA, ->] (\yshift-1.8, \xshift-0.5) -- (\yshift-1.3, \xshift) -- (\yshift-1.8, \xshift+0.5);
\draw[very thick, \colB, ->] (\yshift-1.5, \xshift+0.8) -- (\yshift-0.9, \xshift+0.2) -- (\yshift-0.3, \xshift+0.2);
\draw[very thick, \colC, ->] (\yshift-0.3, \xshift-0.2) -- (\yshift-0.9, \xshift-0.2) -- (\yshift-1.5, \xshift-0.8);

\node at (\yshift-2, \xshift) {$l'_1$};
\node at (\yshift-0.8, \xshift+0.7) {$l'_3$};
\node at (\yshift-0.8, \xshift-0.7) {$l'_2$};

\end{tikzpicture}
}
\end{center}
\caption{Transformations of the plabic graph under the two-moves at black and white vertices.}
\label{fig:twomove}
\end{figure}
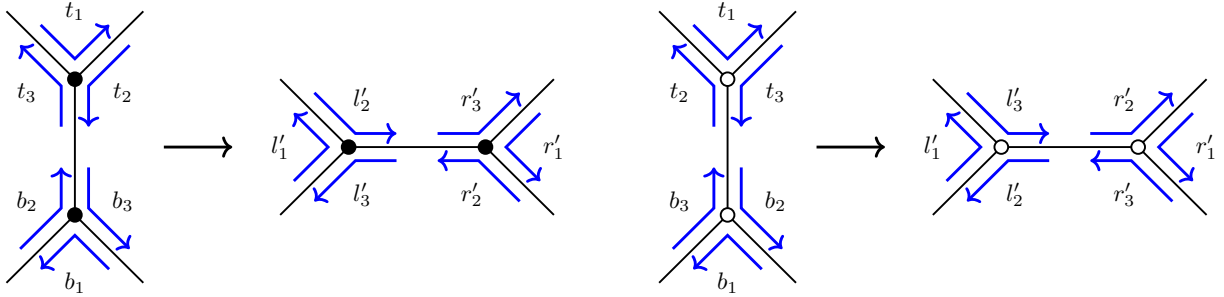

\begin{figure}
\begin{center}
\scalebox{0.9}{
\begin{tikzpicture}

\tikzmath{\xshift=0;\yshift=0;};

\draw[thick] (\xshift-1.5,\yshift+1.5)--(\xshift-0.75,\yshift+0.75)--(\xshift-0.75,\yshift-0.75)--(\xshift-1.5,\yshift-1.5);

\draw[thick] (\xshift+1.5,\yshift+1.5)--(\xshift+0.75,\yshift+0.75)--(\xshift+0.75,\yshift-0.75)--(\xshift+1.5,\yshift-1.5);

\draw[thick] (\xshift+0.75,\yshift+0.75)--(\xshift-0.75,\yshift+0.75);
\draw[thick] (\xshift+0.75,\yshift-0.75)--(\xshift-0.75,\yshift-0.75);

\draw[whiteCircle] (\xshift+0.75,\yshift+0.75) circle;
\draw[blackCircle] (\xshift-0.75,\yshift+0.75) circle;
\draw[whiteCircle] (\xshift-0.75,\yshift-0.75) circle;
\draw[blackCircle] (\xshift+0.75,\yshift-0.75) circle;

\draw[very thick, \colA, ->] (\xshift-0.15,\yshift+0.55) -- (\xshift-0.55,\yshift+0.55) -- (\xshift-0.55,\yshift+0.15);
\draw[very thick, \colB, ->] (\xshift- 0.95,\yshift+1.25) -- (\xshift-0.65,\yshift+0.95) -- (\xshift-0.15,\yshift+0.95);
\draw[very thick, \colC, ->] (\xshift-0.95,\yshift+0.15) -- (\xshift-0.95,\yshift+0.65) -- (\xshift-1.25,\yshift+0.95);

\draw[very thick, \colA, ->] (\xshift+0.55,\yshift+0.15) -- (\xshift+0.55,\yshift+0.55) -- (\xshift+0.15,\yshift+0.55);
\draw[very thick, \colB, ->] (\xshift+1.25,\yshift+0.95) -- (\xshift+0.95,\yshift+0.65) -- (\xshift+0.95,\yshift+0.15);
\draw[very thick, \colC, ->] (\xshift+0.15,\yshift+0.95) -- (\xshift+0.65,\yshift+0.95) -- (\xshift+0.95,\yshift+1.25);

\draw[very thick, \colA, ->] (\xshift+0.15,\yshift-0.55) -- (\xshift+0.55,\yshift-0.55) -- (\xshift+0.55,\yshift-0.15);
\draw[very thick, \colB, ->] (\xshift+0.95,\yshift-1.25) -- (\xshift+0.65,\yshift-0.95) --(\xshift+0.15,\yshift-0.95);
\draw[very thick, \colC, ->] (\xshift+0.95,\yshift-0.15) -- (\xshift+0.95,\yshift-0.65) -- (\xshift+1.25,\yshift-0.95);

\draw[very thick, \colA, ->] (\xshift-0.55,\yshift-0.15) -- (\xshift-0.55,\yshift-0.55) -- (\xshift-0.15,\yshift-0.55);
\draw[very thick, \colB, ->] (\xshift-1.25,\yshift-0.95) -- (\xshift-0.95,\yshift-0.65) -- (\xshift-0.95,\yshift-0.15);
\draw[very thick, \colC, ->] (\xshift-0.15,\yshift-0.95) -- (\xshift-0.65,\yshift-0.95) --(\xshift-0.95,\yshift-1.25);

\node at (\xshift-0.2,\yshift+0.3) {$a_1$};
\node at (\xshift-1.2,\yshift+0.5) {$a_2$};
\node at (\xshift-0.5,\yshift+1.2) {$a_3$};

\node at (\xshift+0.3,\yshift+0.3) {$b_1$};
\node at (\xshift+0.5,\yshift+1.2) {$b_3$};
\node at (\xshift+1.2,\yshift+0.5) {$b_2$};

\node at (\xshift+0.3,\yshift-0.35) {$c_1$};
\node at (\xshift+1.2,\yshift-0.5) {$c_2$};
\node at (\xshift+0.5,\yshift-1.2) {$c_3$};

\node at (\xshift-0.2,\yshift-0.3) {$d_1$};
\node at (\xshift-0.5,\yshift-1.2) {$d_3$};
\node at (\xshift-1.2,\yshift-0.5) {$d_2$};

\draw[very thick, ->] (2.1,0)--(3.1,0);

\tikzmath{\xshift=5;\yshift=0;};

\draw[thick] (\xshift-1.5,\yshift+1.5)--(\xshift-0.75,\yshift+0.75)--(\xshift-0.75,\yshift-0.75)--(\xshift-1.5,\yshift-1.5);

\draw[thick] (\xshift+1.5,\yshift+1.5)--(\xshift+0.75,\yshift+0.75)--(\xshift+0.75,\yshift-0.75)--(\xshift+1.5,\yshift-1.5);

\draw[thick] (\xshift+0.75,\yshift+0.75)--(\xshift-0.75,\yshift+0.75);
\draw[thick] (\xshift+0.75,\yshift-0.75)--(\xshift-0.75,\yshift-0.75);

\draw[blackCircle] (\xshift+0.75,\yshift+0.75) circle;
\draw[whiteCircle] (\xshift-0.75,\yshift+0.75) circle;
\draw[blackCircle] (\xshift-0.75,\yshift-0.75) circle;
\draw[whiteCircle] (\xshift+0.75,\yshift-0.75) circle;

\draw[very thick, \colA, ->] (\xshift-0.15,\yshift+0.55) -- (\xshift-0.55,\yshift+0.55) -- (\xshift-0.55,\yshift+0.15);
\draw[very thick, \colB, ->] (\xshift- 0.95,\yshift+1.25) -- (\xshift-0.65,\yshift+0.95) -- (\xshift-0.15,\yshift+0.95);
\draw[very thick, \colC, ->] (\xshift-0.95,\yshift+0.15) -- (\xshift-0.95,\yshift+0.65) -- (\xshift-1.25,\yshift+0.95);

\draw[very thick, \colA, ->] (\xshift+0.55,\yshift+0.15) -- (\xshift+0.55,\yshift+0.55) -- (\xshift+0.15,\yshift+0.55);
\draw[very thick, \colB, ->] (\xshift+1.25,\yshift+0.95) -- (\xshift+0.95,\yshift+0.65) -- (\xshift+0.95,\yshift+0.15);
\draw[very thick, \colC, ->] (\xshift+0.15,\yshift+0.95) -- (\xshift+0.65,\yshift+0.95) -- (\xshift+0.95,\yshift+1.25);

\draw[very thick, \colA, ->] (\xshift+0.15,\yshift-0.55) -- (\xshift+0.55,\yshift-0.55) -- (\xshift+0.55,\yshift-0.15);
\draw[very thick, \colB, ->] (\xshift+0.95,\yshift-1.25) -- (\xshift+0.65,\yshift-0.95) --(\xshift+0.15,\yshift-0.95);
\draw[very thick, \colC, ->] (\xshift+0.95,\yshift-0.15) -- (\xshift+0.95,\yshift-0.65) -- (\xshift+1.25,\yshift-0.95);

\draw[very thick, \colA, ->] (\xshift-0.55,\yshift-0.15) -- (\xshift-0.55,\yshift-0.55) -- (\xshift-0.15,\yshift-0.55);
\draw[very thick, \colB, ->] (\xshift-1.25,\yshift-0.95) -- (\xshift-0.95,\yshift-0.65) -- (\xshift-0.95,\yshift-0.15);
\draw[very thick, \colC, ->] (\xshift-0.15,\yshift-0.95) -- (\xshift-0.65,\yshift-0.95) --(\xshift-0.95,\yshift-1.25);

\node at (\xshift-0.2,\yshift+0.3) {$a'_1$};
\node at (\xshift-1.2,\yshift+0.5) {$a'_2$};
\node at (\xshift-0.5,\yshift+1.2) {$a'_3$};

\node at (\xshift+0.3,\yshift+0.3) {$b'_1$};
\node at (\xshift+0.5,\yshift+1.2) {$b'_3$};
\node at (\xshift+1.2,\yshift+0.5) {$b'_2$};

\node at (\xshift+0.3,\yshift-0.35) {$c'_1$};
\node at (\xshift+1.2,\yshift-0.5) {$c'_2$};
\node at (\xshift+0.5,\yshift-1.2) {$c'_3$};

\node at (\xshift-0.2,\yshift-0.3) {$d'_1$};
\node at (\xshift-0.5,\yshift-1.2) {$d'_3$};
\node at (\xshift-1.2,\yshift-0.5) {$d'_2$};


\tikzmath{\xs=-1;\ys=-3.75;}

\draw[styleQuiverEdge, gray] (\xs+0,\ys+1) -- (\xs+1,\ys+2);
\draw[styleQuiverEdge, gray] (\xs+2,\ys+1) -- (\xs+1,\ys+0);
\draw[styleQuiverEdge, gray] (\xs+0,\ys+1) -- (\xs+1,\ys+0);
\draw[styleQuiverEdge, gray] (\xs+2,\ys+1) -- (\xs+1,\ys+2);

\draw[styleQuiverEdge] (\xs+1,\ys+1) -- (\xs+0,\ys+1);
\draw[styleQuiverEdge] (\xs+1,\ys+0) -- (\xs+1,\ys+1);
\draw[styleQuiverEdge] (\xs+1,\ys+2) -- (\xs+1,\ys+1);
\draw[styleQuiverEdge] (\xs+1,\ys+1) -- (\xs+2,\ys+1);

\draw[quiverVertex] (\xs+0,\ys+1) circle;
\draw[quiverVertex] (\xs+1,\ys+0) circle;
\draw[quiverVertex] (\xs+1,\ys+2) circle;
\draw[quiverVertex] (\xs+2,\ys+1) circle;
\draw[quiverVertex] (\xs+1,\ys+1) circle;

\draw[->, very thick] (\xs+3.1,\ys+1) -- (\xs+3.1+1,\ys+1);

\draw[styleQuiverEdge] (\xs+5,\ys+1) -- (\xs+6,\ys+1);
\draw[styleQuiverEdge] (\xs+6,\ys+1) -- (\xs+6,\ys+0);
\draw[styleQuiverEdge] (\xs+6,\ys+1) -- (\xs+6,\ys+2);
\draw[styleQuiverEdge] (\xs+7,\ys+1) -- (\xs+6,\ys+1);

\draw[styleQuiverEdge, gray] (\xs+6,\ys+2) -- (\xs+7,\ys+1);
\draw[styleQuiverEdge, gray] (\xs+6,\ys+0) -- (\xs+5,\ys+1);
\draw[styleQuiverEdge, gray] (\xs+6,\ys+0) -- (\xs+7,\ys+1);
\draw[styleQuiverEdge, gray] (\xs+6,\ys+2) -- (\xs+5,\ys+1);

\draw[quiverVertex] (\xs+5,\ys+1) circle;
\draw[quiverVertex] (\xs+6,\ys+0) circle;
\draw[quiverVertex] (\xs+6,\ys+2) circle;
\draw[quiverVertex] (\xs+7,\ys+1) circle;
\draw[quiverVertex] (\xs+6,\ys+1) circle;


\tikzmath{\xshift=8.5;\yshift=0;};

\draw[thick] (\xshift-1,\yshift+1.25)--(\xshift+1,\yshift+1.25);
\draw[thick] (\xshift,\yshift+1.25)--(\xshift,\yshift-1.25);
\draw[thick] (\xshift-1,\yshift-1.25)--(\xshift+1,\yshift-1.25);

\draw[blackCircle] (\xshift,\yshift+1.25) circle;
\draw[whiteCircle] (\xshift,\yshift-1.25) circle;

\tikzmath{\xshift=12;\yshift=0;};

\draw[thick] (\xshift-1,\yshift+1.25)--(\xshift+1,\yshift+1.25);
\draw[thick] (\xshift,\yshift+1.25)--(\xshift,\yshift-1.25);
\draw[thick] (\xshift-1,\yshift-1.25)--(\xshift+1,\yshift-1.25);
\draw[thick, fill = white] (\xshift,\yshift) circle (0.5cm);

\draw[blackCircle] (\xshift,\yshift+1.25) circle;
\draw[blackCircle] (\xshift,\yshift+0.5) circle;
\draw[whiteCircle] (\xshift,\yshift-0.5) circle;
\draw[whiteCircle] (\xshift,\yshift-1.25) circle;

\draw[->, very thick] (\xshift-1.25,\yshift) -- (\xshift-2.25,\yshift);
\draw[->, very thick] (\xshift+1.25,\yshift) -- (\xshift+2.25,\yshift);

\tikzmath{\xshift=15.5;\yshift=0;};

\draw[thick] (\xshift-1,\yshift+1.25)--(\xshift+1,\yshift+1.25);
\draw[thick] (\xshift-1,\yshift-1.25)--(\xshift+1,\yshift-1.25);


\tikzmath{\xs=7.5;\ys=-3.75;}

\draw[styleQuiverEdge, gray] (\xs+0.25,\ys+1) -- (\xs+1,\ys+2);
\draw[styleQuiverEdge, gray] (\xs+1,\ys+0) -- (\xs+1.75,\ys+1);
\draw[styleQuiverEdge, gray] (\xs+0.25,\ys+1) -- (\xs+1,\ys+0);
\draw[styleQuiverEdge, gray] (\xs+1,\ys+2) -- (\xs+1.75,\ys+1);
\draw[styleQuiverEdge] (\xs+1.75,\ys+1) -- (\xs+0.25,\ys+1);

\draw[quiverVertex] (\xs+0.25,\ys+1) circle;
\draw[quiverVertex] (\xs+1,\ys+0) circle;
\draw[quiverVertex] (\xs+1,\ys+2) circle;
\draw[quiverVertex] (\xs+1.75,\ys+1) circle;

\tikzmath{\xs=11;\ys=-3.75;}

\draw[styleQuiverEdge, gray] (\xs+0.25,\ys+1) -- (\xs+1,\ys+2);
\draw[styleQuiverEdge, gray] (\xs+1,\ys+0) -- (\xs+1.75,\ys+1);
\draw[styleQuiverEdge, gray] (\xs+0.25,\ys+1) -- (\xs+1,\ys+0);
\draw[styleQuiverEdge, gray] (\xs+1,\ys+2) -- (\xs+1.75,\ys+1);
\draw[styleQuiverEdge] (\xs+1.75,\ys+1) -- (\xs+1,\ys+1);
\draw[styleQuiverEdge] (\xs+1,\ys+1) -- (\xs+0.25,\ys+1);

\draw[quiverVertex] (\xs+0.25,\ys+1) circle;
\draw[quiverVertex] (\xs+1,\ys+0) circle;
\draw[quiverVertex] (\xs+1,\ys+1) circle;
\draw[quiverVertex] (\xs+1,\ys+2) circle;
\draw[quiverVertex] (\xs+1.75,\ys+1) circle;

\node at (\xs-0.75, \ys+2.5) {$(a)$};
\node at (\xs+2.75, \ys+2.5) {$(b)$};
\draw[->, very thick] (\xs-0.25,\ys+1) -- (\xs-1.25,\ys+1);
\draw[->, very thick] (\xs+2.25,\ys+1) -- (\xs+3.25,\ys+1);

\tikzmath{\xs=14.5;\ys=-3.75;}

\draw[quiverVertex] (\xs+1,\ys+2) circle;
\draw[quiverVertex] (\xs+1,\ys+1) circle;
\draw[quiverVertex] (\xs+1,\ys+0) circle;


\end{tikzpicture}

}
\end{center}
\caption{Left, top: change of bipartite graph under the spider-move. Left, bottom: changes in the quiver. Grey arrows are for the entries $\pm 1/2$ of exchange matrix $\varepsilon$. Right, top: two ways for parallel bigon reduction. Right, bottom: changes in the quiver.}
\label{fig:fourumove}
\end{figure}
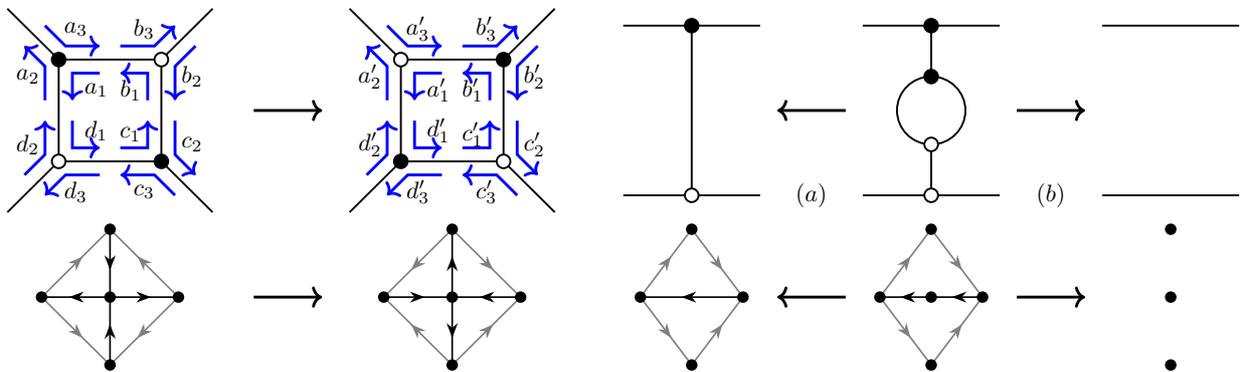

There are two well-known basic local transformations of perfect networks, which preserve both partition function of flows on them and Poisson structure: two-move shown in Fig.~\ref{fig:twomove} and four-move (also known as spider move or urban renewal or square move) shown in Fig.~\ref{fig:fourumove}. The choosing of perfect orientation is inessential here. For the two-move either face variables and quiver stay the same, while under the four-move they change as under mutation \cite{P, GK:2011}. Here we present formulas for transformation of corner variables under this moves, which will be required in Section \ref{s:tetra}.\\

For both two- and four- moves we derive mapping of the corner variables from reasonable monomial ans\"atze using three requirements
\begin{enumerate}
\item Transfer matrix of flows has to be preserved
\item Poisson brackets of new corner variables have to be consistent with the transformed plabic graph
\item Mapping has to respect symmetries of plabic graph
\end{enumerate}

It is easy to see that the unique monomial transformation rule under the black two-move for corner variables labelled in Fig.~\ref{fig:twomove}, left, satisfying this requirements is
\begin{equation}
\label{eq:twoMoveBlack}
l'_1 = t_3 b_2,~~~
l'_2 = b_3(t_1 b_1)^{\frac{1}{2}},~~~
l'_3 = t_2(t_1 b_1)^{\frac{1}{2}},~~~
r'_1 = b_3 t_2,~~~
r'_2 = t_3(t_1 b_1)^{\frac{1}{2}},~~~
r'_3 = b_2(t_1 b_1)^{\frac{1}{2}}.
\end{equation}
Under white two-move variables labelled in Fig.~\ref{fig:twomove}, right, transform as
\begin{equation}
\label{eq:twoMoveWhite}
l'_1 = t_2 b_3,~~~
l'_2 = t_3(t_1 b_1)^{\frac{1}{2}},~~~
l'_3 = b_2(t_1 b_1)^{\frac{1}{2}},~~~
r'_1 = b_2 t_3,~~~
r'_2 = b_3(t_1 b_1)^{\frac{1}{2}},~~~
r'_3 = t_2(t_1 b_1)^{\frac{1}{2}}.
\end{equation}
Exchange matrix $\varepsilon$ does not change under these transformations.

For the four-move there is a family of transformations, parametrized by $\spA,\spB,\spC$, which acts on corner variables by
\begin{equation}
\label{eq:fourMove}
\begin{array}{lll}
a_1' = b_2 d_3 \cdot
m_2^{-\spC+\spB} m_3^{-\spC-\spB},
&
a_2' = d_2\cdot (1+\x^{-1})^{-\frac{1}{2}}
m_1^{-\frac{1}{4}+\spA} m_2^{-\spB} m_3^{\spC} ,
&
a_3' = b_3\cdot (1+\x)^{\frac{1}{2}}
m_1^{-\frac{1}{4}-\spA} m_2^{\spC} m_3^{\spB},
\\
b_1' = a_2 c_3 \cdot
m_2^{\spC+\spB} m_3^{-\spC+\spB},
&
b_2' = c_2\cdot(1+\x^{-1})^{-\frac{1}{2}}
m_1^{\frac{1}{4}-\spA} m_2^{-\spB} m_3^{\spC},
&
b_3' = a_3\cdot (1+\x)^{\frac{1}{2}}
m_1^{\frac{1}{4}+\spA} m_2^{-\spC} m_3^{-\spB},
\\
c_1' = d_2 b_3 \cdot
m_2^{\spC-\spB} m_3^{\spC+\spB},
&
c_2' = b_2\cdot (1+\x^{-1})^{-\frac{1}{2}}
m_1^{-\frac{1}{4}+\spA} m_2^{\spB} m_3^{-\spC},
&
c_3' = d_3\cdot (1+\x)^{\frac{1}{2}}
m_1^{-\frac{1}{4}-\spA} m_2^{-\spC} m_3^{-\spB},
\\
d_1' = c_2 a_3 \cdot
m_2^{-\spC-\spB} m_3^{\spC-\spB},
&
d_2' = a_2\cdot (1+\x^{-1})^{-\frac{1}{2}}
m_1^{\frac{1}{4}-\spA} m_2^{\spB} m_3^{-\spC} ,
&
d_3' = c_3\cdot (1+\x)^{\frac{1}{2}}
m_1^{\frac{1}{4}+\spA} m_2^{\spC} m_3^{\spB},
\end{array}
\end{equation}
where
\begin{equation}
\x = a_1 b_1 c_1 d_1,
~~~
m_1 = \dfrac{a_1 c_1}{b_1 d_1},
~~~
m_2 = \dfrac{a_2 d_2}{b_2 c_2},
~~~
m_3 = \dfrac{a_3 b_3}{c_3 d_3}.
\end{equation}

Quivers encoding exchange matrix $\varepsilon$ before and after transformation are drawn in Fig.~\ref{fig:fourumove}, left, bottom. Whole family gives usual transformation rules for face variables, and are equivalent for our purposes, however choosing $\spA=\spB=0,~\spC=-\frac{1}{2}$ strangely makes formulas simpler.

The most subtle transformation is so-called parallel bigon reduction shown in Fig.~\ref{fig:fourumove}, right. Recall that the zig-zags are paths, which turn right at each black vertex, and turn left on each white one. Parallel bigon is a pairs of zig-zags which have such pair of intersection points, that disk(s) bounded by their segments between intersection points cannot be oriented in a way, consistent with orientation of segments.

The subtlety of parallel bigon reduction is that there are two different ways to perform it, both of which are bad. One of them, labelled by (a) in Fig.~\ref{fig:fourumove}, change topology of zig-zag paths which will be unwanted for us in the following, but preserves transfer matrix of flows and acts as cluster transformation (mutation supplied by forgetting of one variable) on cluster variables. Another one, labelled by (b), does not change topology of zig-zags, however, its action on cluster variables is ill-defined and it changes partition function of flows on plabic network. In the following, we will either assume that the network does not contain parallel bigons, or reduce first all its parallel bigons with transformation (b), before considering any flows.

\section{Tetrahedron equation from cluster algebra}

\label{s:tetra}
The claim of this section is that transfer matrices for both plabic graphs shown in Fig.~\ref{fig:fourgon}, left, coincide with Bazhanov-Sergeev solution of tetrahedron equation. Moreover, we will show that tetrahedron transformation is the result of sequence of four spider-moves.

\subsection{Lax operators}
As only paths which got both ends on the external edges of bipartite graph contribute to the transfer matrices of flows, we need only path variables $\gamma_i$ shown in Fig.~\ref{fig:fourgon}, left. 
For both graphs Poisson brackets of variables are
\begin{equation}
\label{eq:poissonFour}
\begin{array}{c}
\{ \gamma_1, \gamma_2 \} = -\frac{1}{2} \gamma_1 \gamma_2, ~~
\{ \gamma_2, \gamma_3 \} = \frac{1}{2} \gamma_2 \gamma_3, ~~
\{ \gamma_3, \gamma_4 \} = -\frac{1}{2} \gamma_3 \gamma_4, ~~
\{ \gamma_4, \gamma_1 \} = \frac{1}{2} \gamma_4 \gamma_1, \\
\{ \gamma_1, \gamma_3 \} = 0, ~~
\{ \gamma_2, \gamma_4 \} = 0.
\end{array}
\end{equation}
All the paths contributing to the transfer matrices are drawn in Fig.~\ref{fig:fourgon}, right.  Note that the only difference between the cases is in non-equivalent perfect orientation. Two plabic graphs are related by one spider move.
\begin{figure}[h!]
\begin{center}
\scalebox{0.9}{
\begin{tikzpicture}

\tikzmath{\d=1.3;};

\tikzmath{\xshift=-4.7;\yshift=-0.15;};

\draw[thick, styleDirectedEdge] (\xshift+0.5*\d,\yshift)--(\xshift+1*\d,\yshift);
\draw[thick, styleDirectedEdge] (\xshift+1*\d,\yshift)--(\xshift+2*\d,\yshift);
\draw[thick, styleDirectedEdge] (\xshift+2*\d,\yshift)--(\xshift+2.5*\d,\yshift);

\draw[thick, styleDirectedEdge] (\xshift+0.5*\d,\yshift+\d)--(\xshift+1*\d,\yshift+\d);
\draw[thick, styleDirectedEdge] (\xshift+1*\d,\yshift+\d)--(\xshift+2*\d,\yshift+\d);
\draw[thick, styleDirectedEdge] (\xshift+2*\d,\yshift+\d)--(\xshift+2.5*\d,\yshift+\d);

\draw[thick, styleDirectedEdge] (\xshift+\d,\yshift+\d)--(\xshift+\d,\yshift);
\draw[thick, styleDirectedEdge] (\xshift+2*\d,\yshift)--(\xshift+2*\d,\yshift+\d);

\draw[blackCircle] (\xshift+\d,\yshift+\d) circle;
\draw[whiteCircle] (\xshift+\d,\yshift+0) circle;
\draw[blackCircle] (\xshift+2*\d,\yshift+0) circle;
\draw[whiteCircle] (\xshift+2*\d,\yshift+\d) circle;

\draw[very thick, \colA, <-] (\xshift+0.5*\d,\yshift+\d-0.1*\d)--(\xshift+\d-0.4*\d,\yshift+\d-0.1*\d)..controls(\xshift+\d-0.1*\d,\yshift+\d-0.1*\d)..(\xshift+\d-0.1*\d,\yshift+\d-0.4*\d)--(\xshift+\d-0.1*\d,\yshift+0.4*\d)..controls(\xshift+\d-0.1*\d,\yshift+0.1*\d)..(\xshift+\d-0.4*\d,\yshift+0.1*\d)--(\xshift+0.5*\d, \yshift+0.1*\d);

\draw[very thick, \colB, ->] (\xshift+2.5*\d,\yshift+\d-0.1*\d)--(\xshift+2.4*\d,\yshift+\d-0.1*\d)..controls(\xshift+2.1*\d,\yshift+\d-0.1*\d)..(\xshift+2.1*\d,\yshift+\d-0.4*\d)--(\xshift+2.1*\d,\yshift+0.4*\d)..controls(\xshift+2.1*\d,\yshift+0.1*\d)..(\xshift+2.4*\d,\yshift+0.1*\d)--(\xshift+2.5*\d,\yshift+0.1*\d);

\draw[very thick, \colC, <-] (\xshift+0.5*\d,\yshift-0.1*\d) -- (\xshift+2.5*\d,\yshift-0.1*\d);

\draw[very thick, \colD, ->] (\xshift+0.5*\d,\yshift+1.1*\d) -- (\xshift+2.5*\d,\yshift+1.1*\d);

\node at (\xshift+1.5*\d,\yshift+1.3*\d) {$\gamma_1$};
\node at (\xshift+2.55*\d,\yshift+0.5*\d) {$\gamma_2$};
\node at (\xshift+1.5*\d,\yshift-0.3*\d) {$\gamma_3$};
\node at (\xshift+0.5*\d,\yshift+0.5*\d) {$\gamma_4$};

\node at (\xshift+0.2*\d,\yshift+1*\d) {$1$};
\node at (\xshift+0.2*\d,\yshift) {$2$};
\node at (\xshift+2.8*\d,\yshift) {$1'$};
\node at (\xshift+2.8*\d,\yshift+1*\d) {$2'$};


\begin{scope}

\tikzmath{\shift=0;}
\draw[thick] (\shift-0.25,1)--(\shift+1.25,1);
\draw[thick] (\shift-0.25,0)--(\shift+1.25,0);
\draw[thick] (\shift,1)--(\shift,0);
\draw[thick] (\shift+1,1)--(\shift+1,0);
\node at (\shift+0.5, -0.5) {$1$};

\tikzmath{\shift=2;}
\draw[thick] (\shift-0.25,1)--(\shift+1.25,1);
\draw[thick] (\shift-0.25,0)--(\shift+1.25,0);
\draw[thick] (\shift,1)--(\shift,0);
\draw[thick] (\shift+1,1)--(\shift+1,0);
\draw[blue, line width=0.5mm] (\shift-0.25,1)--(\shift,1);
\draw[styleArrowDimer] (\shift,1)--(\shift+1,1);
\draw[blue, line width=0.5mm] (\shift+1,1)--(\shift+1.25,1);
\node at (\shift+0.5, -0.5) {$\gamma_1$};

\tikzmath{\shift=4;}
\draw[thick] (\shift-0.25,1)--(\shift+1.25,1);
\draw[thick] (\shift-0.25,0)--(\shift+1.25,0);
\draw[thick] (\shift,1)--(\shift,0);
\draw[thick] (\shift+1,1)--(\shift+1,0);
\draw[blue, line width=0.5mm] (\shift-0.25,1)--(\shift,1)--(\shift,0);
\draw[styleArrowDimer] (\shift,0)--(\shift+1,0);
\draw[blue, line width=0.5mm] (\shift+1,0)--(\shift+1,1)--(\shift+1.25,1);
\node at (\shift+0.5, -0.5) {$\frac{1}{\gamma_2\gamma_3\gamma_4}$};

\tikzmath{\shift=6;}
\draw[thick] (\shift-0.25,1)--(\shift+1.25,1);
\draw[thick] (\shift-0.25,0)--(\shift+1.25,0);
\draw[thick] (\shift,1)--(\shift,0);
\draw[thick] (\shift+1,1)--(\shift+1,0);
\draw[styleArrowDimer] (\shift-0.25,0)--(\shift+1.25,0);
\node at (\shift+0.5, -0.5) {$\frac{1}{\gamma_3}$};

\tikzmath{\shift=8;}
\draw[thick] (\shift-0.25,1)--(\shift+1.25,1);
\draw[thick] (\shift-0.25,0)--(\shift+1.25,0);
\draw[thick] (\shift,1)--(\shift,0);
\draw[thick] (\shift+1,1)--(\shift+1,0);
\draw[blue, line width=0.5mm] (\shift-0.25,1)--(\shift,1)--(\shift,0);
\draw[styleArrowDimer] (\shift,0)--(\shift+1,0);
\draw[blue, line width=0.5mm] (\shift+1,0)--(\shift+1.25,0);
\node at (\shift+0.5, -0.5) {$\frac{1}{\gamma_3\gamma_4}$};

\tikzmath{\shift=10;}
\draw[thick] (\shift-0.25,1)--(\shift+1.25,1);
\draw[thick] (\shift-0.25,0)--(\shift+1.25,0);
\draw[thick] (\shift,1)--(\shift,0);
\draw[thick] (\shift+1,1)--(\shift+1,0);
\draw[blue, line width=0.5mm] (\shift-0.25,0)--(\shift,0);
\draw[styleArrowDimer] (\shift,0)--(\shift+1,0);
\draw[blue, line width=0.5mm] (\shift+1,0)--(\shift+1,1)--(\shift+1.25,1);
\node at (\shift+0.5, -0.5) {$\frac{1}{\gamma_2\gamma_3}$};
\end{scope}

\tikzmath{\shift=12;}
\draw[thick] (\shift-0.25,1)--(\shift+1.25,1);
\draw[thick] (\shift-0.25,0)--(\shift+1.25,0);
\draw[thick] (\shift,1)--(\shift,0);
\draw[thick] (\shift+1,1)--(\shift+1,0);
\draw[blue, line width=0.5mm] (\shift-0.25,1)--(\shift,1);
\draw[styleArrowDimer] (\shift,1)--(\shift+1,1);
\draw[blue, line width=0.5mm] (\shift+1,1)--(\shift+1.25,1);
\draw[blue, line width=0.5mm] (\shift-0.25,0)--(\shift,0);
\draw[styleArrowDimer] (\shift,0)--(\shift+1,0);
\draw[blue, line width=0.5mm] (\shift+1,0)--(\shift+1.25,0);
\node at (\shift+0.5, -0.5) {$\frac{\gamma_1}{\gamma_3}$};
\end{tikzpicture}
}
\scalebox{0.9}{
\begin{tikzpicture}

\tikzmath{\d=1.3;};

\tikzmath{\xshift=-4.7;\yshift=-0.15;};

\draw[thick, styleDirectedEdge] (\xshift+0.5*\d,\yshift)--(\xshift+1*\d,\yshift);
\draw[thick, styleDirectedEdge] (\xshift+1*\d,\yshift)--(\xshift+2*\d,\yshift);
\draw[thick, styleDirectedEdge] (\xshift+2*\d,\yshift)--(\xshift+2.5*\d,\yshift);

\draw[thick, styleDirectedEdge] (\xshift+0.5*\d,\yshift+\d)--(\xshift+1*\d,\yshift+\d);
\draw[thick, styleDirectedEdge] (\xshift+1*\d,\yshift+\d)--(\xshift+2*\d,\yshift+\d);
\draw[thick, styleDirectedEdge] (\xshift+2*\d,\yshift+\d)--(\xshift+2.5*\d,\yshift+\d);

\draw[thick, styleDirectedEdge] (\xshift+\d,\yshift)--(\xshift+\d,\yshift+\d);
\draw[thick, styleDirectedEdge] (\xshift+2*\d,\yshift+\d)--(\xshift+2*\d,\yshift);

\draw[whiteCircle] (\xshift+\d,\yshift+\d) circle;
\draw[blackCircle] (\xshift+\d,\yshift+0) circle;
\draw[whiteCircle] (\xshift+2*\d,\yshift+0) circle;
\draw[blackCircle] (\xshift+2*\d,\yshift+\d) circle;

\draw[very thick, \colA, <-] (\xshift+0.5*\d,\yshift+\d-0.1*\d)--(\xshift+\d-0.4*\d,\yshift+\d-0.1*\d)..controls(\xshift+\d-0.1*\d,\yshift+\d-0.1*\d)..(\xshift+\d-0.1*\d,\yshift+\d-0.4*\d)--(\xshift+\d-0.1*\d,\yshift+0.4*\d)..controls(\xshift+\d-0.1*\d,\yshift+0.1*\d)..(\xshift+\d-0.4*\d,\yshift+0.1*\d)--(\xshift+0.5*\d, \yshift+0.1*\d);

\draw[very thick, \colB, ->] (\xshift+2.5*\d,\yshift+\d-0.1*\d)--(\xshift+2.4*\d,\yshift+\d-0.1*\d)..controls(\xshift+2.1*\d,\yshift+\d-0.1*\d)..(\xshift+2.1*\d,\yshift+\d-0.4*\d)--(\xshift+2.1*\d,\yshift+0.4*\d)..controls(\xshift+2.1*\d,\yshift+0.1*\d)..(\xshift+2.4*\d,\yshift+0.1*\d)--(\xshift+2.5*\d,\yshift+0.1*\d);

\draw[very thick, \colC, <-] (\xshift+0.5*\d,\yshift-0.1*\d) -- (\xshift+2.5*\d,\yshift-0.1*\d);

\draw[very thick, \colD, ->] (\xshift+0.5*\d,\yshift+1.1*\d) -- (\xshift+2.5*\d,\yshift+1.1*\d);

\node at (\xshift+1.5*\d,\yshift+1.3*\d) {$\gamma_2$};
\node at (\xshift+2.55*\d,\yshift+0.5*\d) {$\gamma_3$};
\node at (\xshift+1.5*\d,\yshift-0.3*\d) {$\gamma_4$};
\node at (\xshift+0.5*\d,\yshift+0.5*\d) {$\gamma_1$};

\node at (\xshift+0.2*\d,\yshift+1*\d) {$1$};
\node at (\xshift+0.2*\d,\yshift) {$2$};
\node at (\xshift+2.8*\d,\yshift) {$1'$};
\node at (\xshift+2.8*\d,\yshift+1*\d) {$2'$};


\tikzmath{\shift=0;}
\draw[thick] (\shift-0.25,1)--(\shift+1.25,1);
\draw[thick] (\shift-0.25,0)--(\shift+1.25,0);
\draw[thick] (\shift,1)--(\shift,0);
\draw[thick] (\shift+1,1)--(\shift+1,0);
\node at (\shift+0.5, -0.5) {$1$};

\tikzmath{\shift=2;}
\draw[thick] (\shift-0.25,1)--(\shift+1.25,1);
\draw[thick] (\shift-0.25,0)--(\shift+1.25,0);
\draw[thick] (\shift,1)--(\shift,0);
\draw[thick] (\shift+1,1)--(\shift+1,0);
\draw[blue, line width=0.5mm] (\shift-0.25,1)--(\shift,1);
\draw[styleArrowDimer] (\shift,1)--(\shift+1,1);
\draw[blue, line width=0.5mm] (\shift+1,1)--(\shift+1.25,1);
\node at (\shift+0.5, -0.5) {$\gamma_2$};

\tikzmath{\shift=4;}
\draw[thick] (\shift-0.25,1)--(\shift+1.25,1);
\draw[thick] (\shift-0.25,0)--(\shift+1.25,0);
\draw[thick] (\shift,1)--(\shift,0);
\draw[thick] (\shift+1,1)--(\shift+1,0);
\draw[styleArrowDimer] (\shift-0.25,0)--(\shift+1.25,0);
\node at (\shift+0.5, -0.5) {$\frac{1}{\gamma_4}$};

\tikzmath{\shift=6;}
\draw[thick] (\shift-0.25,1)--(\shift+1.25,1);
\draw[thick] (\shift-0.25,0)--(\shift+1.25,0);
\draw[thick] (\shift,1)--(\shift,0);
\draw[thick] (\shift+1,1)--(\shift+1,0);
\draw[blue, line width=0.5mm] (\shift-0.25,0)--(\shift,0)--(\shift,1);
\draw[styleArrowDimer] (\shift,1)--(\shift+1,1);
\draw[blue, line width=0.5mm] (\shift+1,1)--(\shift+1,0)--(\shift+1.25,0);
\node at (\shift+0.5, -0.5) {$\gamma_1\gamma_2\gamma_3$};

\tikzmath{\shift=8;}
\draw[thick] (\shift-0.25,1)--(\shift+1.25,1);
\draw[thick] (\shift-0.25,0)--(\shift+1.25,0);
\draw[thick] (\shift,1)--(\shift,0);
\draw[thick] (\shift+1,1)--(\shift+1,0);
\draw[blue, line width=0.5mm] (\shift-0.25,1)--(\shift,1);
\draw[styleArrowDimer] (\shift,1)--(\shift+1,1);
\draw[blue, line width=0.5mm] (\shift+1,1)--(\shift+1,0)--(\shift+1.25,0);
\node at (\shift+0.5, -0.5) {$\gamma_2\gamma_3$};

\tikzmath{\shift=10;}
\draw[thick] (\shift-0.25,1)--(\shift+1.25,1);
\draw[thick] (\shift-0.25,0)--(\shift+1.25,0);
\draw[thick] (\shift,1)--(\shift,0);
\draw[thick] (\shift+1,1)--(\shift+1,0);
\draw[blue, line width=0.5mm] (\shift-0.25,0)--(\shift,0)--(\shift,1);
\draw[styleArrowDimer] (\shift,1)--(\shift+1,1);
\draw[blue, line width=0.5mm] (\shift+1,1)--(\shift+1.25,1);
\node at (\shift+0.5, -0.5) {$\gamma_1\gamma_2$};

\tikzmath{\shift=12;}
\draw[thick] (\shift-0.25,1)--(\shift+1.25,1);
\draw[thick] (\shift-0.25,0)--(\shift+1.25,0);
\draw[thick] (\shift,1)--(\shift,0);
\draw[thick] (\shift+1,1)--(\shift+1,0);
\draw[blue, line width=0.5mm] (\shift-0.25,1)--(\shift,1);
\draw[styleArrowDimer] (\shift,1)--(\shift+1,1);
\draw[blue, line width=0.5mm] (\shift+1,1)--(\shift+1.25,1);
\draw[blue, line width=0.5mm] (\shift-0.25,0)--(\shift,0);
\draw[styleArrowDimer] (\shift,0)--(\shift+1,0);
\draw[blue, line width=0.5mm] (\shift+1,0)--(\shift+1.25,0);
\node at (\shift+0.5, -0.5) {$\frac{\gamma_2}{\gamma_4}$};

\end{tikzpicture}
}
\end{center}
\caption{Left. Four-gonal pieces of bipartite graphs whose transfer matrices define Lax operators. Right. Paths contributing to transfer matrices.}
\label{fig:fourgon}
\end{figure}

The transfer matrices for upper and lower networks in the basis $\mathbb{C}^2 \otimes \mathbb{C}^2 = \langle e_{+}\otimes e_{+}, \, e_{+}\otimes e_{-}, \, e_{-}\otimes e_{+}, \, e_{-}\otimes e_{-} \rangle$ are respectively
\begin{equation}
\label{eq:LaxCL}
\LCL(\gamma)=
\left(
\begin{matrix}
\gamma_1  \gamma_3^{-1} &  &  &  \\
 & (\gamma_3 \gamma_4)^{-1} & \gamma_3^{-1} & \\
 & \gamma_1 + (\gamma_2 \gamma_3 \gamma_4)^{-1} & (\gamma_2 \gamma_3)^{-1} & \\
 &  &  & 1
\end{matrix}
\right),
~~~
\LCLm(\gamma)=
\left(
\begin{matrix}
\gamma_2  \gamma_4^{-1} &  &  &  \\
 & \gamma_2 \gamma_3 & \gamma_4^{-1} + \gamma_1 \gamma_2 \gamma_3  & \\
 & \gamma_2 & \gamma_1 \gamma_2 & \\
 &  &  & 1
\end{matrix}
\right).
\end{equation}
Matrix $\LCL$ coincides with Bazhanov-Sergeev Lax operator (\ref{eq:LaxBS}) after conjugation
\begin{equation}
\LBS = (\sigma_1 \otimes \sigma_1 \circ \P) \circ \LCL \circ (\sigma_1 \otimes \sigma_1 \circ \P)
\end{equation}
where $\P$ is a permutation matrix $\P(u\otimes v) = v\otimes u$, and after identification of variables
\begin{equation}
\label{eq:BStoClustVar}
x = \gamma_1^{-1}, ~~~
y = \gamma_1 + (\gamma_2 \gamma_3 \gamma_4)^{-1}, ~~~
\lambda = -i \sqrt{\dfrac{\gamma_1 \gamma_4}{\gamma_2 \gamma_3}}, ~~~
\mu = -i \sqrt{\dfrac{\gamma_1 \gamma_2}{\gamma_3 \gamma_4}},~~~
k = \dfrac{i}{\sqrt{\gamma_1 \gamma_2 \gamma_3 \gamma_4}}.
\end{equation}
The Poisson brackets (\ref{eq:BSPoisson}) follow from (\ref{eq:poissonFour}). Matrix $\LCLm$ can be mapped to $\LCL$ by conjugation with $\P$ and replacement
\begin{equation}
\gamma_1 \mapsto \gamma_4^{-1}, ~~~
\gamma_2 \mapsto \gamma_3^{-1}, ~~~
\gamma_3 \mapsto \gamma_2^{-1}, ~~~
\gamma_4 \mapsto \gamma_1^{-1}.
\end{equation}
In the following we will be dealing with matrix $\LCL$ only.
\subsection{Tetrahedron transformation}
\label{ss:tetra}
\begin{figure}[h!]
\begin{center}
\scalebox{0.95}{
\begin{tikzpicture}

\tikzmath{\d=1.5;};

\tikzmath{\ybot=0*\d;\xshift=0;\yshift=\ybot+0*\d;};

\node at (-0.3*\d, 2*\d+\yshift) {$1$};
\node at (-0.3*\d, \d+\yshift) {$2$};
\node at (-0.3*\d, \yshift) {$3$};

\draw[thick, styleDirectedEdge] (0*\d,2*\d+\yshift)--(1*\d,2*\d+\yshift);
\draw[thick, styleDirectedEdge] (1*\d,2*\d+\yshift)--(2*\d,2*\d+\yshift);
\draw[thick, styleDirectedEdge] (2*\d,2*\d+\yshift)--(4*\d,2*\d+\yshift);
\node[cross, very thick, gray] at (4*\d,2*\d+\yshift) {};
\draw[thick, styleDirectedEdge] (4*\d,2*\d+\yshift)--(6*\d,2*\d+\yshift);
\draw[thick, styleDirectedEdge] (6*\d,2*\d+\yshift)--(7*\d,2*\d+\yshift);
\draw[thick, styleDirectedEdge] (7*\d,2*\d+\yshift)--(8*\d,2*\d+\yshift);

\draw[thick, styleDirectedEdge] (0*\d,\d+\yshift)--(1*\d,\d+\yshift);
\draw[thick, styleDirectedEdge] (1*\d,\d+\yshift)--(2*\d,\d+\yshift);
\draw[thick, styleDirectedEdge] (2*\d,\d+\yshift)--(2.75*\d,\d+\yshift);
\node[cross, very thick, gray] at (2.75*\d,\d+\yshift) {};
\draw[thick, styleDirectedEdge] (2.75*\d,\d+\yshift)--(3.5*\d,\d+\yshift);
\draw[thick, styleDirectedEdge] (3.5*\d,\d+\yshift)--(4.5*\d,\d+\yshift);
\draw[thick, styleDirectedEdge] (4.5*\d,\d+\yshift)--(5.25*\d,\d+\yshift);
\node[cross, very thick, gray] at (5.25*\d,\d+\yshift) {};
\draw[thick, styleDirectedEdge] (5.25*\d,\d+\yshift)--(6*\d,\d+\yshift);
\draw[thick, styleDirectedEdge] (6*\d,\d+\yshift)--(7*\d,\d+\yshift);
\draw[thick, styleDirectedEdge] (7*\d,\d+\yshift)--(8*\d,\d+\yshift);

\draw[thick] (0*\d,\yshift)--(1*\d,\yshift);
\draw[thick, styleDirectedEdge] (1*\d,\yshift)--(2*\d,\yshift);
\draw[thick] (2*\d,\yshift)--(3.5*\d,\yshift);
\draw[thick, styleDirectedEdge] (3.5*\d,\yshift)--(4.5*\d,\yshift);
\draw[thick] (4.5*\d,\yshift)--(6*\d,\yshift);
\draw[thick, styleDirectedEdge] (6*\d,\yshift)--(7*\d,\yshift);
\draw[thick] (7*\d,\yshift)--(8*\d,\yshift);

\node at (8.3*\d, 2*\d+\yshift) {$3'$};
\node at (8.3*\d, \d+\yshift) {$2'$};
\node at (8.3*\d, \yshift) {$1'$};

\tikzmath{\xshift=0;\yshift=\ybot+\d;};

\draw[thick, styleDirectedEdge] (\xshift+\d,\yshift+\d)--(\xshift+\d,\yshift);
\draw[thick, styleDirectedEdge] (\xshift+2*\d,\yshift)--(\xshift+2*\d,\yshift+\d);

\draw[blackCircle] (\xshift+\d,\yshift+\d) circle;
\draw[whiteCircle] (\xshift+\d,\yshift+0) circle;
\draw[blackCircle] (\xshift+2*\d,\yshift+0) circle;
\draw[whiteCircle] (\xshift+2*\d,\yshift+\d) circle;

\draw[very thick, \colA, <-] (\xshift+0.5*\d,\yshift+\d-0.1*\d)--(\xshift+\d-0.4*\d,\yshift+\d-0.1*\d)..controls(\xshift+\d-0.1*\d,\yshift+\d-0.1*\d)..(\xshift+\d-0.1*\d,\yshift+\d-0.4*\d)--(\xshift+\d-0.1*\d,\yshift+0.4*\d)..controls(\xshift+\d-0.1*\d,\yshift+0.1*\d)..(\xshift+\d-0.4*\d,\yshift+0.1*\d)--(\xshift+0.5*\d, \yshift+0.1*\d);

\draw[very thick, \colB, ->] (\xshift+2.5*\d,\yshift+\d-0.1*\d)--(\xshift+2.4*\d,\yshift+\d-0.1*\d)..controls(\xshift+2.1*\d,\yshift+\d-0.1*\d)..(\xshift+2.1*\d,\yshift+\d-0.4*\d)--(\xshift+2.1*\d,\yshift+0.4*\d)..controls(\xshift+2.1*\d,\yshift+0.1*\d)..(\xshift+2.4*\d,\yshift+0.1*\d)--(\xshift+2.5*\d,\yshift+0.1*\d);

\draw[very thick, \colC, <-] (\xshift+0.5*\d,\yshift-0.1*\d) -- (\xshift+2.5*\d,\yshift-0.1*\d);

\draw[very thick, \colD, ->] (\xshift+0.5*\d,\yshift+1.1*\d) -- (\xshift+2.5*\d,\yshift+1.1*\d);

\node at (\xshift+1.5*\d,\yshift+1.3*\d) {$\gamma_{c,1}$};
\node at (\xshift+2.55*\d,\yshift+0.5*\d) {$\gamma_{c,2}$};
\node at (\xshift+1.5*\d,\yshift-0.3*\d) {$\gamma_{c,3}$};
\node at (\xshift+0.5*\d,\yshift+0.5*\d) {$\gamma_{c,4}$};

\node at (\xshift+0*\d,\yshift+0.5*\d) {$\x_1$};
\node at (\xshift+1.5*\d,\yshift+0.5*\d) {$\x_2$};
\node at (\xshift+2.3*\d,\yshift-0.5*\d) {$\x_6$};

\tikzmath{\xshift=2.5*\d;\yshift=\ybot;};

\draw[thick, styleDirectedEdge] (\xshift+\d,\yshift+\d)--(\xshift+\d,\yshift);
\draw[thick, styleDirectedEdge] (\xshift+2*\d,\yshift)--(\xshift+2*\d,\yshift+\d);

\draw[blackCircle] (\xshift+\d,\yshift+\d) circle;
\draw[whiteCircle] (\xshift+\d,\yshift+0) circle;
\draw[blackCircle] (\xshift+2*\d,\yshift+0) circle;
\draw[whiteCircle] (\xshift+2*\d,\yshift+\d) circle;

\draw[very thick, \colA, <-] (\xshift+0.5*\d,\yshift+\d-0.1*\d)--(\xshift+\d-0.4*\d,\yshift+\d-0.1*\d)..controls(\xshift+\d-0.1*\d,\yshift+\d-0.1*\d)..(\xshift+\d-0.1*\d,\yshift+\d-0.4*\d)--(\xshift+\d-0.1*\d,\yshift+0.4*\d)..controls(\xshift+\d-0.1*\d,\yshift+0.1*\d)..(\xshift+\d-0.4*\d,\yshift+0.1*\d)--(\xshift+0.5*\d, \yshift+0.1*\d);

\draw[very thick, \colB, ->] (\xshift+2.5*\d,\yshift+\d-0.1*\d)--(\xshift+2.4*\d,\yshift+\d-0.1*\d)..controls(\xshift+2.1*\d,\yshift+\d-0.1*\d)..(\xshift+2.1*\d,\yshift+\d-0.4*\d)--(\xshift+2.1*\d,\yshift+0.4*\d)..controls(\xshift+2.1*\d,\yshift+0.1*\d)..(\xshift+2.4*\d,\yshift+0.1*\d)--(\xshift+2.5*\d,\yshift+0.1*\d);

\draw[very thick, \colC, <-] (\xshift+0.5*\d,\yshift-0.1*\d) -- (\xshift+2.5*\d,\yshift-0.1*\d);

\draw[very thick, \colD, ->] (\xshift+0.5*\d,\yshift+1.1*\d) -- (\xshift+2.5*\d,\yshift+1.1*\d);

\node at (\xshift+1.5*\d,\yshift+1.3*\d) {$\gamma_{b,1}$};
\node at (\xshift+2.55*\d,\yshift+0.5*\d) {$\gamma_{b,2}$};
\node at (\xshift+1.5*\d,\yshift-0.3*\d) {$\gamma_{b,3}$};
\node at (\xshift+0.5*\d,\yshift+0.5*\d) {$\gamma_{b,4}$};

\node at (\xshift+1.5*\d,\yshift+2.3*\d) {$\x_0$};
\node at (\xshift+1.5*\d,\yshift+1.6*\d) {$\x_3$};
\node at (\xshift+1.5*\d,\yshift+0.5*\d) {$\x_7$};
\node at (\xshift+1.5*\d,\yshift-0.7*\d) {$\x_9$};

\tikzmath{\xshift=5*\d;\yshift=\ybot+\d;};

\draw[thick, styleDirectedEdge] (\xshift+\d,\yshift+\d)--(\xshift+\d,\yshift);
\draw[thick, styleDirectedEdge] (\xshift+2*\d,\yshift)--(\xshift+2*\d,\yshift+\d);

\draw[blackCircle] (\xshift+\d,\yshift+\d) circle;
\draw[whiteCircle] (\xshift+\d,\yshift+0) circle;
\draw[blackCircle] (\xshift+2*\d,\yshift+0) circle;
\draw[whiteCircle] (\xshift+2*\d,\yshift+\d) circle;

\draw[very thick, \colA, <-] (\xshift+0.5*\d,\yshift+\d-0.1*\d)--(\xshift+\d-0.4*\d,\yshift+\d-0.1*\d)..controls(\xshift+\d-0.1*\d,\yshift+\d-0.1*\d)..(\xshift+\d-0.1*\d,\yshift+\d-0.4*\d)--(\xshift+\d-0.1*\d,\yshift+0.4*\d)..controls(\xshift+\d-0.1*\d,\yshift+0.1*\d)..(\xshift+\d-0.4*\d,\yshift+0.1*\d)--(\xshift+0.5*\d, \yshift+0.1*\d);

\draw[very thick, \colB, ->] (\xshift+2.5*\d,\yshift+\d-0.1*\d)--(\xshift+2.4*\d,\yshift+\d-0.1*\d)..controls(\xshift+2.1*\d,\yshift+\d-0.1*\d)..(\xshift+2.1*\d,\yshift+\d-0.4*\d)--(\xshift+2.1*\d,\yshift+0.4*\d)..controls(\xshift+2.1*\d,\yshift+0.1*\d)..(\xshift+2.4*\d,\yshift+0.1*\d)--(\xshift+2.5*\d,\yshift+0.1*\d);

\draw[very thick, \colC, <-] (\xshift+0.5*\d,\yshift-0.1*\d) -- (\xshift+2.5*\d,\yshift-0.1*\d);

\draw[very thick, \colD, ->] (\xshift+0.5*\d,\yshift+1.1*\d) -- (\xshift+2.5*\d,\yshift+1.1*\d);

\node at (\xshift+1.5*\d,\yshift+1.3*\d) {$\gamma_{a,1}$};
\node at (\xshift+2.55*\d,\yshift+0.5*\d) {$\gamma_{a,2}$};
\node at (\xshift+1.5*\d,\yshift-0.3*\d) {$\gamma_{a,3}$};
\node at (\xshift+0.5*\d,\yshift+0.5*\d) {$\gamma_{a,4}$};

\node at (\xshift+3*\d,\yshift+0.5*\d) {$\x_5$};
\node at (\xshift+1.5*\d,\yshift+0.5*\d) {$\x_4$};
\node at (\xshift+0.7*\d,\yshift-0.5*\d) {$\x_8$};


\draw[thick, ->] (3.96*\d,\yshift-2*\d)--(3.96*\d,\yshift-2.5*\d);
\node at (4.3*\d,\yshift-2.25*\d) {$\Rmut$};


\tikzmath{\ybot = - 4.5*\d;\xshift=0;\yshift=\ybot+0*\d;};

\node at (-0.3*\d, 2*\d+\yshift) {$1$};
\node at (-0.3*\d, \d+\yshift) {$2$};
\node at (-0.3*\d, \yshift) {$3$};

\draw[thick, styleDirectedEdge] (0*\d,\yshift)--(1*\d,\yshift);
\draw[thick, styleDirectedEdge] (1*\d,\yshift)--(2*\d,\yshift);
\draw[thick, styleDirectedEdge] (2*\d,\yshift)--(4*\d,\yshift);
\node[cross, very thick, gray] at (4*\d,\yshift) {};
\draw[thick, styleDirectedEdge] (4*\d,\yshift)--(6*\d,\yshift);
\draw[thick, styleDirectedEdge] (6*\d,\yshift)--(7*\d,\yshift);
\draw[thick, styleDirectedEdge] (7*\d,\yshift)--(8*\d,\yshift);

\draw[thick, styleDirectedEdge] (0*\d,\d+\yshift)--(1*\d,\d+\yshift);
\draw[thick, styleDirectedEdge] (1*\d,\d+\yshift)--(2*\d,\d+\yshift);
\draw[thick, styleDirectedEdge] (2*\d,\d+\yshift)--(2.75*\d,\d+\yshift);
\node[cross, very thick, gray] at (2.75*\d,\d+\yshift) {};
\draw[thick, styleDirectedEdge] (2.75*\d,\d+\yshift)--(3.5*\d,\d+\yshift);
\draw[thick, styleDirectedEdge] (3.5*\d,\d+\yshift)--(4.5*\d,\d+\yshift);
\draw[thick, styleDirectedEdge] (4.5*\d,\d+\yshift)--(5.25*\d,\d+\yshift);
\node[cross, very thick, gray] at (5.25*\d,\d+\yshift) {};
\draw[thick, styleDirectedEdge] (5.25*\d,\d+\yshift)--(6*\d,\d+\yshift);
\draw[thick, styleDirectedEdge] (6*\d,\d+\yshift)--(7*\d,\d+\yshift);
\draw[thick, styleDirectedEdge] (7*\d,\d+\yshift)--(8*\d,\d+\yshift);

\draw[thick] (0*\d,2*\d+\yshift)--(1*\d,2*\d+\yshift);
\draw[thick, styleDirectedEdge] (1*\d,2*\d+\yshift)--(2*\d,2*\d+\yshift);
\draw[thick] (2*\d,2*\d+\yshift)--(3.5*\d,2*\d+\yshift);
\draw[thick, styleDirectedEdge] (3.5*\d,2*\d+\yshift)--(4.5*\d,2*\d+\yshift);
\draw[thick] (4.5*\d,2*\d+\yshift)--(6*\d,2*\d+\yshift);
\draw[thick, styleDirectedEdge] (6*\d,2*\d+\yshift)--(7*\d,2*\d+\yshift);
\draw[thick] (7*\d,2*\d+\yshift)--(8*\d,2*\d+\yshift);

\node at (8.3*\d, 2*\d+\yshift) {$3'$};
\node at (8.3*\d, \d+\yshift) {$2'$};
\node at (8.3*\d, \yshift) {$1'$};

\tikzmath{\xshift=0;\yshift=\ybot+0*\d;};


\draw[thick] (\xshift+\d,\yshift+\d)--(\xshift+\d,\yshift);
\draw[thick] (\xshift+2*\d,\yshift+\d)--(\xshift+2*\d,\yshift);

\draw[blackCircle] (\xshift+\d,\yshift+\d) circle;
\draw[whiteCircle] (\xshift+\d,\yshift+0) circle;
\draw[blackCircle] (\xshift+2*\d,\yshift+0) circle;
\draw[whiteCircle] (\xshift+2*\d,\yshift+\d) circle;

\draw[very thick, \colA, <-] (\xshift+0.5*\d,\yshift+\d-0.1*\d)--(\xshift+\d-0.4*\d,\yshift+\d-0.1*\d)..controls(\xshift+\d-0.1*\d,\yshift+\d-0.1*\d)..(\xshift+\d-0.1*\d,\yshift+\d-0.4*\d)--(\xshift+\d-0.1*\d,\yshift+0.4*\d)..controls(\xshift+\d-0.1*\d,\yshift+0.1*\d)..(\xshift+\d-0.4*\d,\yshift+0.1*\d)--(\xshift+0.5*\d, \yshift+0.1*\d);

\draw[very thick, \colB, ->] (\xshift+2.5*\d,\yshift+\d-0.1*\d)--(\xshift+2.4*\d,\yshift+\d-0.1*\d)..controls(\xshift+2.1*\d,\yshift+\d-0.1*\d)..(\xshift+2.1*\d,\yshift+\d-0.4*\d)--(\xshift+2.1*\d,\yshift+0.4*\d)..controls(\xshift+2.1*\d,\yshift+0.1*\d)..(\xshift+2.4*\d,\yshift+0.1*\d)--(\xshift+2.5*\d,\yshift+0.1*\d);

\draw[very thick, \colC, <-] (\xshift+0.5*\d,\yshift-0.1*\d) -- (\xshift+2.5*\d,\yshift-0.1*\d);

\draw[very thick, \colD, ->] (\xshift+0.5*\d,\yshift+1.1*\d) -- (\xshift+2.5*\d,\yshift+1.1*\d);

\node at (\xshift+1.5*\d,\yshift+1.3*\d) {$\gamma'_{a,1}$};
\node at (\xshift+2.55*\d,\yshift+0.5*\d) {$\gamma'_{a,2}$};
\node at (\xshift+1.5*\d,\yshift-0.3*\d) {$\gamma'_{a,3}$};
\node at (\xshift+0.5*\d,\yshift+0.5*\d) {$\gamma'_{a,4}$};

\node at (\xshift+0*\d,\yshift+0.5*\d) {$\x'_6$};
\node at (\xshift+1.5*\d,\yshift+0.5*\d) {$\x'_2$};

\tikzmath{\xshift=2.5*\d;\yshift=\ybot+1*\d;};

\draw[thick] (\xshift+\d,\yshift+\d)--(\xshift+\d,\yshift);
\draw[thick] (\xshift+2*\d,\yshift+\d)--(\xshift+2*\d,\yshift);


\draw[blackCircle] (\xshift+\d,\yshift+\d) circle;
\draw[whiteCircle] (\xshift+\d,\yshift+0) circle;
\draw[blackCircle] (\xshift+2*\d,\yshift+0) circle;
\draw[whiteCircle] (\xshift+2*\d,\yshift+\d) circle;

\draw[very thick, \colA, <-] (\xshift+0.5*\d,\yshift+\d-0.1*\d)--(\xshift+\d-0.4*\d,\yshift+\d-0.1*\d)..controls(\xshift+\d-0.1*\d,\yshift+\d-0.1*\d)..(\xshift+\d-0.1*\d,\yshift+\d-0.4*\d)--(\xshift+\d-0.1*\d,\yshift+0.4*\d)..controls(\xshift+\d-0.1*\d,\yshift+0.1*\d)..(\xshift+\d-0.4*\d,\yshift+0.1*\d)--(\xshift+0.5*\d, \yshift+0.1*\d);

\draw[very thick, \colB, ->] (\xshift+2.5*\d,\yshift+\d-0.1*\d)--(\xshift+2.4*\d,\yshift+\d-0.1*\d)..controls(\xshift+2.1*\d,\yshift+\d-0.1*\d)..(\xshift+2.1*\d,\yshift+\d-0.4*\d)--(\xshift+2.1*\d,\yshift+0.4*\d)..controls(\xshift+2.1*\d,\yshift+0.1*\d)..(\xshift+2.4*\d,\yshift+0.1*\d)--(\xshift+2.5*\d,\yshift+0.1*\d);

\draw[very thick, \colC, <-] (\xshift+0.5*\d,\yshift-0.1*\d) -- (\xshift+2.5*\d,\yshift-0.1*\d);

\draw[very thick, \colD, ->] (\xshift+0.5*\d,\yshift+1.1*\d) -- (\xshift+2.5*\d,\yshift+1.1*\d);

\node at (\xshift+1.5*\d,\yshift+1.3*\d) {$\gamma'_{b,1}$};
\node at (\xshift+2.55*\d,\yshift+0.5*\d) {$\gamma'_{b,2}$};
\node at (\xshift+1.5*\d,\yshift-0.3*\d) {$\gamma'_{b,3}$};
\node at (\xshift+0.5*\d,\yshift+0.5*\d) {$\gamma'_{b,4}$};

\node at (\xshift-0.5*\d,\yshift+0.5*\d) {$\x'_1$};
\node at (\xshift+1.5*\d,\yshift+1.7*\d) {$\x'_0$};
\node at (\xshift+1.5*\d,\yshift+0.5*\d) {$\x'_3$};
\node at (\xshift+3.5*\d,\yshift+0.5*\d) {$\x'_5$};
\node at (\xshift+1.5*\d,\yshift-0.7*\d) {$\x'_7$};
\node at (\xshift+1.5*\d,\yshift-1.3*\d) {$\x'_9$};

\tikzmath{\xshift=5*\d;\yshift=\ybot+0*\d;};

\draw[thick] (\xshift+\d,\yshift+\d)--(\xshift+\d,\yshift);
\draw[thick] (\xshift+2*\d,\yshift+\d)--(\xshift+2*\d,\yshift);


\draw[blackCircle] (\xshift+\d,\yshift+\d) circle;
\draw[whiteCircle] (\xshift+\d,\yshift+0) circle;
\draw[blackCircle] (\xshift+2*\d,\yshift+0) circle;
\draw[whiteCircle] (\xshift+2*\d,\yshift+\d) circle;

\draw[very thick, \colA, <-] (\xshift+0.5*\d,\yshift+\d-0.1*\d)--(\xshift+\d-0.4*\d,\yshift+\d-0.1*\d)..controls(\xshift+\d-0.1*\d,\yshift+\d-0.1*\d)..(\xshift+\d-0.1*\d,\yshift+\d-0.4*\d)--(\xshift+\d-0.1*\d,\yshift+0.4*\d)..controls(\xshift+\d-0.1*\d,\yshift+0.1*\d)..(\xshift+\d-0.4*\d,\yshift+0.1*\d)--(\xshift+0.5*\d, \yshift+0.1*\d);

\draw[very thick, \colB, ->] (\xshift+2.5*\d,\yshift+\d-0.1*\d)--(\xshift+2.4*\d,\yshift+\d-0.1*\d)..controls(\xshift+2.1*\d,\yshift+\d-0.1*\d)..(\xshift+2.1*\d,\yshift+\d-0.4*\d)--(\xshift+2.1*\d,\yshift+0.4*\d)..controls(\xshift+2.1*\d,\yshift+0.1*\d)..(\xshift+2.4*\d,\yshift+0.1*\d)--(\xshift+2.5*\d,\yshift+0.1*\d);

\draw[very thick, \colC, <-] (\xshift+0.5*\d,\yshift-0.1*\d) -- (\xshift+2.5*\d,\yshift-0.1*\d);

\draw[very thick, \colD, ->] (\xshift+0.5*\d,\yshift+1.1*\d) -- (\xshift+2.5*\d,\yshift+1.1*\d);

\node at (\xshift+1.5*\d,\yshift+1.3*\d) {$\gamma'_{c,1}$};
\node at (\xshift+2.55*\d,\yshift+0.5*\d) {$\gamma'_{c,2}$};
\node at (\xshift+1.5*\d,\yshift-0.3*\d) {$\gamma'_{c,3}$};
\node at (\xshift+0.5*\d,\yshift+0.5*\d) {$\gamma'_{c,4}$};

\node at (\xshift+3*\d,\yshift+0.5*\d) {$\x'_8$};
\node at (\xshift+1.5*\d,\yshift+0.5*\d) {$\x'_4$};

\end{tikzpicture}
}
\scalebox{0.8}{
\begin{tikzpicture}

\tikzmath{\xs=0;\ys=0;}

\draw[styleQuiverEdge] (\xs+1,\ys+2) -- (\xs+0,\ys+1);
\draw[styleQuiverEdge] (\xs+1,\ys+2) -- (\xs+2,\ys+1);
\draw[styleQuiverEdge, gray] (\xs-1,\ys+1) -- (\xs+1,\ys+2);
\draw[styleQuiverEdge, gray] (\xs+3,\ys+1) -- (\xs+1,\ys+2);

\draw[styleQuiverEdge] (\xs,\ys+1) -- (\xs-1,\ys+1);
\draw[styleQuiverEdge] (\xs+0,\ys+1) -- (\xs+1,\ys+1);
\draw[styleQuiverEdge] (\xs+2,\ys+1) -- (\xs+1,\ys+1);
\draw[styleQuiverEdge] (\xs+2,\ys+1) -- (\xs+3,\ys+1);

\draw[styleQuiverEdge, gray] (\xs+3,\ys+1) -- (\xs+2,\ys);
\draw[styleQuiverEdge, gray] (\xs-1,\ys+1) -- (\xs,\ys);

\draw[styleQuiverEdge] (\xs+1,\ys+1) -- (\xs+1,\ys+0);
\draw[styleQuiverEdge] (\xs+1,\ys+1) -- (\xs+1,\ys+2);
\draw[styleQuiverEdge] (\xs+1,\ys+0) -- (\xs+2,\ys+0);
\draw[styleQuiverEdge] (\xs+1,\ys+0) -- (\xs+0,\ys+0);
\draw[styleQuiverEdge] (\xs+2,\ys+0) -- (\xs+2,\ys+1);
\draw[styleQuiverEdge] (\xs+0,\ys+0) -- (\xs+0,\ys+1);

\draw[styleQuiverEdge, gray] (\xs,\ys) -- (\xs+1,\ys-1);
\draw[styleQuiverEdge, gray] (\xs+2,\ys) -- (\xs+1,\ys-1);

\draw[styleQuiverEdge] (\xs+1,\ys-1) -- (\xs+1,\ys);

\draw[quiverVertex] (\xs+1,\ys+2) circle;
\draw[quiverVertex] (\xs-1,\ys+1) circle;
\draw[quiverVertex] (\xs+0,\ys+1) circle;
\draw[quiverVertex] (\xs+1,\ys+1) circle;
\draw[quiverVertex] (\xs+2,\ys+1) circle;
\draw[quiverVertex] (\xs+3,\ys+1) circle;
\draw[quiverVertex] (\xs+0,\ys+0) circle;
\draw[quiverVertex] (\xs+1,\ys+0) circle;
\draw[quiverVertex] (\xs+2,\ys+0) circle;
\draw[quiverVertex] (\xs+1,\ys-1) circle;


\draw[->, very thick] (\xs+1,\ys-3) -- (\xs+1,\ys-3.75);


\tikzmath{\xs=0;\ys=-7.75;}

\draw[styleQuiverEdge] (\xs+1,\ys-1) -- (\xs+0,\ys);
\draw[styleQuiverEdge] (\xs+1,\ys-1) -- (\xs+2,\ys);
\draw[styleQuiverEdge, gray] (\xs-1,\ys) -- (\xs+1,\ys-1);
\draw[styleQuiverEdge, gray] (\xs+3,\ys) -- (\xs+1,\ys-1);

\draw[styleQuiverEdge] (\xs,\ys) -- (\xs-1,\ys);
\draw[styleQuiverEdge] (\xs+0,\ys) -- (\xs+1,\ys);
\draw[styleQuiverEdge] (\xs+2,\ys) -- (\xs+1,\ys);
\draw[styleQuiverEdge] (\xs+2,\ys) -- (\xs+3,\ys);

\draw[styleQuiverEdge, gray] (\xs+3,\ys) -- (\xs+2,\ys+1);
\draw[styleQuiverEdge, gray] (\xs-1,\ys) -- (\xs,\ys+1);

\draw[styleQuiverEdge] (\xs+1,\ys+0) -- (\xs+1,\ys+1);
\draw[styleQuiverEdge] (\xs+1,\ys+0) -- (\xs+1,\ys-1);
\draw[styleQuiverEdge] (\xs+1,\ys+1) -- (\xs+2,\ys+1);
\draw[styleQuiverEdge] (\xs+1,\ys+1) -- (\xs+0,\ys+1);
\draw[styleQuiverEdge] (\xs+2,\ys+1) -- (\xs+2,\ys+0);
\draw[styleQuiverEdge] (\xs+0,\ys+1) -- (\xs+0,\ys+0);

\draw[styleQuiverEdge, gray] (\xs,\ys+1) -- (\xs+1,\ys+2);
\draw[styleQuiverEdge, gray] (\xs+2,\ys+1) -- (\xs+1,\ys+2);

\draw[styleQuiverEdge] (\xs+1,\ys+2) -- (\xs+1,\ys+1);

\draw[quiverVertex] (\xs+1,\ys-1) circle;
\draw[quiverVertex] (\xs-1,\ys+0) circle;
\draw[quiverVertex] (\xs+0,\ys+0) circle;
\draw[quiverVertex] (\xs+1,\ys+0) circle;
\draw[quiverVertex] (\xs+2,\ys+0) circle;
\draw[quiverVertex] (\xs+3,\ys+0) circle;
\draw[quiverVertex] (\xs+0,\ys+1) circle;
\draw[quiverVertex] (\xs+1,\ys+1) circle;
\draw[quiverVertex] (\xs+2,\ys+1) circle;
\draw[quiverVertex] (\xs+1,\ys+2) circle;

\draw[white] (0,-10)--(1,-10);
\end{tikzpicture}
}
\end{center}
\caption{Left. Tetrahedron equation realized as equality of transfer matrices of perfect networks. Grey crosses are for the points of gluing of four-gonal building blocks. Graphs are related by sequence of four spider-moves $\Rmut = \mu_{7}\mu_{4}\mu_{2}\mu_{3}$. Right. Corresponding quiver before and after mutation.}
\label{fig:tetrahedron}
\end{figure}
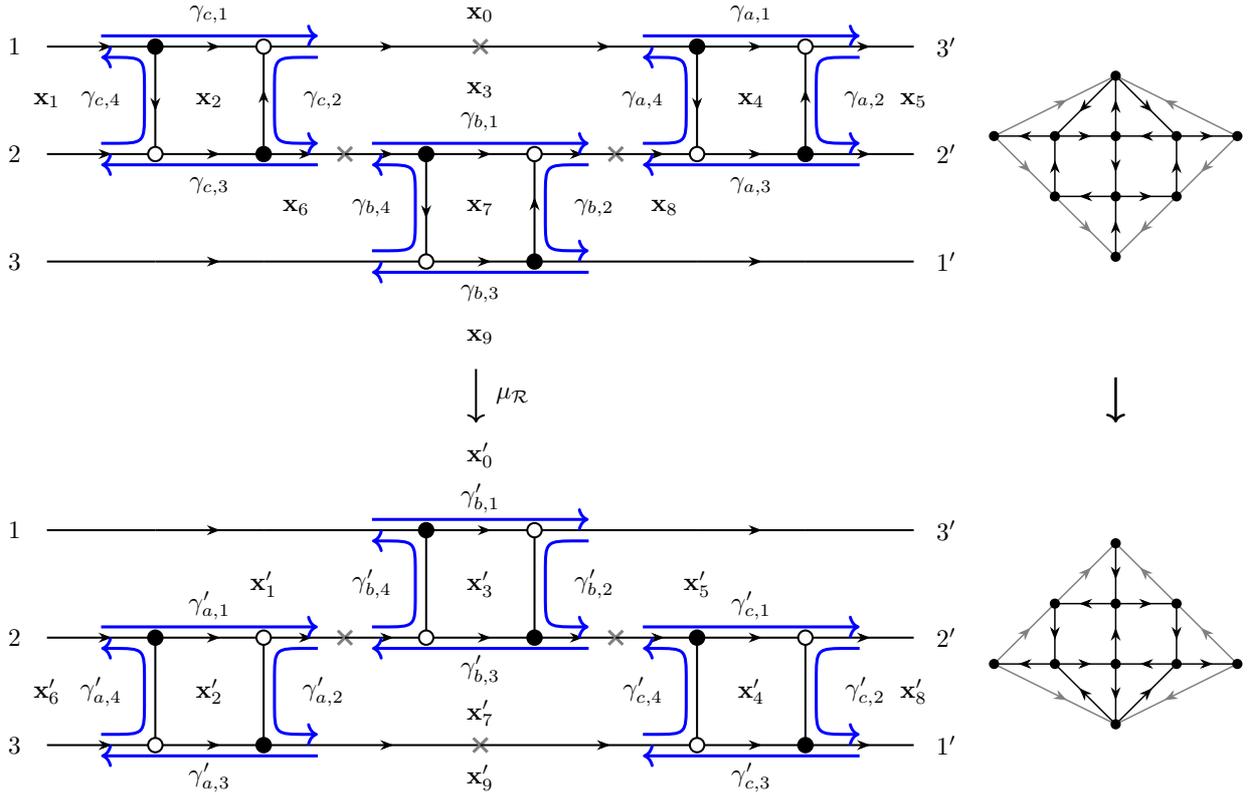

Tetrahedron transformation (\ref{eq:BSTetra}) for the Lax operators $\LBS$ itself recasts into the relation
\begin{equation}
\LCL^{23}(\gamma_a)\LCL^{13}(\gamma_b)\LCL^{12}(\gamma_c)
=
\LCL^{12}(\gamma'_c)\LCL^{13}(\gamma'_b)\LCL^{23}(\gamma'_a) 
\end{equation}
for the transfer matrices of perfect networks. Gluing left and right sides of this equation from the blocks shown in Fig.~\ref{fig:fourgon} gives equality for networks as drawn in Fig.~\ref{fig:tetrahedron}. Note that as in Fig.~\ref{fig:fourgon}, each Lax operator 'permutes' vector spaces. The networks are related by sequence of four spider-moves $\Rmut = \mu_{7}\mu_{4}\mu_{2}\mu_{3}$ supported by two-moves, detailed sequence is shown in Fig.~\ref{fig:tetraMovesDetailed}. Mapping (\ref{eq:BSvarTrans}) being rewritten in $\gamma$-variables using (\ref{eq:BStoClustVar}) results in 
\begin{equation}
\label{eq:tetraClustGamma}
\begin{array}{lll}
\gamma'_{a,1} = \dfrac{\gamma _{a,1} \gamma _{a,4}}{\gamma_{b,4} \gamma_{c,3} [\x_3^{-1},\x_2^{-1}]}\sqrt{\dfrac{\x_4}{\x_2 \x_3} A},
&
\gamma'_{b,1} = \dfrac{\gamma _{a,1} \gamma _{c,1}}{[\x_3^{-1}]},
&
\gamma'_{c,1} = \dfrac{\gamma _{c,1} \gamma _{c,2}}{\gamma_{a,3} \gamma_{b,2} [\x_3^{-1},\x_4^{-1}]} \sqrt{\dfrac{\x_2}{\x_4 \x_3} A},
\\
\gamma'_{a,2} = \dfrac{\gamma _{a,2} \gamma _{b,4} \gamma _{c,3} }{\gamma _{a,4}} \x_2 \x_3 [\x_3^{-1},\x_2^{-1}],
&
\gamma'_{b,2} = \dfrac{\gamma_{b,2} [\x_3^{-1}]}{\gamma _{a,1}\gamma _{a,4} \gamma_{c,1} \gamma _{c,2}} \sqrt{\dfrac{\x_3}{\x_2 \x_4}\dfrac{1}{A}},
&
\gamma'_{c,2} = \gamma _{b,2} \gamma _{a,3} \x_4 \x_3 [\x_3^{-1},\x_4^{-1}],
\\
\gamma'_{a,3} = \dfrac{\gamma _{a,3} \gamma _{a,4}}{\gamma _{b,4} \gamma _{c,3} [\x_3^{-1},\x_2^{-1}]} \sqrt{\dfrac{\x_4}{\x_2 \x_3} A},
&
\gamma'_{b,3} = \dfrac{\gamma _{b,3} \gamma _{a,1} \gamma _{a,4} \gamma _{c,1} \gamma _{c,2}}{\x_3[\x_3^{-1}]},
&
\gamma'_{c,3} = \dfrac{\gamma _{c,3} \gamma _{c,2}}{\gamma _{a,3} \gamma _{b,2} [\x_3^{-1},\x_4^{-1}]} \sqrt{\dfrac{\x_2}{\x_4 \x_3} A},
\\
\gamma'_{a,4} = \gamma_{b,4} \gamma_{c,3} \x_2 \x_3 [\x_3^{-1},\x_2^{-1}],
&
\gamma'_{b,4} = \dfrac{\gamma _{b,4} [\x_3^{-1}]}{ \gamma _{a,1} \gamma _{a,4} \gamma _{c,1} \gamma _{c,2}} \sqrt{\dfrac{\x_3}{\x_2 \x_4}\dfrac{1}{A}},
&
\gamma'_{c,4} = \dfrac{\gamma_{a,3} \gamma _{b,2} \gamma _{c,4} }{\gamma _{c,2}} \x_4 \x_3 [\x_3^{-1},\x_4^{-1}]
\end{array}
\end{equation}
where $A = 1 + \x_3^{-1} + \x_7^{-1}[\x_3^{-1},\x_2^{-1}][\x_3^{-1},\x_4^{-1}]$, $[x]=1+x,~[x,y] = 1 + x(1+y)$ and locations of face variables $\x_i$ are shown in Fig.~\ref{fig:tetrahedron}.
Their explicit expressions in terms of \(\gamma\)-variables are
\begin{equation}\begin{gathered}
\label{eq:1}
\mathbf{x}_2=\frac1{\gamma_{c,1}\gamma_{c,4}\gamma_{c,3}\gamma_{c,2}},\quad
\mathbf{x}_3=\gamma_{b,1}\gamma_{a,4}\gamma_{c,2},\quad
\mathbf{x}_4=\frac1{\gamma_{a,1}\gamma_{a,4}\gamma_{a,3}\gamma_{a,2}},\quad
\mathbf{x}_7=\frac1{\gamma_{b,1}\gamma_{b,4}\gamma_{b,3}\gamma_{b,2}},\\
\mathbf{x}_1=\gamma_{c,4}\times (\text{weights of other boundaries}),\quad
\mathbf{x}_5=\gamma_{a,2}\times (\text{weights of other boundaries}),\\
\mathbf{x}_6=\gamma_{b,4}\times (\text{weights of other boundaries}),\quad
\mathbf{x}_8=\gamma_{b,2}\times (\text{weights of other boundaries}),
\end{gathered}\end{equation}
where by ``weights of the other boundaries'' we denote a product of the \(\gamma\)-variables that correspond to the other boundaries of the face corresponding to given \(\mathbf{x}\)-variable, which are unimportant as neither transform under four- and two-moves, nor contribute into the transfer matrices of flows.

It is easy to check that formulas (\ref{eq:tetraClustGamma}) are consistent with the mapping of $\mathcal{X}$-cluster variables
\begin{equation}
\label{eq:tetraXtrans}
\begin{array}{c}
\x'_0 = \x_0 \dfrac{1}{[\x_3^{-1}]},
\\
\x'_1 =  \x_1 \dfrac{[\x_3^{-1}]}{[\x_3^{-1},\x_2^{-1}]},
~~~
\x'_2 = \dfrac{1}{\x_2 \x_3}\dfrac{1}{A},
~~~
\x'_3 = \x_3 \x_2 \x_4 \x_7 A,
~~~
\x'_4 = \dfrac{1}{\x_3 \x_4}\dfrac{1}{A},
~~~
\x'_5 = \x_5 \dfrac{[\x_3^{-1}]}{[\x_3^{-1},\x_4^{-1}]},
\\
\x'_6 = \x_6 \cdot \x_3 \x_2 [\x_3^{-1},\x_2^{-1}],
~~~
\x'_7 = \dfrac{1}{\x_7}\dfrac{[\x_3^{-1},\x_2^{-1}][\x_3^{-1},\x_4^{-1}]}{[\x_3^{-1}]},
~~~
\x'_8 = \x_8 \cdot \x_3 \x_4[\x_3^{-1},\x_4^{-1}],
\\
\x'_9 = \x_9 \dfrac{\x_7}{[\x_3^{-1},\x_2^{-1}][\x_3^{-1},\x_4^{-1}]}A,
\end{array}
\end{equation}
under $\Rmut$ which follows from (\ref{eq:mutX}). Trying to recover formulas  (\ref{eq:tetraClustGamma}) using 'refined' formulas (\ref{eq:twoMoveBlack}) -- (\ref{eq:fourMove}) for transformation of corner variables one faces problems. In Appendix A we explain how these problems can be treated successfully.

\section{Integrable system for arbitrary Newton polygon}
\label{s:clustint}

In this section we give explicit construction for bi-coloured graph $G$ defining integrable system with arbitrary Newton polygon. It will turn out that for symmetric Newton polygon Lax operator is 'patchworked' by contraction of '$\mathrm{XXZ}$ spin chain' rectangular blocks (see Fig.~\ref{fig:generalNewt}), which are made from tetrahedron Lax operators (\ref{eq:LaxCL}). This extends results of \cite{BS} and \cite{MS} to the case of non-rectangular Newton polygons.

Then we will show, how our constructions come out in the approach to cluster integrable systems via double Bruhat cells in $\GLb(N)$. Tetrahedron Lax operator will be identified with generator $s_i \bar{s}_i$ of diagonal sub-group $W(A_{N-1}^{(1)}) \subset W(A_{N-1}^{(1)} \times A_{N-1}^{(1)})$, and tetrahedron transformation --- with the Coxeter relation there. Embedding of commuting subgroups $\GLb(a_1)\times ... \times \GLb(a_n) \subset \GLb(N),~N = a_1+...+a_n$, will provide natural framework for the construction of Bruhat cell for arbitrary symmetric Newton polygon. Finally, we will construct double Bruhat cells for non-symmetric Newton polygons via triangular decomposition of Lax operators, discuss additional freedom, coming from Newton polygons with sides, containing internal integral points, and prove classification theorem for perfect networks on torus.

\subsection{Spectral curve and perfect network on torus}

To the moment we were considering bi-coloured graphs on disks only. Integrable system appears once we consider network on torus: due to \cite{GK:2011}, spectral curve, which is generating function of Hamiltonians of integrable system
\begin{equation}
\label{eq:spectalGK}
S = \{ (\lambda,\mu) \in \mathbb{C}^* \times \mathbb{C}^* ~|~ S(\lambda,\mu) = \sum_{(i,j)\in \mathbb{Z}^2} \lambda^{i}\mu^{j} H_{ij} = 0 \},
\end{equation}
is equal to the partition function of flows on perfect network $N=(G,w)$ on torus $\T2$. In this subsection we are going to explain how spectral parameters $(\lambda,\mu)$ and Hamiltonians of the system appear.\\

There are two major differences in structure of $\XG$ for the network $N=(G,w)$ on torus $\T2$ compared to the case of disk. First, there are no open faces, so $\H1(G,\partial G) = \H1(G)$, and second, not any path on $G$ can be decomposed as a sum of paths along boundaries of faces, one has to take also representatives of $\H1(\T2)$. Bringing this together we can uniquely decompose any closed path $\gamma \in \H1(G)$ into
\begin{equation}
\label{eq:torusDecomp}
\gamma = n_A(\gamma) \gamma_A + n_B(\gamma) \gamma_B + \sum_{f_i \in \F} n_i(\gamma) \partial f_i,
\end{equation}
where $\F$ is the set of faces of graph $G$ embedded into $\T2$, and $\gamma_A, \gamma_B$ is fixed pair of paths on $G$, which represent two classes in homologies of torus with non-trivial intersection. The best choices for $\gamma_{A,B}$ are zig-zag paths $\Zig$ (those oriented paths which turn left at each white vertex and turn right at each black one) because, as it is easy to see, all face variables $\x_i$ and all zig-zag variables $\z_\alpha = w_{\zig_\alpha},~\zig_\alpha \in \Zig$ are Poisson-commuting
\begin{equation}
\{ \x_i, \z_\alpha \} = 0, ~~~ \{ \z_\alpha, \z_\beta \} = 0.
\end{equation}
with respect to the bracket (\ref{eq:poissonClust}), so they are good candidates for the role of 'spectral parameters'. For further convenience, we fix trivialization $\H1(\T2)=\mathbb{Z}^2$ by choosing a pair of cycles on torus $\ha,\hb$ with simple intersection $<\ha ,\, \hb > = 1$, so one can assign a vector $\vec{u}_\alpha = (a_\alpha,b_\alpha) \in \H1(\T2)$ to each zig-zag $[\zig_\alpha] = a_\alpha [\ha] + b_\alpha [\hb]$. It often happens that the lattice $[\Zig]$, generated by classes $\vec{u}_{\alpha}$ of all zig-zags, does not generate entire lattice $\mathbb{Z}^2=\H1(\T2)$, but some sub-lattice of finite index $|\H1(\T2)/[\Zig]| = d$ instead. In those cases there is no way to choose any pair of zig-zags $z_{\alpha},z_{\beta} \in \Zig$ to be 'basic cycles' $\gamma_A = z_{\alpha}$ and $\gamma_B = z_{\beta}$, and express all classes in homologies as their integral combination. In such cases one has to make coefficients $n_A, n_B, n_i$ in (\ref{eq:torusDecomp}) rational numbers with denominators being divisors of $d$ instead. So we get an embedding of finite index $\H1(G)\subset \mathbb{Z}^{2}\oplus \frac{1}{d}\mathbb{Z}^{|F|}$ which implies decomposition for the space of functions
\begin{equation}
\label{eq:XGpoiss}
\XG = \mathbb{C}[(\H1(G))^*] \subset \mathbb{C}[\lambda^{\pm1}, \mu^{\pm1}] \otimes \mathbb{C}[\x_i^{\pm1/d}]_{i\in \F},
\end{equation}
where $\lambda = (\zeta_\alpha)^{k_{A,\alpha}}(\zeta_\beta)^{k_{A,\beta}}$, $\mu = (\zeta_\alpha)^{k_{B,\alpha}}(\zeta_\beta)^{k_{B,\beta}}$ will have powers chosen so that
\begin{equation}
k_{A,\alpha} \vec{u}_{\alpha} + k_{A,\beta} \vec{u}_{\beta} = (1,0),
~~~
k_{B,\alpha} \vec{u}_{\alpha} + k_{B,\beta} \vec{u}_{\beta} = (0,1),
\end{equation}
so that $\lambda, \mu$ are variables corresponding to generators $(1,0)$ and $(0,1)$ of homologies, and will play the role of 'spectral parameters' in the following. Now, spectral curve of cluster integrable system defined by perfect network $N = (G,w)$ can be calculated as partition function of flows
\begin{equation}
\label{eq:spectTor}
S(\lambda,\mu) = \Z_{\T2} = \sum_{p\in \mathcal{F}_{\T2}} w_{p} = \sum_{p\in \mathcal{F}_{\T2}} \lambda^{<\,[p],\, \hb >}\mu^{-<\, [p],\, \ha >} \prod_{f_i \in \F}\x_i^{n_i(p)} = \sum_{(i,j) \in \Newt} \lambda^{i} \mu^{j} H_{ij},
\end{equation}
where $\mathcal{F}_{\T2}$ is the set of flows on torus, Hamiltonians $H_{ij} = H_{ij}(\{ \x_a \})$ depend only on face variables $\x_i$, which are $\mathcal{X}$-cluster coordinates, and set $\Newt\subset \mathbb{Z}^2$ is convex envelope of those $(i,j)\in \mathbb{Z}^2$ for which $H_{ij}$ are non-zero, and is called Newton polygon of curve $S$. It was proved in \cite{GK:2011} that for minimal bi-coloured graphs there exist special perfect orientations, called $\alpha$-orientations (we will give both definitions in a moment)\footnote{Actually, logic of \cite{GSV:2009} and \cite{GSTV:2014} suggests that similar statement holds for any perfect orientation, however the understanding of this point is still missing in the literature.} for which following theorem holds.

\paragraph*{Theorem (\cite{GK:2011}).}{Let $N = (G,w)$ be $\alpha$-orientated perfect network on torus with minimal bi-coloured graph and (\ref{eq:spectTor}) be partition function of flows on it. Then:
\begin{enumerate}
\item Hamiltonians corresponding to boundary points of $\Newt$ are Casimir functions.\footnote{Spectral parameters $(\lambda, \mu)$ are obviously Casimirs as well, as they are expressible via zig-zag variables only.}
\item Hamiltonians corresponding to internal points are algebraically independent and in involution
\begin{equation}
\{ H_{ij},H_{kl} \}=0.
\end{equation}
\item Number of Hamiltonians (which are not Casimirs) is exactly half of the dimension of symplectic leaf.
\item The Newton polygon $\Newt$ is the unique (up to permutation of collinear vectors) convex polygon whose set of primitive oriented boundary intervals is $\{\vec{u}_k\}_{k=1}^{|\Zig|}$.\footnote{It might be so because there is a pair of zig-zags which travel in two opposite directions along each edge of graph, so $\sum_k [\zig_k] = \sum_k \vec{u}_k = 0$.}
\end{enumerate}
}
Together statements 1-3 imply integrability of the system. Statement 4 gives simple way to predict shape of the Newton polygon without computation of entire spectral curve. We will use it intensively in the following. By deforming slightly zig-zag paths from the graph, so that they cross edges only at grey vertices, and erasing graph itself, one obtains so-called wiring diagram. This operation is invertible: it is easy to recover the graph from its wiring diagram \cite{GK:2011}.

Partition function of flows $\Z_{\T2}$ on toric network $N = (G,w)$, $G \subset \T2$ can be obtained by gluing of sides of the disk with network $\tilde{N} = (\tilde{G},w)$. To do this, divide boundary of the disk into four clockwise oriented segments $\ell_a, \ell_b, \ell_c, \ell_d$ with no sources or sinks at the points of contact of segments. The gluing is possible if one can find such continuous monotonic map $j_1:\ell_a\to \ell_c$ that puts beginning of $\ell_a$ to the end of $\ell_c$, end of $\ell_a$ to the beginning of $\ell_c$, sources to sinks and sinks to sources, and similar map for $j_2:\ell_b \to \ell_d$. If one found $j_{1,2}$, then the partition function of flows on perfect network $N$ on torus is related to $\tilde{N}$ on the disk, from which the torus is glued with $j_{1,2}$, by
\begin{equation}
\Z_{\T2} = \sum_{A\subset I_a} \sum_{B\subset I_b} \sum_{C\subset I_c} \sum_{D\subset I_d} \Z_{\tilde{N}}(A \cup B \cup C \cup D \to j_1(A) \cup j_2(B) \cup (j_1)^{-1}(C) \cup (j_2)^{-1}(D))
\end{equation}
where $I_k$ and $J_k$ are sets of sinks and sources on $\ell_k$ for $k\in \{a,b,c,d\}$, and we use identifications $j_1(I_a)=J_c, ~ j_2(I_b)=J_d, ~ j_1^{-1}(I_c)=J_a, ~ j_2^{-1}(I_d)=J_b$. Term with chosen subsets $(A,B,C,D)$ contributes to Hamiltonian $H_{|C|-|A|,|D|-|B|}$, if generators of $\H1(\T2)$ are chosen to be $\ha = -\ell_b = \ell_d$ and $\hb = \ell_a = - \ell_c$ respectively. To obtain the same partition function using transfer matrix of flows, one can 'take trace' of transfer matrix by contraction of spaces whose boundary points are glued by $j_{1,2}$. With explicit dependence on $\lambda$ and $\mu$ (which are not $\lambda,\mu$ from (\ref{eq:XGpoiss}), but just generating parameters, keeping trace of classes in homologies) incorporated it looks
\begin{equation}
\label{eq:partOnTorFromT}
\Z_{\T2} =
\mathrm{Tr}_{j_1,j_2}
\left(
T_{N} \circ \lambda^{\hat{P}_{J_a} - \hat{P}_{J_c}} \mu^{\hat{P}_{J_b} - \hat{P}_{J_d}}
\right),
~~~
\hat{P}_{X} = 
\sum\limits_{i\in X} \frac{1}{2}(\mathrm{1}+\hat{\sigma}_{z,i}),
\end{equation}
where $\hat{\sigma}_{z,i}=\mathrm{1}\otimes ... \otimes \sigma_z \otimes ... \otimes \mathrm{1}$ is operator acting by $\sigma_z$-matrix in space $i$, and by unity in all other spaces.
\\

Now, it remains to construct special orientation for network on torus, for which Hamiltonians are involutive. We construct it using so-called dominant orientation for network on disk. In the following we will be considering only graphs called minimal graphs, for which zig-zags do not have self-intersections, there are no closed zig-zags (i.e. those isotopic to $S^1$) and no parallel bigons of zig-zags. For minimal planar graphs, we can label zig-zags by their staring points.

Take any linear order $\leqslant$ on the set of zig-zags, i.e. for any pair of zig-zags $z_1$ and $z_2$ set $z_1 \leqslant z_2$ or $z_2 \leqslant z_1$. For intersecting zig-zags order must be strict, those zig-zags which do not have intersection points could be equal in this order. Take any black or white vertex $v$, and let zig-zags which pass it are $z_{a} < z_{a-1} < ... < z_1$, where $a$ is the degree of the vertex. We say that $z_a$ is the lowest zig-zag at $v$ and $z_1$ is the highest zig-zag at $v$. The order is said to be consistent at $v$ if it satisfies the following requirements:
\begin{itemize}
\item If zig-zag $z_1$ is highest at $v$, then it is highest in the next vertex along $z_1$ if the next vertex is black, and in the previous vertex along $z_1$ if the previous vertex is white. Note, that the both cases could occur at the same vertex, as we do not demand graph to be bipartite.
\item Any other zig-zags $z_i,~i = 2,...,a$ is not the highest in the next vertex along $z_i$ if the next vertex is black, and in the previous vertex along $z_i$, if the previous vertex is white. 
\end{itemize}
The order is consistent if it is consistent at all vertices. To construct perfect orientation on the graph by ordering on zig-zags, define first orientation on fans of all internal vertices. For any black vertex the only incoming half-edge is those, along which the highest zig-zag come to the vertex, and all the other are outgoing. For any white vertex the only outgoing half-edge is those, along which the highest zig-zag leave the vertex, and all the other are incoming. It is easy to see that if the order on zig-zags is consistent, then orientations of halves of all the internal edges are consistent. We do not give explicit description of the set of all consistent orders on zig-zags, however make the following

\paragraph*{Conjecture.}{All perfect orientations without oriented loops for graphs on disks are orientations constructed from some consistent orders on zig-zags.} \\

If one glue pair of disks $D_1$ and $D_2$, each equipped with dominant orientation, the dominant orientation on $D_1\cup_{\ell} D_2$ can also be obtained, once the orders on zig-zags are concerted, and consistency condition at boundary vertices holds (note that all the gluings in Section \ref{ss:tetra} was so). The same is true also for gluing disk into torus. Now, construct $\alpha$-ordering by taking any zig-zag to be the highest among all, and other zig-zags to be ordered according to counter-clockwise order of their classes in $\H1(\T2, \mathbb{Z})$ considered as vectors in $\mathbb{Z}^2$. As it was claimed, for orientation built from such ordering, Hamiltonians $H_{ij}$ are involutive.

\subsection{Integrable system with symmetric Newton polygon}
\label{ss:generalNewt}
In this sub-section using four-gonal block from Fig.~(\ref{fig:fourgon}) we construct cluster integrable systems with arbitrary symmetric Newton polygon. As it was discussed in the previous sub-section, for this it is enough to construct such bi-coloured graph, that collection of homology classes of its zig-zags coincides with the set of oriented boundary intervals of the Newton polygon.\\

\begin{figure}[!ht]
\begin{center}
\begin{tikzpicture}

\tikzmath{\xs=0;\ys=0;\d=1;\Lx=4;\Ly=4;};


\foreach \x in {0,...,\Lx}
	\foreach \y in {0,...,\Ly}
		\draw[fill] (\xs+\d*\x,\ys+\d*\y) circle[radius=0.05];

\draw[red, very thick, styleArrowShort] (\xs, \ys+\d) -- (\xs+\d, \ys);
\draw[green, very thick, styleArrowShort] (\xs+\d, \ys) -- (\xs+4*\d, \ys + 2*\d);
\draw[blue, very thick, styleArrowShort] (\xs+4*\d, \ys+2*\d) -- (\xs+4*\d, \ys + 3*\d);
\draw[red, very thick, styleArrowShort] (\xs+4*\d, \ys+3*\d) -- (\xs+3*\d, \ys+4*\d);
\draw[green, very thick, styleArrowShort] (\xs+3*\d, \ys+4*\d) -- (\xs, \ys + 2*\d);
\draw[blue, very thick, styleArrowShort] (\xs, \ys+2*\d) -- (\xs, \ys+\d);

\node[anchor = west] at (4.5*\d,3*\d) {$\vec{u}_1 = (1,-1)$};
\node[anchor = west] at (4.5*\d,2*\d) {$\vec{u}_2 = (3,2)$};
\node[anchor = west] at (4.5*\d,1*\d) {$\vec{u}_3 = (0,1)$};

\draw[white] (\xs+7.5*\d,\ys+4.2*\d) -- (\xs+7.5*\d,\ys+3.2*\d);

\end{tikzpicture}
\begin{tikzpicture}

\tikzmath{\xs=0;\ys=0;\d=1;\Lx=5;\Ly=6;\ds=0.1;};

\clip (0*\d - \ds, 1*\d - \ds) rectangle (4*\d + \ds, 5*\d+\ds);

\draw[step=\d, shift={(0,0)}, dashed, thin]
	(\xs-1*\d,\ys) grid (\xs+\d*\Lx, \ys+\d*\Ly);


\draw [very thick, red]
	(-1*\d, 4.5*\d) -- (0*\d, 4.5*\d)
	.. controls (0.25*\d, 4.5*\d) and (0.5*\d, 4.25*\d) ..
	(0.5*\d, 4*\d) -- (0.5*\d, 0*\d);


\draw [very thick, blue] (3.5*\d, 0*\d) -- (3.5*\d, 6*\d);


\draw [very thick, green] (-1*\d, 3.5*\d) -- (1*\d, 3.5*\d);
\draw [very thick, green] (3*\d, 3.5*\d) -- (5*\d, 3.5*\d);

\draw [very thick, green] (-1*\d, 2.5*\d) -- (1*\d, 2.5*\d);
\draw [very thick, green] (3*\d, 2.5*\d) -- (5*\d, 2.5*\d);

\draw [very thick, green] (-1*\d, 1.5*\d) -- (1*\d, 1.5*\d);
\draw [very thick, green] (3*\d, 1.5*\d) -- (5*\d, 1.5*\d);

\draw [very thick, green] (1.5*\d, 0*\d) -- (1.5*\d, 1*\d);
\draw [very thick, green] (1.5*\d, 4*\d) -- (1.5*\d, 6*\d);

\draw [very thick, green] (2.5*\d, 0*\d) -- (2.5*\d, 1*\d);
\draw [very thick, green] (2.5*\d, 4*\d) -- (2.5*\d, 6*\d);

\draw [very thick, green]
	(1*\d, 3.5*\d)
	.. controls (1.25*\d, 3.5*\d) and (1.5*\d, 3.75*\d) ..
	(1.5*\d, 4*\d);

\draw [very thick, green]
	(1*\d, 2.5*\d)
	.. controls (2*\d, 2.5*\d) and (2.5*\d, 3*\d) ..
	(2.5*\d, 4*\d);

\draw [very thick, green]
	(1*\d, 1.5*\d)
	.. controls (1.25*\d, 1.5*\d) and (1.3*\d, 1.75*\d) ..
	(1.5*\d, 2*\d)
	.. controls (1.7*\d, 2.25*\d) ..
	(2*\d, 2.5*\d)
	.. controls (2.3*\d, 2.75*\d) ..
	(2.5*\d, 3*\d)
	.. controls (2.7*\d, 3.25*\d) and (2.75*\d, 3.5*\d) ..
	(3*\d, 3.5*\d);

\draw [very thick, green]
	(1.5*\d, 1*\d)
	.. controls (1.5*\d, 2*\d) and (2*\d, 2.5*\d) ..
	(3*\d, 2.5*\d);

\draw [very thick, green]
	(2.5*\d, 1*\d)
	.. controls (2.5*\d, 1.25*\d) and (2.75*\d, 1.5*\d) ..
	(3*\d, 1.5*\d);
	

\draw [very thick, red]
	(0.5*\d, 6*\d) -- (0.5*\d, 5*\d)
	.. controls (0.5*\d, 4.75*\d) and (0.75*\d, 4.5*\d) ..
	(1*\d, 4.5*\d) -- (5*\d, 4.5*\d);

\draw[thick] (-1*\d,5*\d) rectangle (5*\d,5*\d);
\draw[thick] (-1*\d,4*\d) rectangle (5*\d,4*\d);
\draw[thick] (-1*\d,1*\d) rectangle (5*\d,1*\d);

\draw[thick] (0*\d,0*\d) rectangle (0*\d,6*\d);
\draw[thick] (1*\d,0*\d) rectangle (1*\d,6*\d);
\draw[thick] (3*\d,0*\d) rectangle (3*\d,6*\d);
\draw[thick] (4*\d,0*\d) rectangle (4*\d,6*\d);

\tikzmath{\op=0;};

\node[draw, very thick, diamond, fill=black!\op] at (0.5*\d,1.5*\d) {};
\node[draw, very thick, diamond, fill=black!\op] at (0.5*\d,2.5*\d) {};
\node[draw, very thick, diamond, fill=black!\op] at (0.5*\d,3.5*\d) {};

\node[draw, very thick, diamond, fill=black!\op] at (1.5*\d,4.5*\d) {};
\node[draw, very thick, diamond, fill=black!\op] at (2.5*\d,4.5*\d) {};
\node[draw, very thick, diamond, fill=black!\op] at (3.5*\d,4.5*\d) {};
\node[draw, very thick, diamond, fill=black!\op] at (3.5*\d,3.5*\d) {};
\node[draw, very thick, diamond, fill=black!\op] at (3.5*\d,2.5*\d) {};
\node[draw, very thick, diamond, fill=black!\op] at (3.5*\d,1.5*\d) {};

\end{tikzpicture}
\begin{tikzpicture}

\tikzmath{\xshift=0;\yshift=0;\d=1.2;};

\draw[white] (-2.5*\d,0*\d) -- (-2.5*\d,1*\d);

\draw[very thick] (\xshift-1.5*\d,\yshift)--(\xshift-0.5*\d,\yshift);
\draw[very thick] (\xshift-0.5*\d,\yshift)--(\xshift,\yshift+\d);
\draw[very thick] (\xshift-0.5*\d,\yshift)--(\xshift,\yshift-\d);

\draw[very thick] (\xshift+0.5*\d,\yshift)--(\xshift,\yshift+\d);
\draw[very thick] (\xshift+0.5*\d,\yshift)--(\xshift,\yshift-\d);
\draw[very thick] (\xshift+1.5*\d,\yshift)--(\xshift+0.5*\d,\yshift);

\draw[very thick] (\xshift,\yshift+\d)--(\xshift,\yshift+1.5*\d);
\draw[very thick] (\xshift,\yshift-\d)--(\xshift,\yshift-1.5*\d);

\draw[blackCircle] (\xshift - 0.5*\d,\yshift) circle;
\draw[blackCircle] (\xshift + 0.5*\d,\yshift) circle;
\draw[whiteCircle] (\xshift,\yshift + \d) circle;
\draw[whiteCircle] (\xshift,\yshift - \d) circle;

\draw[very thick, \colZ, ->] (\xshift-0.25*\d,\yshift-1.5*\d) -- (\xshift-0.25*\d,\yshift+1.5*\d);
\draw[very thick, \colZ, ->] (\xshift+0.25*\d,\yshift+1.5*\d) -- (\xshift+0.25*\d,\yshift-1.5*\d);

\draw[very thick, \colZ, <-]
	(\xshift-1.5*\d,\yshift-0.25*\d)
	..controls (\xshift-1*\d,\yshift-0.25*\d) and (\xshift-0.75*\d,\yshift+0.5*\d)..
	(\xshift-0.25*\d,\yshift+0.5*\d)--(\xshift+0.25*\d,\yshift+0.5*\d)
	..controls (\xshift+0.75*\d,\yshift+0.5*\d) and (\xshift+1*\d,\yshift-0.25*\d)..
	(\xshift+1.5*\d,\yshift-0.25*\d);

\draw[very thick, \colZ, ->]
	(\xshift-1.5*\d,\yshift+0.25*\d)
	..controls (\xshift-1*\d,\yshift+0.25*\d) and (\xshift-0.75*\d,\yshift-0.5*\d)..
	(\xshift-0.25*\d,\yshift-0.5*\d)--(\xshift+0.25*\d,\yshift-0.5*\d)
	..controls (\xshift+0.75*\d,\yshift-0.5*\d) and (\xshift+1*\d,\yshift+0.25*\d)..
	(\xshift+1.5*\d,\yshift+0.25*\d);


\end{tikzpicture}
\end{center}
\caption{Left. Example of the Newton polygon. Center. Schematic drawing of the graph on torus. Four-gonal blocks are drawn in details on the right panel. Edges are coloured according to the colours of zig-zags going along them, by colours from the left panel. Right. Detailed view on graph and on wiring diagram of zig-zags at the intersection points.}
\label{fig:generalNewt}
\end{figure}
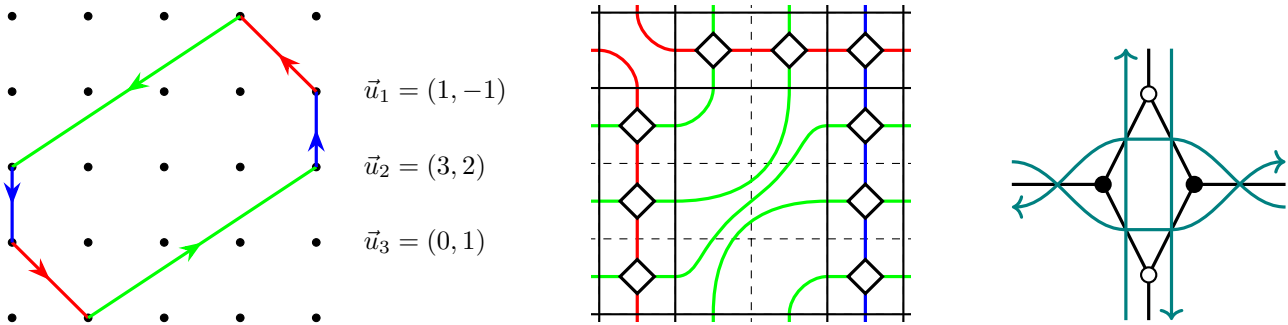

We say that Newton polygon is 'symmetric' if it is invariant under the central symmetry $(i,j)\mapsto (-i,-j)$, see e.g. Fig.~\ref{fig:generalNewt}. Due to the symmetry, it always has even number of vertices --- it is $2n$-gonal. Let's select any point with the minimal $i$-coordinate. Starting from this point, we enumerate all oriented boundary intervals $\vec{u}_1=(a_1,b_1),\vec{u}_2=(a_2,b_2),...,\vec{u}_n=(a_n,b_n)$ in counter-clockwise direction, until the point, which is symmetric to the initial one. Since we started from the left-most point, then all $a_i>0$. We assume also that all intervals are primitive, i.e. $\mathrm{gcd}(a_i,b_i)=1$. Opposite half of polygon, which starts at rightmost point, and ends at leftmost, consists of vectors with coordinates $-\vec{u}_1,...,-\vec{u}_n$.

Decompose fundamental domain of torus into grid of $n\times n$ rectangular blocks. Diagonal block at $i$-th position has $a_i$ sources on its left side, $a_i$ sinks on its right side, $|b_i|$ sinks and sources on upper and lower sides respectively if $b_i>0$, or visa versa if $b_i<0$. Edges are non-intersecting, and if $b_i > 0$, then graph is constructed by iterative connection of closest non-connected sources with sinks by edges starting from top-left corner, while if $b_i < 0$, the process of connection starts from bottom-left corner, see example in Fig.~\ref{fig:generalNewt}, center. Non-diagonal block at row $i$ (counting from the top) and column $j$ (counting from the left) is $a_i \times |b_j|$ 'fence net' bipartite graphs, which is rectangular grid glued from four-gonal blocks. As it is shown in Fig.~\ref{fig:generalNewt}, right, at each four-gonal block zig-zag paths are going without changing of direction, so it is easy to convince yourself that the classes of zig-zags in $\H1(\T2)$ are precisely $\vec{u}_1,...,\vec{u}_n, -\vec{u}_1,..., -\vec{u}_n$ as required.

\paragraph{Remark.}{ Bi-coloured graphs on torus obtained in this way might be not simple because of parallel bigons. The evidence for this, is that the graph constructed by proposed recipe, for each pair of boundary intervals $\vec{u}_i$ and $\vec{u}_j$, has $|a_i b_j| + |a_j b_i|$ four-gonal blocks at their intersection points, which is not $\mathrm{SA}(2,\mathbb{Z})$-invariant quantity. Obtaining minimal graph, which is necessary for integrability theorem, requires additional spider-moves and parallel bigon reductions. As an illustration, interested reader can try to construct graph and reduce it for the Newton polygon obtained from the one drawn in Fig.~\ref{fig:generalNewt} by transformation $x\mapsto x+y, y\mapsto y$.
}\\

Transfer matrix of each four-gonal block is $\LCL$ from (\ref{eq:LaxCL}), which we identified with the solution of tetrahedron equation. If all blocks are oriented as in Fig.~\ref{fig:fourgon}, top, then the global orientation turns out to be the $\alpha$-orientation, so does not have oriented cycles. It is known since \cite{BS}, and redirived in the context of cluster integrable systems in \cite{MS}, that 'fence net' $a\times b$ block being glued by pairs of opposite sides to cylinder defines either Lax operator of $\mathfrak{gl}_a$ classical $\mathrm{XXZ}$ spin chain on $b$ sites or $\mathfrak{gl}_b$ chain on $a$ sites, depending on pair of sides chosen to be glued. As we remarked in (\ref{eq:LaxXXZ}), it was noted in \cite{BS} that the result of contraction of tetrahedron Lax operators decomposes into direct sum of Lax operators for $\mathrm{XXZ}$ chain with auxiliary space being sum of all fundamental representations of $\mathfrak{gl}_a$
\begin{equation}
(\mathbb{C}^{2})^{\otimes a} = \bigoplus\limits_{i=0}^{a} \mathbb{C}^{{a\choose i}}
~ \Rightarrow ~
T(\mu) = \bigoplus\limits_{i=0}^{a} \mathcal{L}_{\Lambda^i \mathbb{C}^a}(\mu).
\end{equation}
In our approach this is the result of the natural grading by the number of paths which cross cylinder from the left to the right, and implication\footnote{This is not LGV lemma itself, as we deal with cylinder. Some subtleties with spectral parameter and its signs appear because of the closed paths which go around cylinder. For discussions see \cite{GSV:2009-1} and \cite{LP}.} of LGV lemma \cite{GV, L}. Dependence on spectral parameter $\mu$ comes from the paths which cross horizontal boundary of fundamental domain, and formula (\ref{eq:partOnTorFromT}).

Cylindric transfer matrix of the system with arbitrary Newton polygon can be obtained by cutting of graph drawn in Fig.~\ref{fig:generalNewt} by vertical line between any pair of columns of four-gons. Due to the chosen orientation, all the sources are located on the left, and all sinks are located on the right side of cylinder. The transfer matrix by cylindric LGV lemma again provides Lax operator acting in direct sum $\bigoplus_{i=0}^{r} \Lambda^{i} \mathbb{C}^{r}$, $r=a_1+...+a_n$. The  first fundamental Lax operator $\mathcal{L}_{\mathbb{C}^r}(\mu)$ satisfies $r$-matrix Poisson bracket (\ref{eq:rLLintro}), as it was proved in \cite{GSV:2009-1}. 

One can keep decomposing cylinder by vertical cuts, up to separating transfer matrix into product of $n$ transfer matrices, each corresponding to flows passing one column in the array of fence-nets. We will clarify how this cylindric blocks are related to combinatorics of affine Weyl groups below. However, we want to stress here, that only the toroidal representation of the system makes $\mathrm{SA}(2,\mathbb{Z})$ covariance explicit.

\subsection{Integrable systems on Poisson-Lie group}

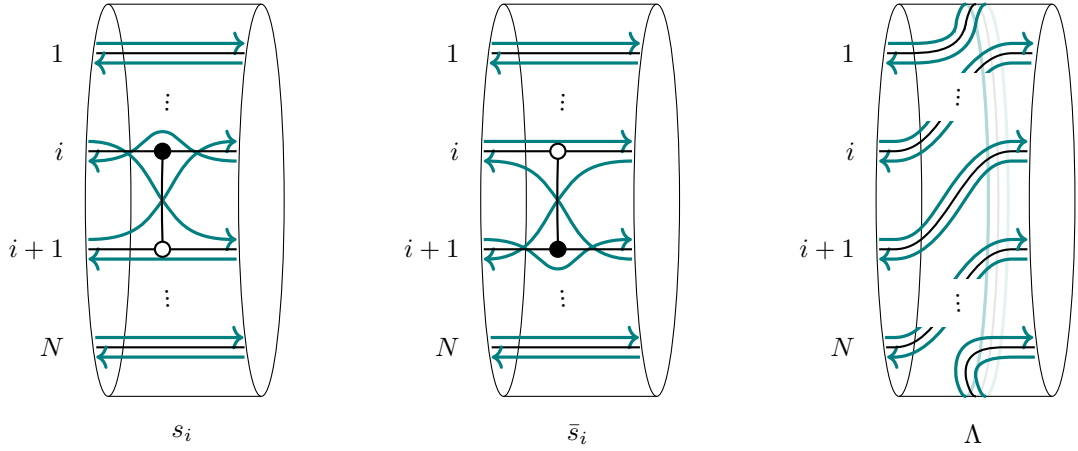
\begin{figure}[h!]
\begin{center}
\scalebox{1}{
\begin{tikzpicture}

\tikzmath{\d=1.3;};


\tikzmath{\xshift=0;\yshift=0;};

\draw
	(\xshift+0.45*\d,\yshift+2.5*\d) --
	(\xshift+2*\d,\yshift+2.5*\d);
\draw
	(\xshift+0.45*\d,\yshift-1.5*\d) --
	(\xshift+2*\d,\yshift-1.5*\d);

\draw (\xshift+0.45*\d, \yshift+0.5*\d) ellipse (0.3 and 2*\d);
\draw (\xshift+2*\d, \yshift+0.5*\d) ellipse (0.3 and 2*\d);


\draw[very thick, \colZ, ->] (\xshift+0.32*\d,\yshift-0.9*\d) -- (\xshift+1.82*\d,\yshift-0.9*\d);
\draw[thick] (\xshift+0.33*\d,\yshift-1*\d)--(\xshift+1.83*\d,\yshift-1*\d);
\draw[very thick, \colZ, <-] (\xshift+0.33*\d,\yshift-1.1*\d) -- (\xshift+1.83*\d,\yshift-1.1*\d);


\node[rotate=94] at (\xshift+1.05*\d,\yshift-0.5*\d) {$...$};


\draw[very thick, \colZ, <-]
	(\xshift+0.25*\d,\yshift+0.9*\d)
	..controls (\xshift+0.8*\d,\yshift+0.9*\d) and (\xshift+0.8*\d,\yshift+1.2*\d)..
	(\xshift+1*\d,\yshift+1.2*\d)
	..controls (\xshift+1.2*\d,\yshift+1.2*\d) and (\xshift+1.2*\d,\yshift+0.9*\d)..
	(\xshift+1.75*\d,\yshift+0.9*\d);

\draw[very thick, \colZ, ->] 
	(\xshift+0.25*\d,\yshift+1.1*\d)
	..controls (\xshift+0.75*\d,\yshift+1.1*\d) and (\xshift+0.85*\d,\yshift+0.7*\d)..
	(\xshift+1*\d,\yshift+0.5*\d)
	..controls (\xshift+1.15*\d,\yshift+0.3*\d) and (\xshift+1.25*\d,\yshift+0.1*\d)..
	(\xshift+1.75*\d,\yshift+0.1*\d);
	
\draw[very thick, \colZ, ->] 
	(\xshift+0.25*\d,\yshift+0.1*\d)
	..controls (\xshift+0.75*\d,\yshift+0.1*\d) and (\xshift+0.85*\d,\yshift+0.3*\d)..
	(\xshift+1*\d,\yshift+0.5*\d)
	..controls (\xshift+1.15*\d,\yshift+0.7*\d) and (\xshift+1.25*\d,\yshift+1.1*\d)..
	(\xshift+1.75*\d,\yshift+1.1*\d);
	
\draw[very thick, \colZ, <-] (\xshift+0.25*\d,\yshift-0.1*\d) -- (\xshift+1.75*\d,\yshift-0.1*\d);


\draw[thick] (\xshift+0.25*\d,\yshift)--(\xshift+1*\d,\yshift);
\draw[thick] (\xshift+1*\d,\yshift)--(\xshift+1.75*\d,\yshift);

\draw[thick] (\xshift+0.25*\d,\yshift+\d)--(\xshift+1*\d,\yshift+\d);
\draw[thick] (\xshift+1*\d,\yshift+\d)--(\xshift+1.75*\d,\yshift+\d);

\begin{scope}
	\clip (\xshift+0.9*\d,\yshift+\d) rectangle (\xshift+1.1*\d,\yshift);
	\draw[thick] (\xshift+1.225*\d, \yshift+0.5*\d) ellipse (0.3 and 2*\d);
\end{scope}

\draw[blackCircle] (\xshift+\d,\yshift+\d) circle;
\draw[whiteCircle] (\xshift+\d,\yshift+0) circle;


\node[rotate=86] at (\xshift+1.05*\d,\yshift+1.5*\d) {$...$};


\draw[very thick, \colZ, ->] (\xshift+0.33*\d,\yshift+2.1*\d) -- (\xshift+1.83*\d,\yshift+2.1*\d);
\draw[thick] (\xshift+0.33*\d,\yshift+2*\d)--(\xshift+1.83*\d,\yshift+2*\d);
\draw[very thick, \colZ, <-] (\xshift+0.31*\d,\yshift+1.9*\d) -- (\xshift+1.81*\d,\yshift+1.9*\d);


\node[anchor = east] at (\xshift+0.1*\d,\yshift+2*\d) {$1$};
\node[anchor = east] at (\xshift+0.1*\d,\yshift+1*\d) {$i$};
\node[anchor = east] at (\xshift+0.1*\d,\yshift) {$i+1$};
\node[anchor = east] at (\xshift+0.1*\d,\yshift-1*\d) {$N$};

\node[anchor = north] at (\xshift+1.2*\d,\yshift-1.7*\d) {$s_i$};


\tikzmath{\xshift=4*\d;\yshift=0;};

\draw
	(\xshift+0.45*\d,\yshift+2.5*\d) --
	(\xshift+2*\d,\yshift+2.5*\d);
\draw
	(\xshift+0.45*\d,\yshift-1.5*\d) --
	(\xshift+2*\d,\yshift-1.5*\d);

\draw (\xshift+0.45*\d, \yshift+0.5*\d) ellipse (0.3 and 2*\d);
\draw (\xshift+2*\d, \yshift+0.5*\d) ellipse (0.3 and 2*\d);


\draw[very thick, \colZ, ->] (\xshift+0.32*\d,\yshift-0.9*\d) -- (\xshift+1.82*\d,\yshift-0.9*\d);
\draw[thick] (\xshift+0.33*\d,\yshift-1*\d)--(\xshift+1.83*\d,\yshift-1*\d);
\draw[very thick, \colZ, <-] (\xshift+0.33*\d,\yshift-1.1*\d) -- (\xshift+1.83*\d,\yshift-1.1*\d);


\node[rotate=94] at (\xshift+1.05*\d,\yshift-0.5*\d) {$...$};


\draw[very thick, \colZ, ->]
	(\xshift+0.25*\d,\yshift+0.1*\d)
	..controls (\xshift+0.8*\d,\yshift+0.1*\d) and (\xshift+0.8*\d,\yshift-0.2*\d)..
	(\xshift+1*\d,\yshift-0.2*\d)
	..controls (\xshift+1.2*\d,\yshift-0.2*\d) and (\xshift+1.2*\d,\yshift+0.1*\d)..
	(\xshift+1.75*\d,\yshift+0.1*\d);

\draw[very thick, \colZ, <-] 
	(\xshift+0.25*\d,\yshift-0.1*\d)
	..controls (\xshift+0.75*\d,\yshift-0.1*\d) and (\xshift+0.85*\d,\yshift+0.3*\d)..
	(\xshift+1*\d,\yshift+0.5*\d)
	..controls (\xshift+1.15*\d,\yshift+0.7*\d) and (\xshift+1.25*\d,\yshift+0.9*\d)..
	(\xshift+1.75*\d,\yshift+0.9*\d);
	
\draw[very thick, \colZ, <-] 
	(\xshift+0.25*\d,\yshift+0.9*\d)
	..controls (\xshift+0.75*\d,\yshift+0.9*\d) and (\xshift+0.85*\d,\yshift+0.7*\d)..
	(\xshift+1*\d,\yshift+0.5*\d)
	..controls (\xshift+1.15*\d,\yshift+0.3*\d) and (\xshift+1.25*\d,\yshift-0.1*\d)..
	(\xshift+1.75*\d,\yshift-0.1*\d);
	
\draw[very thick, \colZ, ->]
	(\xshift+0.25*\d,\yshift+1.1*\d) --
	(\xshift+1.75*\d,\yshift+1.1*\d);


\node[rotate=86] at (\xshift+1.05*\d,\yshift+1.5*\d) {$...$};

\draw[thick] (\xshift+0.25*\d,\yshift)--(\xshift+1*\d,\yshift);
\draw[thick] (\xshift+1*\d,\yshift)--(\xshift+1.75*\d,\yshift);

\draw[thick] (\xshift+0.25*\d,\yshift+\d)--(\xshift+1*\d,\yshift+\d);
\draw[thick] (\xshift+1*\d,\yshift+\d)--(\xshift+1.75*\d,\yshift+\d);

\begin{scope}
	\clip (\xshift+0.9*\d,\yshift+\d) rectangle (\xshift+1.1*\d,\yshift);
	\draw[thick] (\xshift+1.225*\d, \yshift+0.5*\d) ellipse (0.3 and 2*\d);
\end{scope}

\draw[whiteCircle] (\xshift+\d,\yshift+\d) circle;
\draw[blackCircle] (\xshift+\d,\yshift+0) circle;


\draw[very thick, \colZ, ->] (\xshift+0.33*\d,\yshift+2.1*\d) -- (\xshift+1.83*\d,\yshift+2.1*\d);
\draw[thick] (\xshift+0.33*\d,\yshift+2*\d)--(\xshift+1.83*\d,\yshift+2*\d);
\draw[very thick, \colZ, <-] (\xshift+0.31*\d,\yshift+1.9*\d) -- (\xshift+1.81*\d,\yshift+1.9*\d);


\node[anchor = east] at (\xshift+0.1*\d,\yshift+2*\d) {$1$};
\node[anchor = east] at (\xshift+0.1*\d,\yshift+1*\d) {$i$};
\node[anchor = east] at (\xshift+0.1*\d,\yshift) {$i+1$};
\node[anchor = east] at (\xshift+0.1*\d,\yshift-1*\d) {$N$};

\node[anchor = north] at (\xshift+1.2*\d,\yshift-1.7*\d) {$\bar{s}_i$};


\tikzmath{\xshift=8*\d;\yshift=0;\op = 30;};


\begin{scope}
	\clip (\xshift+1.125*\d,\yshift-2*\d) rectangle (\xshift+3*\d,\yshift+3*\d);
	\draw[very thick, \colZ!\op] (\xshift+1.125*\d, \yshift+0.5*\d) ellipse (0.3 and 2*\d);
\end{scope}


\begin{scope}
	\clip (\xshift+1.225*\d,\yshift-2*\d) rectangle (\xshift+3*\d,\yshift+3*\d);
	\draw[thick, black!10] (\xshift+1.225*\d, \yshift+0.5*\d) ellipse (0.3 and 2*\d);
\end{scope}


\begin{scope}
	\clip (\xshift+1.325*\d,\yshift-2*\d) rectangle (\xshift+3*\d,\yshift+3*\d);
	\draw[very thick, \colZ!10] (\xshift+1.325*\d, \yshift+0.5*\d) ellipse (0.3 and 2*\d);
\end{scope}

\draw
	(\xshift+0.45*\d,\yshift+2.5*\d) --
	(\xshift+2*\d,\yshift+2.5*\d);
\draw
	(\xshift+0.45*\d,\yshift-1.5*\d) --
	(\xshift+2*\d,\yshift-1.5*\d);

\draw (\xshift+0.45*\d, \yshift+0.5*\d) ellipse (0.3 and 2*\d);
\draw (\xshift+2*\d, \yshift+0.5*\d) ellipse (0.3 and 2*\d);


\draw[very thick, \colZ, <-]
	(\xshift+1.81*\d,\yshift-0.9*\d) --
	(\xshift+1.425*\d,\yshift-0.9*\d)
	.. controls (\xshift+1.3*\d,\yshift-0.9*\d) and (\xshift+0.98*\d,\yshift-0.9*\d) ..
	(\xshift+1.03*\d,\yshift-1.32*\d);

\draw[thick]
	(\xshift+1.825*\d,\yshift-1*\d) --
	(\xshift+1.525*\d,\yshift-1*\d)
	.. controls (\xshift+1.4*\d,\yshift-1*\d) and (\xshift+1.08*\d,\yshift-1*\d) ..
	(\xshift+1.13*\d,\yshift-1.32*\d);

\draw[very thick, \colZ]
	(\xshift+1.825*\d,\yshift-1.1*\d) --
	(\xshift+1.625*\d,\yshift-1.1*\d)
	.. controls (\xshift+1.5*\d,\yshift-1.1*\d) and (\xshift+1.18*\d,\yshift-1.1*\d) ..
	(\xshift+1.23*\d,\yshift-1.32*\d);


\begin{scope}

\clip (\xshift, \yshift-0.8*\d) rectangle (\xshift+2*\d, \yshift-2*\d);

\draw[very thick, \colZ, <-]
	(\xshift+0.33*\d,\yshift-1.1*\d) --
	(\xshift+0.4*\d,\yshift-1.1*\d)
	.. controls (\xshift+0.65*\d,\yshift-1.1*\d) and (\xshift+0.85*\d,\yshift-0.9*\d) ..
	(\xshift+1.05*\d,\yshift-0.6*\d)
	.. controls (\xshift+1.25*\d,\yshift-0.3*\d) and (\xshift+1.35*\d,\yshift-0.1*\d) ..
	(\xshift+1.6*\d,\yshift-0.1*\d) --
	(\xshift+1.75*\d,\yshift-0.1*\d);

\draw[thick]
	(\xshift+0.32*\d,\yshift-\d) --
	(\xshift+0.4*\d,\yshift-\d)
	.. controls (\xshift+0.65*\d,\yshift-\d) and (\xshift+0.8*\d,\yshift-0.8*\d) ..
	(\xshift+1*\d,\yshift-0.5*\d)
	.. controls (\xshift+1.2*\d,\yshift-0.2*\d) and (\xshift+1.35*\d,\yshift) ..
	(\xshift+1.6*\d,\yshift) --
	(\xshift+1.75*\d,\yshift);
	
\draw[very thick, \colZ, ->]
	(\xshift+0.31*\d,\yshift-0.9*\d) --
	(\xshift+0.4*\d,\yshift-0.9*\d)
	.. controls (\xshift+0.65*\d,\yshift-0.9*\d) and (\xshift+0.75*\d,\yshift-0.7*\d) ..
	(\xshift+0.95*\d,\yshift-0.4*\d)
	.. controls (\xshift+1.15*\d,\yshift-0.1*\d) and (\xshift+1.35*\d,\yshift+0.1*\d) ..
	(\xshift+1.6*\d,\yshift+0.1*\d) --
	(\xshift+1.75*\d,\yshift+0.1*\d);
	
\end{scope}


\node[rotate=94] at (\xshift+1.05*\d,\yshift-0.55*\d) {$...$};


\begin{scope}

\clip (\xshift,\yshift-0.3*\d) rectangle (\xshift+2*\d,\yshift+1.3*\d);

\draw[very thick, \colZ, <-]
	(\xshift+0.25*\d,\yshift-1.1*\d) --
	(\xshift+0.4*\d,\yshift-1.1*\d)
	.. controls (\xshift+0.65*\d,\yshift-1.1*\d) and (\xshift+0.85*\d,\yshift-0.9*\d) ..
	(\xshift+1.05*\d,\yshift-0.6*\d)
	.. controls (\xshift+1.25*\d,\yshift-0.3*\d) and (\xshift+1.35*\d,\yshift-0.1*\d) ..
	(\xshift+1.6*\d,\yshift-0.1*\d) --
	(\xshift+1.75*\d,\yshift-0.1*\d);

\draw[thick]
	(\xshift+0.25*\d,\yshift-\d) --
	(\xshift+0.4*\d,\yshift-\d)
	.. controls (\xshift+0.65*\d,\yshift-\d) and (\xshift+0.8*\d,\yshift-0.8*\d) ..
	(\xshift+1*\d,\yshift-0.5*\d)
	.. controls (\xshift+1.2*\d,\yshift-0.2*\d) and (\xshift+1.35*\d,\yshift) ..
	(\xshift+1.6*\d,\yshift) --
	(\xshift+1.75*\d,\yshift);
	
\draw[very thick, \colZ, ->]
	(\xshift+0.25*\d,\yshift-0.9*\d) --
	(\xshift+0.4*\d,\yshift-0.9*\d)
	.. controls (\xshift+0.65*\d,\yshift-0.9*\d) and (\xshift+0.75*\d,\yshift-0.7*\d) ..
	(\xshift+0.95*\d,\yshift-0.4*\d)
	.. controls (\xshift+1.15*\d,\yshift-0.1*\d) and (\xshift+1.35*\d,\yshift+0.1*\d) ..
	(\xshift+1.6*\d,\yshift+0.1*\d) --
	(\xshift+1.75*\d,\yshift+0.1*\d);


\draw[very thick, \colZ, <-]
	(\xshift+0.25*\d,\yshift-0.1*\d) --
	(\xshift+0.4*\d,\yshift-0.1*\d)
	.. controls (\xshift+0.65*\d,\yshift-0.1*\d) and (\xshift+0.85*\d,\yshift+0.1*\d) ..
	(\xshift+1.05*\d,\yshift+0.4*\d)
	.. controls (\xshift+1.25*\d,\yshift+0.7*\d) and (\xshift+1.35*\d,\yshift+0.9*\d) ..
	(\xshift+1.6*\d,\yshift+0.9*\d) --
	(\xshift+1.75*\d,\yshift+0.9*\d);

\draw[thick]
	(\xshift+0.25*\d,\yshift) --
	(\xshift+0.4*\d,\yshift)
	.. controls (\xshift+0.65*\d,\yshift) and (\xshift+0.8*\d,\yshift+0.2*\d) ..
	(\xshift+1*\d,\yshift+0.5*\d)
	.. controls (\xshift+1.2*\d,\yshift+0.8*\d) and (\xshift+1.35*\d,\yshift+\d) ..
	(\xshift+1.6*\d,\yshift+\d) --
	(\xshift+1.75*\d,\yshift+\d);
	
\draw[very thick, \colZ, ->]
	(\xshift+0.25*\d,\yshift+0.1*\d) --
	(\xshift+0.4*\d,\yshift+0.1*\d)
	.. controls (\xshift+0.65*\d,\yshift+0.1*\d) and (\xshift+0.75*\d,\yshift+0.3*\d) ..
	(\xshift+0.95*\d,\yshift+0.6*\d)
	.. controls (\xshift+1.15*\d,\yshift+0.9*\d) and (\xshift+1.35*\d,\yshift+1.1*\d) ..
	(\xshift+1.6*\d,\yshift+1.1*\d) --
	(\xshift+1.75*\d,\yshift+1.1*\d);


\draw[very thick, \colZ, <-]
	(\xshift+0.25*\d,\yshift+0.9*\d) --
	(\xshift+0.4*\d,\yshift+0.9*\d)
	.. controls (\xshift+0.65*\d,\yshift+0.9*\d) and (\xshift+0.85*\d,\yshift+1.1*\d) ..
	(\xshift+1.05*\d,\yshift+1.4*\d)
	.. controls (\xshift+1.25*\d,\yshift+1.7*\d) and (\xshift+1.35*\d,\yshift+1.9*\d) ..
	(\xshift+1.6*\d,\yshift+1.9*\d) --
	(\xshift+1.75*\d,\yshift+1.9*\d);

\draw[thick]
	(\xshift+0.25*\d,\yshift+\d) --
	(\xshift+0.4*\d,\yshift+\d)
	.. controls (\xshift+0.65*\d,\yshift+\d) and (\xshift+0.8*\d,\yshift+1.2*\d) ..
	(\xshift+1*\d,\yshift+1.5*\d)
	.. controls (\xshift+1.2*\d,\yshift+1.8*\d) and (\xshift+1.35*\d,\yshift+2*\d) ..
	(\xshift+1.6*\d,\yshift+2*\d) --
	(\xshift+1.75*\d,\yshift+2*\d);
	
\draw[very thick, \colZ, ->]
	(\xshift+0.25*\d,\yshift+1.1*\d) --
	(\xshift+0.4*\d,\yshift+1.1*\d)
	.. controls (\xshift+0.65*\d,\yshift+1.1*\d) and (\xshift+0.75*\d,\yshift+1.3*\d) ..
	(\xshift+0.95*\d,\yshift+1.6*\d)
	.. controls (\xshift+1.15*\d,\yshift+1.9*\d) and (\xshift+1.35*\d,\yshift+2.1*\d) ..
	(\xshift+1.6*\d,\yshift+2.1*\d) --
	(\xshift+1.75*\d,\yshift+2.1*\d);

\end{scope}


\begin{scope}

\clip (\xshift,\yshift+1.8*\d) rectangle (\xshift+2*\d,\yshift+3*\d);

\draw[very thick, \colZ]
	(\xshift+0.25*\d,\yshift+0.9*\d) --
	(\xshift+0.4*\d,\yshift+0.9*\d)
	.. controls (\xshift+0.65*\d,\yshift+0.9*\d) and (\xshift+0.85*\d,\yshift+1.1*\d) ..
	(\xshift+1.05*\d,\yshift+1.4*\d)
	.. controls (\xshift+1.25*\d,\yshift+1.7*\d) and (\xshift+1.35*\d,\yshift+1.9*\d) ..
	(\xshift+1.6*\d,\yshift+1.9*\d) --
	(\xshift+1.79*\d,\yshift+1.9*\d);

\draw[thick]
	(\xshift+0.25*\d,\yshift+\d) --
	(\xshift+0.4*\d,\yshift+\d)
	.. controls (\xshift+0.65*\d,\yshift+\d) and (\xshift+0.8*\d,\yshift+1.2*\d) ..
	(\xshift+1*\d,\yshift+1.5*\d)
	.. controls (\xshift+1.2*\d,\yshift+1.8*\d) and (\xshift+1.35*\d,\yshift+2*\d) ..
	(\xshift+1.6*\d,\yshift+2*\d) --
	(\xshift+1.8*\d,\yshift+2*\d);
	
\draw[very thick, \colZ, ->]
	(\xshift+0.25*\d,\yshift+1.1*\d) --
	(\xshift+0.4*\d,\yshift+1.1*\d)
	.. controls (\xshift+0.65*\d,\yshift+1.1*\d) and (\xshift+0.75*\d,\yshift+1.3*\d) ..
	(\xshift+0.95*\d,\yshift+1.6*\d)
	.. controls (\xshift+1.15*\d,\yshift+1.9*\d) and (\xshift+1.35*\d,\yshift+2.1*\d) ..
	(\xshift+1.6*\d,\yshift+2.1*\d) --
	(\xshift+1.81*\d,\yshift+2.1*\d);

\end{scope}


\begin{scope}
	\clip (\xshift+1.02*\d,\yshift-2*\d) rectangle (\xshift+1.125*\d,\yshift+3*\d);
	\draw[very thick, \colZ] (\xshift+1.125*\d, \yshift+0.5*\d) ellipse (0.3 and 2*\d);
\end{scope}


\begin{scope}
	\clip (\xshift+1.12*\d,\yshift-2*\d) rectangle (\xshift+1.225*\d,\yshift+3*\d);
	\draw[thick] (\xshift+1.225*\d, \yshift+0.5*\d) ellipse (0.3 and 2*\d);
\end{scope}


\begin{scope}
	\clip (\xshift+1.22*\d,\yshift-2*\d) rectangle (\xshift+1.325*\d,\yshift+3*\d);
	\draw[very thick, \colZ] (\xshift+1.325*\d, \yshift+0.5*\d) ellipse (0.3 and 2*\d);
\end{scope}


\node[rotate=86] at (\xshift+1.05*\d,\yshift+1.55*\d) {$...$};


\draw[very thick, \colZ, <-] 
	(\xshift+0.32*\d,\yshift+1.9*\d) --
	(\xshift+0.6*\d,\yshift+1.9*\d)
	.. controls (\xshift+0.85*\d,\yshift+1.9*\d) and (\xshift+1.18*\d,\yshift+1.95*\d) ..
	(\xshift+1.23*\d,\yshift+2.32*\d);
	
\draw[thick]
	(\xshift+0.33*\d,\yshift+2*\d) --
	(\xshift+0.5*\d,\yshift+2*\d)
	.. controls (\xshift+0.75*\d,\yshift+2*\d) and (\xshift+1.08*\d,\yshift+2*\d) ..
	(\xshift+1.13*\d,\yshift+2.32*\d);

\draw[very thick, \colZ] 
	(\xshift+0.34*\d,\yshift+2.1*\d) --
	(\xshift+0.6*\d,\yshift+2.1*\d)
	.. controls (\xshift+0.85*\d,\yshift+2.1*\d) and (\xshift+1.03*\d,\yshift+2.15*\d) ..
	(\xshift+1.03*\d,\yshift+2.32*\d);


\node[anchor = east] at (\xshift+0.1*\d,\yshift+2*\d) {$1$};
\node[anchor = east] at (\xshift+0.1*\d,\yshift+1*\d) {$i$};
\node[anchor = east] at (\xshift+0.1*\d,\yshift) {$i+1$};
\node[anchor = east] at (\xshift+0.1*\d,\yshift-1*\d) {$N$};

\node[anchor = north] at (\xshift+1.2*\d,\yshift-1.7*\d) {$\Lambda$};

\end{tikzpicture}
}
\begin{tikzpicture}
\draw[white] (0,0)--(0.5,0);
\end{tikzpicture}
\end{center}
\caption{Basic graphs on cylinder corresponding to generators of Weyl group. Zig-zag paths are drawn by green lines, and generators act on their ends by permutation. Note that zig-zags are drawn so, that in-going and out-going ends of zig-zags alternate along the boundary.}
\label{fig:WeylGen}
\end{figure}

Another origin of cluster coordinates in integrable systems is factorization ans\"atze for elements of Poisson-Lie group $\GLb(N)$ \cite{FG:2005}, \cite{GSV:2009}, \cite{FM:2014}, which appeared in theory of positive matrices \cite{FZ}. In this approach phase spaces of systems are double Bruhat cells $B_{w}\subset \GLb(N)$, which are enumerated by elements $w$ of extended double Weyl group $\widetilde{W}\left(A_{N-1}^{(1)} \times A_{N-1}^{(1)}\right)$, which has presentation
\begin{equation}
\label{eq:WeylGrDef}
\widetilde{W}\left(A_{N-1}^{(1)} \times A_{N-1}^{(1)}\right) = 
\left\langle
\begin{array}{c|l}
s_i,\bar{s}_i,\Lambda
&
s_i s_{i+1} s_i = s_{i+1} s_i s_{i+1}, ~~~
\Lambda s_i = s_{i+1}\Lambda, ~~~
s_i^2=1, ~~~
\bar{s}_i s_j =  s_j\bar{s}_i,
\\
i\in \mathbb{Z}/N\mathbb{Z}
&
\bar{s}_i \bar{s}_{i+1} \bar{s}_i = \bar{s}_{i+1} \bar{s}_i \bar{s}_{i+1}, ~~~
\Lambda \bar{s}_i = \bar{s}_{i+1}\Lambda, ~~~
\bar{s}_i^2=1
\end{array}
\right\rangle .
\end{equation}
Each reduced decomposition of $w$ into product of generators $s_i,\bar{s}_i,\Lambda$ provides open embedding of $\mathcal{X}$-cluster chart in $B_w$: to each generator one assigns certain matrix (namely, transfer matrix in one-path sector of blocks shown in Fig.~ \ref{fig:WeylGen}), depending on $\X$-cluster variables. Product of these matrices in the same order, as letters in the word $w$ are located, provide matrix $g(\lambda)$ parametrizing $B_w$. Cycle $\ha$ is chosen to be interval lying on the 'back' side of cylinder and connecting its left and right boundaries, so the dependence on $\lambda$ comes from generators $s_0, \bar{s}_0$ and $\Lambda$ which contain edges crossing $\ha$, for details see \cite{FM:2014}.

The restriction of $r$-matrix bracket with trigonometric $r$-matrix
\begin{equation}
\label{eq:poissonRmat}
\{ g(\lambda_1)\otimes g(\lambda_2)\} = [r(\lambda_1/\lambda_2),g(\lambda_1)\otimes g(\lambda_2)]
\end{equation} 
to double Bruhat cells, which are Poisson submanifolds, turns out to be compatible with logarithmically constant bracket (\ref{eq:poissonClust}). The simplest way to check this is by checking for each block drawn in Fig~\ref{fig:WeylGen}, and using co-product property of $r$-matrix bracket, that if $g_1$ and $g_2$ satisfy it, then $g_1 g_2$ also satisfies. Exchange matrix $\varepsilon$ can be easily written from the word $w$ by considering graphs, dual to those drawn in Fig.~\ref{fig:WeylGen}, as it was done in Fig.~\ref{fig:fourumove}. Change of reduced decomposition via Coxeter relations $s_i s_{i+1} s_i = s_{i+1} s_i s_{i+1}$ and $\bar{s}_i \bar{s}_{i+1} \bar{s}_i = \bar{s}_{i+1} \bar{s}_i \bar{s}_{i+1}$ amounts in single four-move and pair of two-moves. Relation $\bar{s}_i s_{j} = s_{j} \bar{s}_i$ can be realized as single two-move and does not affect exchange matrix if $i = j\pm 1$, and is single four-move if $i=j$. The relations $s_i^2=1$ and $\bar{s}_i^2=1$ can be done by pair of two-moves followed by parallel bigon reduction of type $(b)$, and therefore are not cluster transformations and do not preserve transfer matrix, however preserves wiring of zig-zags. If one applies parallel bigon reduction of type $(a)$ instead, one gets Weyl semi-group with relation $s_i^2 = s_i$. Below we will assume that we use reduction of type $(b)$.

Spectral curve of integrable system is given by characteristic equation
\begin{equation}
S(\lambda,\mu) = \mathrm{det}(g(\lambda) - \mu).
\end{equation}
Hamiltonians of the system are Ad-invariant functions on Bruhat cells, and so only conjugacy class of word $w$ matters. Taking characteristic equation of $g(\lambda)$ is close relative of gluing torus into cylinder, so the spectral curve coincides with the one given by (\ref{eq:spectTor}) up to transformations $S(\lambda,\mu) \mapsto f(\lambda) S(\lambda,\mu),~ \mu \mapsto g(\lambda) \mu$, where $f,g$ are some rational functions with coefficients depending on Casimirs. 

Important observation, which we will need in the following, is that the building blocks for $s_i,\bar{s}_i, \Lambda$ indeed 'permute' zig-zag paths, which we will sometimes refer as strands. One can see in Fig.~\ref{fig:WeylGen} that zig-zags going from the left to the right along lines $i$ and $i+1$ are permuted after passing $s_i$, while $\bar{s}_i$ permutes those going from the right to the left along $i$ and $i+1$. Note, that the label $i$ of generators $s_i$ and $\bar{s}_i$ is given not by the number of zig-zag, but by the number of horizontal line of bi-coloured graph. Generator $\Lambda$ shifts by $+1$ all zig-zags going from the left to the right, and by $-1$ those going from the right to the left.\\

\paragraph{Weyl group interpretation of tetrahedron equation}

Double Weyl group of $\GLb(N)$ contains diagonal subgroup $W(A^{(1)}_{N-1}) \subset W(A^{(1)}_{N-1} \times A^{(1)}_{N-1})$ generated by $s_i\bar{s}_i$ and $\Lambda$. Comparing Fig.~\ref{fig:fourgon} and Fig.~\ref{fig:WeylGen} one sees that plabic graphs corresponding to Lax operator of Bazhanov and Sergeev coincides with the one presenting word $s_i \bar{s}_i$ in double Weyl group! As we will see below, systems with symmetric Newton polygons can be constructed using diagonal subgroup only, so this again gives construction of integrable system with arbitrary symmetric Newton polygon from contraction of Lax operators (\ref{eq:LaxBS}). Tetrahedron transformation shown in Fig.~\ref{fig:tetrahedron}, can be interpreted just as braiding relation
\begin{equation}
\Rmut:  ~~~ (s_i \bar{s}_i)(s_{i+1} \bar{s}_{i+1})(s_i \bar{s}_i) \mapsto (s_{i+1} \bar{s}_{i+1})(s_i \bar{s}_i)(s_{i+1} \bar{s}_{i+1})
\end{equation}
for diagonal subgroup of $\widetilde{W}(A_{N-1}^{(1)} \times A_{N-1}^{(1)})$. This is the same transformation, which relates two 'positive' parametrizations \cite{FZ} for the largest Bruhat cell $w_0$ in $\mathrm{PGL}(3)$.

The functional tetrahedron equation (\ref{eq:tetraFunc}) recasts into statement, that two ways to identify two different parametrizations for the largest Bruhat cell $w_0$ in $\mathrm{PGL}(4)$ are equivalent
\begin{equation}
w_0 =
(s_1 \bar{s}_1  s_2 \bar{s}_2 s_3 \bar{s}_3) (s_1 \bar{s}_1 s_2 \bar{s}_2) (s_1 \bar{s}_1)
\cong
(s_3 \bar{s}_3 s_2 \bar{s}_2 s_1 \bar{s}_1) (s_3 \bar{s}_3 s_2 \bar{s}_2) (s_3 \bar{s}_3).
\end{equation}

\paragraph{Symmetric Newton polygon}{
Now, we are ready to show how construction from section \ref{ss:generalNewt} for integrable system with symmetric Newton polygon $(\vec{u}_1,...,\vec{u}_r,-\vec{u}_1...,-\vec{u}_r)$ can be reproduced for double Bruhat cell of the group $\GLb(N),~ N = a_1+...+a_n$. Construction comes from consideration of commuting subgroups $\GLb(a_1)\times \GLb(a_2) \times ... \times \GLb(a_n) $ in $\GLb(N)$, similar to those from \cite{FJMM}, and observation that $s_i$ and $\bar{s}_{i}$ act on zig-zag paths by permutations. Consider subgroup $\widetilde{W}_{i,j} = \widetilde{W}(A^{(1)}_{j-i} \times A^{(1)}_{j-i}) \subset \widetilde{W}(A^{(1)}_{N-1} \times A^{(1)}_{N-1})$ which permutes strands from $i$ to $j$ keeping other strands intact. More precisely, generators $s'_i,\Lambda'$ of this subgroup are
\begin{equation}
\begin{array}{l}
s'_a = s_{i+a-1},~~~
\bar{s}'_a = \bar{s}_{i+a-1},~~~ 1\leq a \leq j-i, \\
s'_0= s_{i-1} s_{i-2}...s_{1}s_{0}s_{N-1}...s_{j+1}s_{j}s_{j+1}...s_{N-1}s_{0}s_1...s_{i-2}s_{i-1}, \\
\bar{s}'_0= \bar{s}_{i-1} \bar{s}_{i-2}...\bar{s}_{1}\bar{s}_{0}\bar{s}_{N-1}...\bar{s}_{j+1}\bar{s}_{j}\bar{s}_{j+1}...\bar{s}_{N-1}\bar{s}_{0}\bar{s}_1...\bar{s}_{i-2}\bar{s}_{i-1}, \\
\Lambda'=s_{i-1}\bar{s}_{i-1}s_{i-2}\bar{s}_{i-2}...s_{1}\bar{s}_{1}s_{0}\bar{s}_{0}s_{N-1}\bar{s}_{N-1}...s_{j+1}\bar{s}_{j+1}\Lambda,
\end{array}
\end{equation}
so generators from $s'_1$ to $s'_{j-i-1}$ act on strands $i,...,j$ as usual, while the affine generators are 'skipping' other strands $1,...,i-1, j+1,...,N$. Generator $\Lambda'$ of subgroup $W_{i,j}$ will be referred as $\Lambda_{i,j}$ in the following. Note, that bipartite graph defined by block $\Lambda_{ij}$ is the same stripe of four-gons as a one, which appeared in Section \ref{ss:generalNewt}.

It is always possible using $\mathrm{SA}(2,\mathbb{Z})$ transformation to place Newton polygon in such a way, that it does not have any vertical sides. It is straightforward to check that the Bruhat cell which gives Newton polygon $(\vec{u}_1,...,\vec{u}_n,-\vec{u}_1...,-\vec{u}_n)$ is defined then by element
\begin{equation}
w=({\Lambda}_{1,r_1})^{b_1} ({\Lambda}_{{r_1+1},{r_2}})^{b_2}...({\Lambda}_{{r_{n-1}+1},{r_n}})^{b_n},~~~ r_k = a_1+...+a_k
\end{equation}
in double Weyl group. Side $(a_k,b_k)$ of the Newton polygon is generated by strands $a_{r_{k-1}+1},...,a_{r_k}$. Together they got projection $a_k$ on the generator of homologies oriented along cylinder.  Generator $\Lambda_{r_{k-1}+1,r_{k}}$ mixes only them, and each application of this 'twist' operator increases their common projection on generator of homologies, oriented across cylinder, by $1$. By applying it $b_k$ times and making torus from cylinder, they are gluing into longer strands representing class $(a_k,b_k) \in \H1(\T2,\mathbb{Z})$, so it presents side $\vec{u}_k$ of the Newton polygon. Strands going along the same lines but with the opposite orientations generate class $-\vec{u}_k$.
}

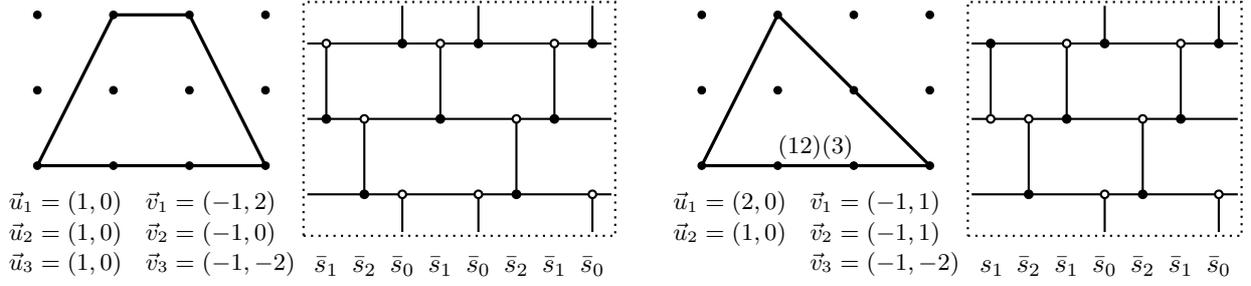
\begin{figure}[!ht]
\begin{center}
\begin{tikzpicture}

\tikzmath{\xs=0;\ys=0;\d=1;\Lx=3;\Ly=2;};


\foreach \x in {0,...,\Lx}
	\foreach \y in {0,...,\Ly}
		\draw[fill] (\xs+\d*\x,\ys+\d*\y) circle[radius=0.05];

\draw[very thick] (\xs, \ys) -- (\xs+3*\d, \ys);
\draw[very thick] (\xs+3*\d, \ys) -- (\xs+2*\d, \ys+2*\d);
\draw[very thick] (\xs+2*\d, \ys+2*\d) -- (\xs+1*\d, \ys+2*\d);
\draw[very thick] (\xs+1*\d, \ys+2*\d) -- (\xs, \ys);

\node[anchor = west, font = \small] at (\xs-0.5*\d,\ys-0.5*\d) {$\vec{u}_1 = (1,0)$};
\node[anchor = west, font = \small] at (\xs-0.5*\d,\ys-0.9*\d) {$\vec{u}_2 = (1,0)$};
\node[anchor = west, font = \small] at (\xs-0.5*\d,\ys-1.3*\d) {$\vec{u}_3 = (1,0)$};

\node[anchor = west, font = \small] at (\xs+1.3*\d,\ys-0.5*\d) {$\vec{v}_1 = (-1,2)$};
\node[anchor = west, font = \small] at (\xs+1.3*\d,\ys-0.9*\d) {$\vec{v}_2 = (-1,0)$};
\node[anchor = west, font = \small] at (\xs+1.3*\d,\ys-1.3*\d) {$\vec{v}_3 = (-1,-2)$};

\tikzmath{\xs=3.8*\d;\ys=-0.38*\d;\d=1;\Lx=3;\Ly=2;\t=0.5*\d;\s=0.05*\d;};


\draw[dotted, thick] (\xs-0.5*\t-\s,\ys-0.5*\d-\s) --(\xs+7.5*\t+\s,\ys-0.5*\d-\s) -- (\xs+7.5*\t+\s,\ys+2.5*\d+\s) -- (\xs-0.5*\t-\s,\ys+2.5*\d+\s) -- (\xs-0.5*\t-\s,\ys-0.5*\d-\s);

\draw[thick] (\xs-0.5*\t, \ys) -- (\xs+7.5*\t, \ys);
\draw[thick] (\xs-0.5*\t, \ys+\d) -- (\xs+7.5*\t, \ys+\d);
\draw[thick] (\xs-0.5*\t, \ys+2*\d) -- (\xs+7.5*\t, \ys+2*\d);


\draw[thick] (\xs, \ys+2*\d) -- (\xs, \ys+\d);
\draw[whiteCircle, radius=0.05] (\xs, \ys+2*\d) circle;
\draw[blackCircle, radius=0.05] (\xs, \ys+\d) circle;
\node[anchor=north] at (\xs, \ys-0.7*\d) {$\bar{s}_{1}$};
\draw[thick] (\xs+\t, \ys+\d) -- (\xs+\t, \ys);
\draw[whiteCircle, radius=0.05] (\xs+\t, \ys+\d) circle;
\draw[blackCircle, radius=0.05] (\xs+\t, \ys) circle;
\node[anchor=north] at (\xs+\t, \ys-0.7*\d) {$\bar{s}_{2}$};

\draw[thick] (\xs+2*\t, \ys) -- (\xs+2*\t, \ys-0.5*\d);
\draw[thick] (\xs+2*\t, \ys+2*\d) -- (\xs+2*\t, \ys+2.5*\d);
\draw[whiteCircle, radius=0.05] (\xs+2*\t, \ys) circle;
\draw[blackCircle, radius=0.05] (\xs+2*\t, \ys+2*\d) circle;
\node[anchor=north] at (\xs+2*\t, \ys-0.7*\d) {$\bar{s}_{0}$};

\draw[thick] (\xs+3*\t, \ys+2*\d) -- (\xs+3*\t, \ys+\d);
\draw[whiteCircle, radius=0.05] (\xs+3*\t, \ys+2*\d) circle;
\draw[blackCircle, radius=0.05] (\xs+3*\t, \ys+\d) circle;
\node[anchor=north] at (\xs+3*\t, \ys-0.7*\d) {$\bar{s}_{1}$};

\draw[thick] (\xs+4*\t, \ys) -- (\xs+4*\t, \ys-0.5*\d);
\draw[thick] (\xs+4*\t, \ys+2*\d) -- (\xs+4*\t, \ys+2.5*\d);
\draw[whiteCircle, radius=0.05] (\xs+4*\t, \ys) circle;
\draw[blackCircle, radius=0.05] (\xs+4*\t, \ys+2*\d) circle;
\node[anchor=north] at (\xs+4*\t, \ys-0.7*\d) {$\bar{s}_{0}$};

\draw[thick] (\xs+5*\t, \ys+\d) -- (\xs+5*\t, \ys);
\draw[whiteCircle, radius=0.05] (\xs+5*\t, \ys+\d) circle;
\draw[blackCircle, radius=0.05] (\xs+5*\t, \ys) circle;
\node[anchor=north] at (\xs+5*\t, \ys-0.7*\d) {$\bar{s}_{2}$};

\draw[thick] (\xs+6*\t, \ys+\d) -- (\xs+6*\t, \ys+2*\d);
\draw[whiteCircle, radius=0.05] (\xs+6*\t, \ys+2*\d) circle;
\draw[blackCircle, radius=0.05] (\xs+6*\t, \ys+\d) circle;
\node[anchor=north] at (\xs+6*\t, \ys-0.7*\d) {$\bar{s}_{1}$};

\draw[thick] (\xs+7*\t, \ys) -- (\xs+7*\t, \ys-0.5*\d);
\draw[thick] (\xs+7*\t, \ys+2*\d) -- (\xs+7*\t, \ys+2.5*\d);
\draw[whiteCircle, radius=0.05] (\xs+7*\t, \ys) circle;
\draw[blackCircle, radius=0.05] (\xs+7*\t, \ys+2*\d) circle;
\node[anchor=north] at (\xs+7*\t, \ys-0.7*\d) {$\bar{s}_{0}$};

\end{tikzpicture}
\begin{tikzpicture}

\tikzmath{\xs=0;\ys=0;\d=1;\Lx=3;\Ly=2;};

\draw[white] (\xs-\d, \ys+1.5*\d)--(\xs-\d, \ys+0.5*\d);


\foreach \x in {0,...,\Lx}
	\foreach \y in {0,...,\Ly}
		\draw[fill] (\xs+\d*\x,\ys+\d*\y) circle[radius=0.05];

\draw[very thick] (\xs, \ys) -- (\xs+3*\d, \ys);
\draw[very thick] (\xs+3*\d, \ys) -- (\xs+1*\d, \ys+2*\d);
\draw[very thick] (\xs+1*\d, \ys+2*\d) -- (\xs, \ys);

\node[font = \small] at (\xs+1.5*\d,\ys+0.25*\d) {$(12)(3)$};

\node[anchor = west, font = \small] at (\xs-0.5*\d,\ys-0.5*\d) {$\vec{u}_1 = (2,0)$};
\node[anchor = west, font = \small] at (\xs-0.5*\d,\ys-0.9*\d) {$\vec{u}_2 = (1,0)$};

\node[anchor = west, font = \small] at (\xs+1.3*\d,\ys-0.5*\d) {$\vec{v}_1 = (-1,1)$};
\node[anchor = west, font = \small] at (\xs+1.3*\d,\ys-0.9*\d) {$\vec{v}_2 = (-1,1)$};
\node[anchor = west, font = \small] at (\xs+1.3*\d,\ys-1.3*\d) {$\vec{v}_3 = (-1,-2)$};

\tikzmath{\xs=2.8*\d;\ys=-0.38*\d;\d=1;\Lx=3;\Ly=2;\t=0.5*\d;\s=0.05*\d;};


\draw[dotted, thick] (\xs+1.5*\t-\s,\ys-0.5*\d-\s) --(\xs+8.5*\t+\s,\ys-0.5*\d-\s) -- (\xs+8.5*\t+\s,\ys+2.5*\d+\s) -- (\xs+1.5*\t-\s,\ys+2.5*\d+\s) -- (\xs+1.5*\t-\s,\ys-0.5*\d-\s);

\draw[thick] (\xs+1.5*\t, \ys) -- (\xs+8.5*\t, \ys);
\draw[thick] (\xs+1.5*\t, \ys+\d) -- (\xs+8.5*\t, \ys+\d);
\draw[thick] (\xs+1.5*\t, \ys+2*\d) -- (\xs+8.5*\t, \ys+2*\d);


\draw[thick] (\xs+2*\t, \ys+2*\d) -- (\xs+2*\t, \ys+\d);
\draw[blackCircle, radius=0.05] (\xs+2*\t, \ys+2*\d) circle;
\draw[whiteCircle, radius=0.05] (\xs+2*\t, \ys+\d) circle;
\node[anchor=north] at (\xs+2*\t+0.03*\t, \ys-0.75*\d) {$s_{1}$};

\draw[thick] (\xs+3*\t, \ys+\d) -- (\xs+3*\t, \ys);
\draw[whiteCircle, radius=0.05] (\xs+3*\t, \ys+\d) circle;
\draw[blackCircle, radius=0.05] (\xs+3*\t, \ys) circle;
\node[anchor=north] at (\xs+3*\t, \ys-0.7*\d) {$\bar{s}_{2}$};

\draw[thick] (\xs+4*\t, \ys+2*\d) -- (\xs+4*\t, \ys+\d);
\draw[whiteCircle, radius=0.05] (\xs+4*\t, \ys+2*\d) circle;
\draw[blackCircle, radius=0.05] (\xs+4*\t, \ys+\d) circle;
\node[anchor=north] at (\xs+4*\t, \ys-0.7*\d) {$\bar{s}_{1}$};

\draw[thick] (\xs+5*\t, \ys) -- (\xs+5*\t, \ys-0.5*\d);
\draw[thick] (\xs+5*\t, \ys+2*\d) -- (\xs+5*\t, \ys+2.5*\d);
\draw[whiteCircle, radius=0.05] (\xs+5*\t, \ys) circle;
\draw[blackCircle, radius=0.05] (\xs+5*\t, \ys+2*\d) circle;
\node[anchor=north] at (\xs+5*\t, \ys-0.7*\d) {$\bar{s}_{0}$};

\draw[thick] (\xs+6*\t, \ys+\d) -- (\xs+6*\t, \ys);
\draw[whiteCircle, radius=0.05] (\xs+6*\t, \ys+\d) circle;
\draw[blackCircle, radius=0.05] (\xs+6*\t, \ys) circle;
\node[anchor=north] at (\xs+6*\t, \ys-0.7*\d) {$\bar{s}_{2}$};

\draw[thick] (\xs+7*\t, \ys+2*\d) -- (\xs+7*\t, \ys+\d);
\draw[whiteCircle, radius=0.05] (\xs+7*\t, \ys+2*\d) circle;
\draw[blackCircle, radius=0.05] (\xs+7*\t, \ys+\d) circle;
\node[anchor=north] at (\xs+7*\t, \ys-0.7*\d) {$\bar{s}_{1}$};

\draw[thick] (\xs+8*\t, \ys) -- (\xs+8*\t, \ys-0.5*\d);
\draw[thick] (\xs+8*\t, \ys+2*\d) -- (\xs+8*\t, \ys+2.5*\d);
\draw[whiteCircle, radius=0.05] (\xs+8*\t, \ys) circle;
\draw[blackCircle, radius=0.05] (\xs+8*\t, \ys+2*\d) circle;
\node[anchor=north] at (\xs+8*\t, \ys-0.7*\d) {$\bar{s}_{0}$};

\end{tikzpicture}
\end{center}
\caption{Left: Example of the double Bruhat cell in $\GLb(3)$ with non-symmetric Newton polygon. The corresponding element in the double Weyl group is $w = (\Lambda^{-}_{1,1})^{-2} (\Lambda^{-}_{3,3})^{2} = (\bar{s}_0 \bar{s}_2 \Lambda)^{-2} (\bar{s}_2 \bar{s}_1 \Lambda)^{2} = \bar{s}_1 \bar{s}_2 \bar{s}_0 \bar{s}_1 \bar{s}_0 \bar{s}_2 \bar{s}_1 \bar{s}_0$, where we used commutation relations of $\bar{s}_i$ with $\Lambda$. The bipartite graphs are drawn on torus, i.e. one has to glue right side with the left one, and upper with the lower. Right: Example of the double Bruhat cell in $\GLb(3)$ with non-trivial wiring of parallel zig-zags, the corresponding element in double Weyl group is $w = s_1 (\Lambda^{-}_{1,1})^{-1} (\Lambda^{-}_{2,2})^{-1} (\Lambda^{-}_{3,3})^{2} = s_1 (\bar{s}_0 \bar{s}_2 \Lambda)^{-1} (\bar{s}_1 \bar{s}_0 \Lambda)^{-1} (\bar{s}_2 \bar{s}_1 \Lambda)^{2} = s_1 \bar{s}_2 \bar{s}_1 \bar{s}_0 \bar{s}_2 \bar{s}_1 \bar{s}_0$.}
\label{fig:assymNewt}
\end{figure}

\paragraph{Non-symmetric Newton polygons}{
For integrable system with non-symmetric Newton polygon it is convenient to present Lax operator in triangular decomposed form. This requires getting out of diagonal subgroup of $\widetilde{W}(A_{N-1}^{(1)} \times A_{N-1}^{(1)})$, and considering separately 'positive' and 'negative' commuting subgroups $\widetilde{W}(A_{a_1-1}^{(1)}) \times ... \times \widetilde{W}(A_{a_n-1}^{(1)})$ and $\widetilde{W}(A_{c_1-1}^{(1)}) \times ... \times \widetilde{W}(A_{c_m-1}^{(1)})$, where $\vec{u}_1 = (a_1, b_1),...,\vec{u}_n = (a_n, b_n)$ are primitive oriented boundary intervals of polygon between the leftmost and rightmost points, $\vec{v}_1 = (-c_1,-d_1),...,\vec{v}_m = (-c_m,-d_m)$ are intervals between the rightmost and leftmost points in counter-clockwise direction. Introducing halves of 'twisting' operators   
\begin{equation}
\label{eq:Lambdadef}
\Lambda_{ij}^{+}=s_{i-1}s_{i-2}...s_{1}s_{0}s_{N-1}...s_{j+1} \Lambda,~~~
\Lambda_{ij}^{-}=\bar{s}_{i-1}\bar{s}_{i-2}...\bar{s}_{1}\bar{s}_{0}\bar{s}_{N-1}...\bar{s}_{j+1} \Lambda,
\end{equation}
where $r_i = a_1+...+a_i$ and $l_i = c_1+...+c_i$, the word in double Weyl group which provides wiring diagram for non-symmetric Newton polygon is
\begin{equation}
w = w^+ w^- \Lambda^{-b_1-...-b_n},
~~~
w^{+} = ({\Lambda}^{+}_{1,r_1})^{b_1} ... ({\Lambda}^{+}_{{r_{n-1}+1},{r_n}})^{b_n},	
~~~
w^{-} = ({\Lambda}^{-}_{1,l_1})^{d_1} ... ({\Lambda}^{-}_{{l_{m-1}+1},{l_m}})^{d_m},
\end{equation}
see example in Fig.~\ref{fig:assymNewt}, left. As far as the shifts $w^- \mapsto \Lambda^{k} w^- \Lambda^{-k}$ preserve Newton polygon, $b_1+...+b_n = d_1+...+d_m$ and only the conjugacy class of word matters, the same Newton polygon is provided by $w = w^- w^+ \Lambda^{-d_1-...-d_n}$. The upper- and lower- diagonal Lax operators defined by $w^{\pm}$ are constructed from hexagonal graph, in contrast to the symmetric case, where the basic building blocks were four-gonal 'fence-net' graph.
}

\paragraph{Wiring of parallel zig-zags}{
It remains to discuss a wiring of parallel zig-zags. Take Newton polygon with integral points on the boundary, which are not at the corners, i.e. those having at least one 'non-simple' side $\vec{u}'_k = h_k \cdot (a_k,b_k)$, with $\mathrm{gcd}(a_k,b_k)=1$ ($h_k > 0$, and let $a'_k > 0$ for certainty). Considering $h_k$ simple boundary intervals $(a_k,b_k)$ separately, one gets $h_k$ commuting sub-groups $(\widetilde{W}(A^{(1)}_{a_k-1}))^{\times h_k}$, whose resulting contribution  into word in double Weyl group by twists is 
\begin{equation}
w_{k,k+h_k} =
\left(\Lambda^{+}_{r_{k-1}+1,r_{k-1}+a_{k}} \right)^{b_{k}}
... 
\left(\Lambda^{+}_{r_{k-1} + (h_k-1)a_{k} + 1,r_{k-1} + h_k a_{k}} \right)^{b_{k}}.
\end{equation}
Alternatively, one can consider this intervals together, which gives group $\widetilde{W}(A^{(1)}_{h_k a_k-1})$ contributing by
\begin{equation}
w_{k,k+h_k} =
\left(\Lambda^{+}_{r_{k-1}+1,r_{k-1}+h_k a_{k}} \right)^{b_{k} h_k}.
\end{equation}
Two choices can be transformed one into another by local moves, however the second ansatz is more reduced compared to the first one, as it involves $(N - a_k h_k + 1)b_k h_k$ generators against $(N - a_k + 1)b_k h_k$ in the first case. Another benefit is that it can be easily extended to involve 'wiring' of parallel strands, by
\begin{equation}
\label{eq:wordkGeneric}
w_{k,k+h_k} =
\left(\Lambda^{+}_{r_{k-1}+1,r_{k-1}+h_k a_{k}} \right)^{h_k b_{k}} \tilde{w}_k,
~~~
\tilde{w}_k \in W(A_{h_k-1}),
\end{equation}
where $W(A_{h_k-1})$ is group acting by permutations of strands $r_{k-1}+1, r_{k-1}+2,..., r_{k-1} + h_k$, see example in Fig.~ \ref{fig:assymNewt}, right. One can assign such 'non-affine' word to each non-simple boundary interval of Newton polygon, however it is more natural not to bring all parallel intervals together, but to join them according to decomposition of $\tilde{w}_k$ into a product of simple cycles.
}

\subsection{Classification of perfect networks on torus}{
\label{ss:classification}
Systematizing examples of previous subsection, we show now that all bi-coloured graphs on torus can be reduced by local moves to 'normal forms', which are enumerated by Newton polygons (containing information about winding of zig-zags on torus), with the sides containing integral internal points partitioned according to the wiring of parallel zig-zags. Normal form attributed to graph is unique, up to $\mathrm{SA}(2,\mathbb{Z})$ transformation of Newton polygon. Similar combinatorics already appeared in \cite{CW} in the description of moduli spaces of monopole walls.

The statement is straightforward consequence of the fact, proved in \cite{FM:2014}, that one can always 'slice' bipartite graph on torus, and put into correspondence to it some conjugacy class in double Weyl group (\ref{eq:WeylGrDef}), and the following
\paragraph{Lemma.}{
Any conjugacy class in double Weyl group (\ref{eq:WeylGrDef}) contains unique element of the form
\begin{equation}
    w = w^+_{1} \cdot ... \cdot w^+_{n} \cdot w^-_{1} \cdot ... \cdot w^-_{m} \cdot \Lambda^{-b_1-...-b_n},
\end{equation}
$$
    w^+_{k} = \left(\Lambda^{+}_{r_{k-1}+1,r_{k}} \right)^{b_{k}} (r_{k-1}+\mathrm{gcd}(a_k,b_k),...,r_{k-1}+1),
    ~~~
    w^-_{k} = \left(\Lambda^{-}_{l_{k-1}+1,l_{k}} \right)^{d_{k}} \overline{(l_{k}+\mathrm{gcd}(c_k,d_k),...,l_{k}+1)},
$$
where
\begin{itemize}
    \item Numbers $a_k,b_k,c_k,d_k$ define ordered set of counter-clockwise oriented, boundary intervals $\vec{u}_k = (a_k, b_k)$, $\vec{v}_k = (-c_k, -d_k)$ with $a_k,c_k > 0$, of some Newton polygon of width $N$. The order starts from the direction $(0,-1)$, 'parallel' vectors (i.e. proportional, with positive rational coefficient)\footnote{Sides of the Newton polygon, containing internal integral points, can be split into pieces in various different ways.} are ordered from the longest to shortest. 
    \item Numbers $r_{k},l_{k}$ are defined by $r_{k} = a_1  + ... + a_k, \, l_{k} = c_1 + ... + c_k$ for $k>0$, $r_{0} = l_{0} = 0$.
    \item Words $\Lambda^{\pm}_{ij}$ are 'subgroup twists' defined by formula (\ref{eq:Lambdadef}).
    \item Words $(j,...,i) = s_{i} s_{i+1} ... s_{j-2} s_{j-1}$ and $\overline{(j,...,i)} = \bar{s}_{i} \bar{s}_{i+1} ... \bar{s}_{j-2} \bar{s}_{j-1}$ are simple cycles\footnote{The name comes from its action as permutation $j \mapsto j-1, ... , i+1 \mapsto i, i \mapsto j$}, $i<j$.
\end{itemize}}

\paragraph{Proof.}{
Any element $w$ of the group $\widetilde{W}(A^{(1)}_{N-1} \times A^{(1)}_{N-1})$ admits decomposition $w = w^{+}w^{-} \Lambda^{-K}$, where $w^{+},w^{-}$ are words, which contain only generators $s_i,\Lambda$ or $\bar{s}_i, \Lambda$ respectively, and total degree of $\Lambda$ in either $w^{+}$ or $w^{-}$ is $K$. Both $w^{\pm}$ belong to sub-groups of $\widetilde{W}(A^{(1)}_{N-1})$ - type, so we will classify conjugacy classes of its elements, and then show, how ambiguity with the distribution of $\Lambda$ can be fixed. Choose for definiteness subgroup generated by $s_i,\Lambda$. There is a structure of semi-direct product
\begin{equation}
\widetilde{W}(A^{(1)}_{N-1}) = \mathbb{Z}^N \rtimes  W(A_{N-1}),
\end{equation}
which comes from presentation $w^{+} = L\cdot g$, where $g$ is element of non-affine Weyl group generated by $s_i$, and $L$ is element of lattice generated by commuting elements $\Lambda_{i,i}^{+}$, as defined in (\ref{eq:Lambdadef}), i.e. those which take strand, wind it up over cylinder, and bring back onto initial place. Writing this as pairs, and using additive notation for elements of lattice $e_i = \Lambda_{i,i}^{+}$, we get product rule
\begin{equation}
(\Lat_1 \,;\, g_1)\cdot (\Lat_2 \,;\, g_2) = (\Lat_1+R_{g_1} (\Lat_2) \,;\, g_1 g_2),
\end{equation}
where $R_{g_1}$ acts on the basis elements of lattice by permutations
\begin{equation}
R_{s_i}(e_i) = e_{i+1},\, R_{s_i}(e_{i+1}) = e_{i},\, R_{s_i}(e_{j}) = e_{j} \text{~if~} i\neq j, j+1 ~ \text{and} ~ R_{g_1} R_{g_2} = R_{g_1 g_2}.
\end{equation}

The conjugacy classes in $\widetilde{W}(A^{(1)}_{N-1})$ are in bijection with the set of pairs $(\vec{q}\, , \lambda )$, where $\lambda = (\lambda_1 \geq ... \geq \lambda_{\ell(\lambda)} > 0)$ is the partition of number $N$, $\vec{q} \in \mathbb{Z}^{\ell(\lambda)}$ and $\ell(\lambda)$ is the number of parts in the partition $\lambda$. Indeed, conjugacy classes of permutations on $N$ elements are enumerated by partitions $\lambda$ of number $N$, each containing representative
\begin{equation}
(p_1,...\, ,1)
(p_2,...\, ,p_1+1)
\,...\,
(p_{\ell(\lambda)},...\, ,p_{\ell(\lambda)-1}+1),
\end{equation}
where $p_0 = 0$, $p_i = \lambda_1+...\, +\lambda_i$, and $(j,...\, ,i) = s_i...\,s_{j-1}$ is cyclic permutation, acting on the lattice by
\begin{equation}
R_{(j,...,i)}: ~~~ 
e_i \mapsto e_{i+1}
~~~,~~~...~~~,~~~
e_{j-1} \mapsto e_{j}
~~~,~~~~
e_{j} \mapsto e_{i}
\end{equation}
for $i<j$. As simple cycles $(p_{k},...,p_{k-1}+1)$ commute for different $k$, and generators of the lattice can be shifted along the cycles
\begin{equation}
\label{eq:LatticeTwist}
[(e_k \,;\, \mathrm{id})\cdot (0 \,;\, (j,...\,,i))] = [(e_l \,;\, \mathrm{id})\cdot (0 \,;\,  (j,...\,,i))],
~~~ \forall ~ i\leq k,l \leq j,
\end{equation}
where $[~~~]$ is taking of conjugacy class, then by moving elements of the lattice to the 'first lines', one gets
\begin{equation}
[w^+] = [ w^+_1 ...\, w^+_{\ell(\lambda)} ] \, ,
~~~
w_k^{+} = (q_k \cdot e_{p_{k-1}+1} \,;\, (p_k,...\, ,p_{k-1}+1)),
\end{equation}
for some $q_k\in \mathbb{Z}$, so the vector $\vec{q} = (q_1,...,q_{\ell(\lambda)})$ is the vector of the 'lengths' of lattice elements. To put conjugacy class in the form of the products of 'twists' $\Lambda_{ij}$, note that
\begin{equation}
\label{eq:twistDecomp}
    \left(\Lambda_{p_{k-1}+1,\, p_{k}}\right)^{q_k} \cdot
    (p_{k-1}+\mathrm{gcd}(\lambda_k, q_k),...\, ,p_{k-1}+1) = 
   (  V_k \,;\, \sigma_k \, ( p_{k}, p_{k-1}+1 ) \, \sigma_k^{-1} ),
\end{equation}
where the lattice element $V_{k} = t'_k ( e_{p_{k-1}+1} + ... + e_{p_{k}}) + e_{p_{k-1}+1} + ... + e_{p_{k-1}+t''_k}$ with $t'_k \in \mathbb{Z}_{\geq 0},~ 0 \leq t''_k < \lambda_k$ is defined by $q_k = t'_k \lambda_k + t''_k$, and comes from decomposition
\begin{equation}
    \left(\Lambda_{p_{k-1}+1,\, p_{k}}\right)^{q_k} = ( V_k \,;\, (p_k,...,p_{k-1}+1)^{q_k} ),
\end{equation}
which can be checked by direct computation, using that $\Lambda^{+}_{ij} = (k, ...\, , i) \cdot \Lambda^{+}_{kk} \cdot (j, ...\, , k) $ for any $i \leq k \leq j$, and non-affine permutation $\sigma_k$ is defined from
\begin{equation}
\label{eq:permDecomp}
    (p_{k}, ... , p_{k-1}+1)^{q_k} \cdot (p_{k-1} + \mathrm{gcd}(\lambda_k, q_k), ... \, ,p_{k-1}+1) = \sigma_k \cdot (p_{k}, ... , p_{k-1}+1) \cdot \sigma_k^{-1}
\end{equation}
which holds, because all orbits of the action of $i \mapsto i + q_k$ on $\mathbb{Z}/\lambda_k \mathbb{Z}$ can be uniquely presented by one of the numbers $1,...,\mathrm{gcd}(\lambda_k,q_k)$, so both sides of (\ref{eq:permDecomp}) got only one orbit. From (\ref{eq:twistDecomp}), using (\ref{eq:LatticeTwist}), for conjugacy classes follows
\begin{equation}
    [\left(\Lambda_{p_{k-1}+1,\, p_{k}}\right)^{q_k} \cdot (p_{k-1}+\mathrm{gcd}(\lambda_k, q_k),...\, ,p_{k-1}+1)] = 
    [ \left( q_k \cdot e_{p_{k-1}+1} \,;\, ( p_{k} , ... , p_{k-1}+1 ) \right) ] = [w^+_k],
\end{equation}
which is almost statement of the Lemma. The $w^{-}$ part can be reduced to the normal form, encoded by $(\vec{\bar{q}},\bar{\lambda})$, in the same way. The only element, which is common for words $w^+$ and $w^-$ is $\Lambda$, which also do not commute with all generators $s_{i}$ and $\bar{s}_i$. However, we initially distributed it in $w = w^{+} w^{-} \Lambda^{-K}$ in a such way, that the total degree of $\Lambda$ inside $w^{-}\Lambda^{-K}$ or $\Lambda^{-K} w^{+}$ is zero, so the treatment of $w^{+}$ or $w^{-}$ is not affected by another part. Finally, conjugating $w^{\pm}$ by suitable permutations from non-affine parts, we can rearrange indices of $s_i,\bar{s}_i$ inside $w_k^{\pm}$ by counter-clockwise order on the directions of vectors $(\lambda_i, q_i)$, $(-\bar{\lambda}_i, -\bar{q}_i)$, starting from the direction $(0,-1)$, and by decrease of lengths for the vectors of the same slope, obtaining numbers $(a_i,b_i)$ and $(c_i,d_i)$. The properties that the sum of vectors is zero, i.e. that they can be composed into the boundary of Newton polygon, and that the width of this polygon is $N$, are guaranteed by $\sum_i \lambda_i = \sum_i \bar{\lambda}_i = N$, $\sum_i q_i = \sum_i \bar{q}_i = K$.
}
}
\section{Discussion}

In this paper we have demonstrated that the Bazhanov-Sergeev solution of the tetrahedron equation appears naturally as the basic building block for the transfer matrix of paths in the theory of cluster integrable systems. We have also shown how the integrable system with arbitrary symmetric Newton polygon can be built using this building block. We have explained how this construction originates from the combinatorics of words in the double affine Weyl groups and used it to explicitly construct bi-coloured graph for the integrable system associated with any Newton polygon. We have also proven the classification Lemma stating that we have constructed all possible systems of such kind.

The following questions seem to be promising for future developments of this topic:

\begin{itemize}

\item As the Poisson brackets on weights of paths are bi-linearly constant, it can be quantized in a straightforward way by \cite{FG:2003}
\begin{equation}
\{w_{\gamma_1}, w_{\gamma_2}\} = \varepsilon(\gamma_1,\gamma_2)w_{\gamma_1}w_{\gamma_2}
~~~ \longrightarrow ~~~
\hat{w}_{\gamma_1} \hat{w}_{\gamma_2} = t^{\frac{1}{2}\varepsilon(\gamma_1,\gamma_2)} \hat{w}_{\gamma_1 + \gamma_2}
\end{equation}
The mutation, which was a canonical transformation classically, in the quantum world becomes a conjugation by quantum dilogarithm. Extension of the arguments presented in this paper to the quantum case will provide a closed formula for the tetrahedron $\Rtet$-matrix $\Rtet_{abc}$ in terms of four quantum dilogarithms. This can clarify the appearance of the product of four functions similar to quantum dilogarithms at the root of unity in the vertex weight of the 3d vertex model \cite{BB1, BB2, K93, K94}, whose solution is known to be a solution of the tetrahedron equation \cite{KMS, SMS}. Such product (outside of the roots of unity) was also noted in \cite{S}. Another promising direction of research is construction of new solution for the tetrahedron equation using cluster algebras with fermionic variables \cite{OS}, as suggested by recent appearance of quivers with fermionic nodes in representations theory of affine algebras \cite{KOS, LS16, LS18, BFM, Ze} and approach of \cite{S09} to super-algebras using tetrahedron equation.

\item Surprisingly, the same quiver and the same cluster transformation as those shown in Fig.~\ref{fig:tetrahedron} have already appeared in the context of the relation between cluster algebras and vertex integrable systems in~\cite{Y16}. The physical origin of these solutions was the 2d $\mathcal{N}=(2,2)$ supersymmetric sigma-model, whose K\"ahler parameters were shown in~\cite{BPZ} to transform as cluster variables under Seiberg dualities. From the other side, the approach to cluster integrable systems which we have used here is suspected to originate from $5d$ $\mathcal{N}=1$ theories~\cite{EFS, BGM, MS}, where cluster variables play the role of Seiberg-Witten curve's moduli. This intriguing coincidence should have some unifying physical origins.

\item The systems we have considered were mainly of ``affine'' type: they live on double Bruhat cells of the affine group $\GLb(N)$ and being rewritten in Darboux variables represent ``closed'' chains of interacting particles~\cite{EFS,M,FM:2014,GSTV:2014}. The integrability theorem, proved in \cite{GK:2011}, assumes that the perfect network on torus is minimal, i.e.\ that its zig-zags do not have self-intersections, and that parallel zig-zags (those, whose classes in $\H1(\T2,\mathbb{Z})$ are proportional with positive coefficient) do not intersect. The cluster description of the ``open'' chains\footnote{Which were historically the first examples of the cluster description of integrable systems~\cite{GSV:2009}.}, which live on double Bruhat cells of the non-affine group $\mathrm{PGL}(N)$, involves networks drawn on a cylinder (or on a cut torus --- this can be treated as a particular case of a ``squashed'' Newton polygon of zero area). So all the intersections of zig-zags are either self- or parallel-, and integrability of such systems is not guaranteed by~\cite{GK:2011}. However it can be proved by other methods.

We have unified these classes of systems by considering the wiring of parallel zig-zags. As it was shown in~\cite{GSV:2009-1}, the Lax operator of any network on a cylinder has an $r$-matrix Poisson bracket with itself, however the general integrability criterion, which allows to compare the number of independent integrals of motion and the dimensions of the symplectic leaves still has to be developed.

\item We have proven the classification theorem for bi-coloured graphs on torus. Graphs which contain wiring of parallel zig-zags cannot be made minimal by local moves, i.e.\ self-intersections of zig-zags are protected by topology. However, they can always be made ``locally minimal'', which means that they become such networks on torus, that being cut by any curves into a disk, they become minimal network on the disk, as follows from the reduction theorem proved in~\cite{Thurston}. In the language of double Weyl groups locally minimal diagrams are those defined by reduced words.

However, in our consideration we allowed to reduce parallel bigons by the use of $s_i^2 = 1$ which is not a cluster transformation. Classification of the normal forms of the locally minimal networks ``up to cluster transformations'' with the bigon reduction relation $s_i^2 = s_i$ seems to be a fruitful direction for further investigations, especially as it might exhibit interesting $\mathrm{SL}(2,\mathbb{Z})$ covariant behaviour~\footnote{However, we conjecture that the set of cyclically irreducible words will be the same for both relations, $s_i^2=s_i$ and $s_i^2=1$.}. The problem of parallel bigons itself is still poorly understood in cluster algebras, and also awaits its solution. We also expect, that the condition that the Newton polygon does not contain vertical sides might be removed and full $\mathrm{SL}(2,\mathbb{Z})$ covariance restored by the replacement of the double affine Weyl group with a certain generalization thereof, originating from toroidal algebras.

\item In this paper we have been discussing continuous time integrability only. However, the cluster integrable systems are known to have rich discrete dynamics. In~\cite{GI} the general structure of the group of discrete transformations generated by spider moves was given. However, it is known that even for quivers coming from bi-coloured graphs there is a much larger group of cluster transformations (sequences of mutations and permutations of quiver vertices) which bring quivers back into itself, which however cannot be represented by a sequence of bi-coloured graph transformations (see e.g.~\cite{ILP} for hexagonal lattice and \cite{MS} for the four-gonal one). These transformations are related to boundary intervals of Newton polygons with integral internal points, and realize permutations of ``parallel'' zig-zags (whose classes in torus homology coincide).

We expect that using the results of this paper, a big piece of the cluster mapping class group containing the sub-group $W(A^{(1)}_{N})$ for each boundary interval with $N$ internal points, and a subgroup described in~\cite{GI}, can be explicitly constructed. The half Dehn-twists $\Rtet$-matrix~\cite{HI13, SS} should also find their natural interpretation in this construction.

\end{itemize}

We want to thank to M. Bershtein, V. Fock, A. Marshakov, A. Shapiro, I. Vilkoviskiy for numerous fruitful discussions. Work of P.G. and M.S. was partially supported by the HSE University Basic Research Program, Russian Academic Excellence Project '5-100'. Work of P.G. was also partially supported by the Russian Science Foundation Grant No. 19-11-00275, in particular, results of Sections~\ref{ss:transform}, \ref{ss:tetra}, \ref{ss:classification} were obtained under support of RSF. Work of M.S. was also supported in part by Young Russian Mathematics award. Y.Z. is partly supported by RFBR grant 19-02-00815, 19-51-50008-YaF, 19-51-18006-Bolg.

\section*{Appendix A. Details on tetrahedron transformation.}

\begin{figure}[h!]
\begin{center}
\scalebox{0.8}{
\begin{tikzpicture}

\tikzmath{\d=1.5;};
%

\tikzmath{\xshift=0;\yshift=0;\w=0.75;};

\draw[thick] (\xshift+0*\d,2*\d+\yshift)--(\xshift+6*\d,2*\d+\yshift);
\draw[thick] (\xshift+0*\d,\d+\yshift)--(\xshift+6*\d,\d+\yshift);
\draw[thick] (\xshift+0*\d,\yshift)--(\xshift+6*\d,\yshift);


\draw[thick] (\xshift+0.5*\d,\yshift+2*\d)--(\xshift+0.5*\d,\yshift+\d);
\draw[thick] (\xshift+1.5*\d,\yshift+\d)--(\xshift+1.5*\d,\yshift+2*\d);

\draw[blackCircle] (\xshift+0.5*\d,\yshift+2*\d) circle;
\draw[whiteCircle] (\xshift+0.5*\d,\yshift+\d) circle;
\draw[blackCircle] (\xshift+1.5*\d,\yshift+\d) circle;
\draw[whiteCircle] (\xshift+1.5*\d,\yshift+2*\d) circle;


\draw[thick] (\xshift+2.5*\d,\yshift+\d)--(\xshift+2.5*\d,\yshift);
\draw[thick] (\xshift+3.5*\d,\yshift)--(\xshift+3.5*\d,\yshift+\d);

\draw[blackCircle] (\xshift+2.5*\d,\yshift+\d) circle;
\draw[whiteCircle] (\xshift+2.5*\d,\yshift+0) circle;
\draw[blackCircle] (\xshift+3.5*\d,\yshift+0) circle;
\draw[whiteCircle] (\xshift+3.5*\d,\yshift+\d) circle;


\draw[thick] (\xshift+4.5*\d,\yshift+2*\d)--(\xshift+4.5*\d,\yshift+1*\d);
\draw[thick] (\xshift+5.5*\d,\yshift+1*\d)--(\xshift+5.5*\d,\yshift+2*\d);

\draw[blackCircle] (\xshift+4.5*\d,\yshift+2*\d) circle;
\draw[whiteCircle] (\xshift+4.5*\d,\yshift+1*\d) circle;
\draw[blackCircle] (\xshift+5.5*\d,\yshift+1*\d) circle;
\draw[whiteCircle] (\xshift+5.5*\d,\yshift+2*\d) circle;


\draw[\colMut, line width=\w mm] (\xshift+1.25*\d, \yshift+1*\d) -- (\xshift+2.75*\d, \yshift+1*\d);
\draw[\colMut, line width=\w mm] (\xshift+1.5*\d, \yshift+1*\d) -- (\xshift+1.5*\d, \yshift+1.25*\d);
\draw[\colMut, line width=\w mm] (\xshift+2.5*\d, \yshift+1*\d) -- (\xshift+2.5*\d, \yshift+0.75*\d);

\draw[\colMut, line width=\w mm] (\xshift+3.25*\d, \yshift+1*\d) -- (\xshift+4.75*\d, \yshift+1*\d);
\draw[\colMut, line width=\w mm] (\xshift+3.5*\d, \yshift+1*\d) -- (\xshift+3.5*\d, \yshift+0.75*\d);
\draw[\colMut, line width=\w mm] (\xshift+4.5*\d, \yshift+1*\d) -- (\xshift+4.5*\d, \yshift+1.25*\d);

\tikzmath{\nx = \xshift + 0.5*\d;\ny = \yshift + 2*\d;}

\draw[very thick, ->, \colP] (\nx - 0.2*\d, \ny + 0.1*\d) -- (\nx + 0.2*\d, \ny + 0.1*\d);
\draw[very thick, ->, \colP] (\nx + 0.22*\d, \ny - 0.02*\d) -- (\nx + 0.02*\d, \ny - 0.22*\d);
\draw[very thick, ->, \colP] (\nx - 0.02*\d, \ny - 0.22*\d) -- (\nx - 0.22*\d, \ny - 0.02*\d);

\node at (\nx, \ny + 0.2*\d) {\small $a_3$};
\node at (\nx+0.28*\d, \ny - 0.19*\d) {\small $a_1$};
\node at (\nx-0.22*\d, \ny - 0.18*\d) {\small $a_2$};

\tikzmath{\nx = \xshift + 0.5*\d;\ny = \yshift + 1*\d;}

\draw[very thick, ->, \colP] (\nx + 0.2*\d, \ny - 0.1*\d) -- (\nx - 0.2*\d, \ny - 0.1*\d);
\draw[very thick, ->, \colP] (\nx - 0.22*\d, \ny + 0.02*\d) -- (\nx - 0.02*\d, \ny + 0.22*\d);
\draw[very thick, ->, \colP] (\nx + 0.02*\d, \ny + 0.22*\d) -- (\nx + 0.22*\d, \ny + 0.02*\d);

\node at (\nx, \ny - 0.2*\d) {\small $b_2$};
\node at (\nx-0.24*\d, \ny + 0.19*\d) {\small $b_3$};
\node at (\nx+0.26*\d, \ny + 0.18*\d) {\small $b_1$};

\tikzmath{\nx = \xshift + 1.5*\d;\ny = \yshift + 1*\d;}

\draw[very thick, ->, \colP] (\nx + 0.2*\d, \ny - 0.1*\d) -- (\nx - 0.2*\d, \ny - 0.1*\d);
\draw[very thick, ->, \colP] (\nx - 0.22*\d, \ny + 0.02*\d) -- (\nx - 0.02*\d, \ny + 0.22*\d);
\draw[very thick, ->, \colP] (\nx + 0.02*\d, \ny + 0.22*\d) -- (\nx + 0.22*\d, \ny + 0.02*\d);

\node at (\nx, \ny - 0.2*\d) {\small $c_3$};
\node at (\nx-0.24*\d, \ny + 0.19*\d) {\small $c_1$};
\node at (\nx+0.26*\d, \ny + 0.18*\d) {\small $c_2$};

\tikzmath{\nx = \xshift + 1.5*\d;\ny = \yshift + 2*\d;}

\draw[very thick, ->, \colP] (\nx - 0.2*\d, \ny + 0.1*\d) -- (\nx + 0.2*\d, \ny + 0.1*\d);
\draw[very thick, ->, \colP] (\nx + 0.22*\d, \ny - 0.02*\d) -- (\nx + 0.02*\d, \ny - 0.22*\d);
\draw[very thick, ->, \colP] (\nx - 0.02*\d, \ny - 0.22*\d) -- (\nx - 0.22*\d, \ny - 0.02*\d);

\node at (\nx, \ny + 0.22*\d) {\small $d_2$};
\node at (\nx+0.28*\d, \ny - 0.19*\d) {\small $d_3$};
\node at (\nx-0.22*\d, \ny - 0.18*\d) {\small $d_1$};

\tikzmath{\nx = \xshift + 2.5*\d;\ny = \yshift + 1*\d;}

\draw[very thick, ->, \colP] (\nx - 0.2*\d, \ny + 0.1*\d) -- (\nx + 0.2*\d, \ny + 0.1*\d);
\draw[very thick, ->, \colP] (\nx + 0.22*\d, \ny - 0.02*\d) -- (\nx + 0.02*\d, \ny - 0.22*\d);
\draw[very thick, ->, \colP] (\nx - 0.02*\d, \ny - 0.22*\d) -- (\nx - 0.22*\d, \ny - 0.02*\d);

\node at (\nx, \ny + 0.22*\d) {\small $e_3$};
\node at (\nx+0.27*\d, \ny - 0.2*\d) {\small $e_1$};
\node at (\nx-0.21*\d, \ny - 0.18*\d) {\small $e_2$};

\tikzmath{\nx = \xshift + 2.5*\d;\ny = \yshift;}

\draw[very thick, ->, \colP] (\nx + 0.2*\d, \ny - 0.1*\d) -- (\nx - 0.2*\d, \ny - 0.1*\d);
\draw[very thick, ->, \colP] (\nx - 0.22*\d, \ny + 0.02*\d) -- (\nx - 0.02*\d, \ny + 0.22*\d);
\draw[very thick, ->, \colP] (\nx + 0.02*\d, \ny + 0.22*\d) -- (\nx + 0.22*\d, \ny + 0.02*\d);

\node at (\nx, \ny - 0.2*\d) {\small $f_2$};
\node at (\nx-0.24*\d, \ny + 0.19*\d) {\small $f_3$};
\node at (\nx+0.26*\d, \ny + 0.18*\d) {\small $f_1$};

\tikzmath{\nx = \xshift + 3.5*\d;\ny = \yshift;}

\draw[very thick, ->, \colP] (\nx + 0.2*\d, \ny - 0.1*\d) -- (\nx - 0.2*\d, \ny - 0.1*\d);
\draw[very thick, ->, \colP] (\nx - 0.22*\d, \ny + 0.02*\d) -- (\nx - 0.02*\d, \ny + 0.22*\d);
\draw[very thick, ->, \colP] (\nx + 0.02*\d, \ny + 0.22*\d) -- (\nx + 0.22*\d, \ny + 0.02*\d);

\node at (\nx, \ny - 0.2*\d) {\small $g_3$};
\node at (\nx-0.24*\d, \ny + 0.19*\d) {\small $g_1$};
\node at (\nx+0.26*\d, \ny + 0.18*\d) {\small $g_2$};

\tikzmath{\nx = \xshift + 3.5*\d;\ny = \yshift + 1*\d;}

\draw[very thick, ->, \colP] (\nx - 0.2*\d, \ny + 0.1*\d) -- (\nx + 0.2*\d, \ny + 0.1*\d);
\draw[very thick, ->, \colP] (\nx + 0.22*\d, \ny - 0.02*\d) -- (\nx + 0.02*\d, \ny - 0.22*\d);
\draw[very thick, ->, \colP] (\nx - 0.02*\d, \ny - 0.22*\d) -- (\nx - 0.22*\d, \ny - 0.02*\d);

\node at (\nx, \ny + 0.22*\d) {\small $h_2$};
\node at (\nx+0.27*\d, \ny - 0.2*\d) {\small $h_3$};
\node at (\nx-0.21*\d, \ny - 0.18*\d) {\small $h_1$};

\tikzmath{\nx = \xshift + 4.5*\d;\ny = \yshift + 2*\d;}

\draw[very thick, ->, \colP] (\nx - 0.2*\d, \ny + 0.1*\d) -- (\nx + 0.2*\d, \ny + 0.1*\d);
\draw[very thick, ->, \colP] (\nx + 0.22*\d, \ny - 0.02*\d) -- (\nx + 0.02*\d, \ny - 0.22*\d);
\draw[very thick, ->, \colP] (\nx - 0.02*\d, \ny - 0.22*\d) -- (\nx - 0.22*\d, \ny - 0.02*\d);

\node at (\nx, \ny + 0.22*\d) {\small $i_3$};
\node at (\nx+0.27*\d, \ny - 0.2*\d) {\small $i_1$};
\node at (\nx-0.21*\d, \ny - 0.18*\d) {\small $i_2$};

\tikzmath{\nx = \xshift + 4.5*\d;\ny = \yshift+1*\d;}

\draw[very thick, ->, \colP] (\nx + 0.2*\d, \ny - 0.1*\d) -- (\nx - 0.2*\d, \ny - 0.1*\d);
\draw[very thick, ->, \colP] (\nx - 0.22*\d, \ny + 0.02*\d) -- (\nx - 0.02*\d, \ny + 0.22*\d);
\draw[very thick, ->, \colP] (\nx + 0.02*\d, \ny + 0.22*\d) -- (\nx + 0.22*\d, \ny + 0.02*\d);

\node at (\nx, \ny - 0.2*\d) {\small $j_2$};
\node at (\nx-0.24*\d, \ny + 0.19*\d) {\small $j_3$};
\node at (\nx+0.26*\d, \ny + 0.18*\d) {\small $j_1$};

\tikzmath{\nx = \xshift + 5.5*\d;\ny = \yshift+1*\d;}

\draw[very thick, ->, \colP] (\nx + 0.2*\d, \ny - 0.1*\d) -- (\nx - 0.2*\d, \ny - 0.1*\d);
\draw[very thick, ->, \colP] (\nx - 0.22*\d, \ny + 0.02*\d) -- (\nx - 0.02*\d, \ny + 0.22*\d);
\draw[very thick, ->, \colP] (\nx + 0.02*\d, \ny + 0.22*\d) -- (\nx + 0.22*\d, \ny + 0.02*\d);

\node at (\nx, \ny - 0.2*\d) {\small $k_3$};
\node at (\nx-0.24*\d, \ny + 0.19*\d) {\small $k_1$};
\node at (\nx+0.26*\d, \ny + 0.18*\d) {\small $k_2$};

\tikzmath{\nx = \xshift + 5.5*\d;\ny = \yshift + 2*\d;}

\draw[very thick, ->, \colP] (\nx - 0.2*\d, \ny + 0.1*\d) -- (\nx + 0.2*\d, \ny + 0.1*\d);
\draw[very thick, ->, \colP] (\nx + 0.22*\d, \ny - 0.02*\d) -- (\nx + 0.02*\d, \ny - 0.22*\d);
\draw[very thick, ->, \colP] (\nx - 0.02*\d, \ny - 0.22*\d) -- (\nx - 0.22*\d, \ny - 0.02*\d);

\node at (\nx, \ny + 0.22*\d) {\small $l_2$};
\node at (\nx+0.27*\d, \ny - 0.2*\d) {\small $l_3$};
\node at (\nx-0.21*\d, \ny - 0.18*\d) {\small $l_1$};


\draw[->, very thick] (\xshift+4.5,\yshift - 0.75) -- (\xshift+4.5,\yshift-1.25);


\tikzmath{\xshift=0;\yshift=-5;};

\draw[thick] (\xshift+0*\d,2*\d+\yshift)--(\xshift+6*\d,2*\d+\yshift);
\draw[thick] (\xshift+0*\d,\d+\yshift)--(\xshift+6*\d,\d+\yshift);
\draw[thick] (\xshift+0*\d,\yshift)--(\xshift+6*\d,\yshift);


\draw[thick] (\xshift+0.5*\d,\yshift+2*\d)--(\xshift+0.5*\d,\yshift+\d);
\draw[thick] (\xshift+2.5*\d,\yshift+\d)--(\xshift+2.5*\d,\yshift+2*\d);

\draw[blackCircle] (\xshift+0.5*\d,\yshift+2*\d) circle;
\draw[whiteCircle] (\xshift+0.5*\d,\yshift+\d) circle;
\draw[blackCircle] (\xshift+2.5*\d,\yshift+\d) circle;
\draw[whiteCircle] (\xshift+2.5*\d,\yshift+2*\d) circle;


\draw[thick] (\xshift+1.5*\d,\yshift+\d)--(\xshift+1.5*\d,\yshift);
\draw[thick] (\xshift+4.5*\d,\yshift)--(\xshift+4.5*\d,\yshift+\d);

\draw[blackCircle] (\xshift+1.5*\d,\yshift+\d) circle;
\draw[whiteCircle] (\xshift+1.5*\d,\yshift+0) circle;
\draw[blackCircle] (\xshift+4.5*\d,\yshift+0) circle;
\draw[whiteCircle] (\xshift+4.5*\d,\yshift+\d) circle;


\draw[thick] (\xshift+3.5*\d,\yshift+2*\d)--(\xshift+3.5*\d,\yshift+1*\d);
\draw[thick] (\xshift+5.5*\d,\yshift+1*\d)--(\xshift+5.5*\d,\yshift+2*\d);

\draw[blackCircle] (\xshift+3.5*\d,\yshift+2*\d) circle;
\draw[whiteCircle] (\xshift+3.5*\d,\yshift+1*\d) circle;
\draw[blackCircle] (\xshift+5.5*\d,\yshift+1*\d) circle;
\draw[whiteCircle] (\xshift+5.5*\d,\yshift+2*\d) circle;

\draw[\colMut, line width=\w mm] (\xshift+2.5*\d, \yshift+1*\d) -- (\xshift+3.5*\d, \yshift+1*\d) -- (\xshift+3.5*\d, \yshift+2*\d) -- (\xshift+2.5*\d, \yshift+2*\d) -- (\xshift+2.5*\d, \yshift+1*\d);

\draw[->, very thick] (\xshift+4.5,\yshift - 0.75) -- (\xshift+4.5,\yshift-1.25);


\tikzmath{\xshift=0;\yshift=-10;};

\draw[thick] (\xshift+0*\d,2*\d+\yshift)--(\xshift+6*\d,2*\d+\yshift);
\draw[thick] (\xshift+0*\d,\d+\yshift)--(\xshift+6*\d,\d+\yshift);
\draw[thick] (\xshift+0*\d,\yshift)--(\xshift+6*\d,\yshift);


\draw[thick] (\xshift+0.5*\d,\yshift+2*\d)--(\xshift+0.5*\d,\yshift+\d);
\draw[thick] (\xshift+2.5*\d,\yshift+\d)--(\xshift+2.5*\d,\yshift+2*\d);

\draw[blackCircle] (\xshift+0.5*\d,\yshift+2*\d) circle;
\draw[whiteCircle] (\xshift+0.5*\d,\yshift+\d) circle;
\draw[whiteCircle] (\xshift+2.5*\d,\yshift+\d) circle;
\draw[blackCircle] (\xshift+2.5*\d,\yshift+2*\d) circle;


\draw[thick] (\xshift+1.5*\d,\yshift+\d)--(\xshift+1.5*\d,\yshift);
\draw[thick] (\xshift+4.5*\d,\yshift)--(\xshift+4.5*\d,\yshift+\d);

\draw[blackCircle] (\xshift+1.5*\d,\yshift+\d) circle;
\draw[whiteCircle] (\xshift+1.5*\d,\yshift+0) circle;
\draw[blackCircle] (\xshift+4.5*\d,\yshift+0) circle;
\draw[whiteCircle] (\xshift+4.5*\d,\yshift+\d) circle;


\draw[thick] (\xshift+3.5*\d,\yshift+2*\d)--(\xshift+3.5*\d,\yshift+1*\d);
\draw[thick] (\xshift+5.5*\d,\yshift+1*\d)--(\xshift+5.5*\d,\yshift+2*\d);

\draw[whiteCircle] (\xshift+3.5*\d,\yshift+2*\d) circle;
\draw[blackCircle] (\xshift+3.5*\d,\yshift+1*\d) circle;
\draw[blackCircle] (\xshift+5.5*\d,\yshift+1*\d) circle;
\draw[whiteCircle] (\xshift+5.5*\d,\yshift+2*\d) circle;

\draw[\colMut, line width=\w mm] (\xshift+0.25*\d, \yshift+2*\d) -- (\xshift+2.75*\d, \yshift+2*\d);
\draw[\colMut, line width=\w mm] (\xshift+0.5*\d, \yshift+2*\d) -- (\xshift+0.5*\d, \yshift+1.75*\d);
\draw[\colMut, line width=\w mm] (\xshift+2.5*\d, \yshift+1.75*\d) -- (\xshift+2.5*\d, \yshift+2*\d);

\draw[\colMut, line width=\w mm] (\xshift+3.25*\d, \yshift+2*\d) -- (\xshift+5.75*\d, \yshift+2*\d);
\draw[\colMut, line width=\w mm] (\xshift+3.5*\d, \yshift+2*\d) -- (\xshift+3.5*\d, \yshift+1.75*\d);
\draw[\colMut, line width=\w mm] (\xshift+5.5*\d, \yshift+1.75*\d) -- (\xshift+5.5*\d, \yshift+2*\d);

\draw[->, very thick] (\xshift+4.5,\yshift - 0.75) -- (\xshift+4.5,\yshift-1.25);


\tikzmath{\xshift=0;\yshift=-15;};

\draw[thick] (\xshift+0*\d,2*\d+\yshift)--(\xshift+6*\d,2*\d+\yshift);
\draw[thick] (\xshift+0*\d,\d+\yshift)--(\xshift+0.5*\d,\d+\yshift);
\draw[thick] (\xshift+2.5*\d,\d+\yshift)--(\xshift+3.5*\d,\d+\yshift);
\draw[thick] (\xshift+5.5*\d,\d+\yshift)--(\xshift+6*\d,\d+\yshift);
\draw[thick] (\xshift+0*\d,\yshift)--(\xshift+6*\d,\yshift);


\draw[thick] (\xshift+1.5*\d,\yshift+1.5*\d)--(\xshift+0.5*\d,\yshift+\d);
\draw[thick] (\xshift+2.5*\d,\yshift+\d)--(\xshift+1.5*\d,\yshift+1.5*\d);
\draw[thick] (\xshift+1.5*\d,\yshift+1.5*\d)--(\xshift+1.5*\d,\yshift+2*\d);

\draw[thick] (\xshift+0.5*\d,\yshift+\d)--(\xshift+1.5*\d,\yshift+0.5*\d);
\draw[thick] (\xshift+2.5*\d,\yshift+\d)--(\xshift+1.5*\d,\yshift+0.5*\d);
\draw[thick] (\xshift+1.5*\d,\yshift+0.5*\d)--(\xshift+1.5*\d,\yshift);

\draw[blackCircle] (\xshift+1.5*\d,\yshift+0.5*\d) circle;
\draw[blackCircle] (\xshift+1.5*\d,\yshift+1.5*\d) circle;
\draw[whiteCircle] (\xshift+0.5*\d,\yshift+\d) circle;
\draw[whiteCircle] (\xshift+2.5*\d,\yshift+\d) circle;
\draw[blackCircle] (\xshift+1.5*\d,\yshift+2*\d) circle;
\draw[whiteCircle] (\xshift+1.5*\d,\yshift+0) circle;


\draw[thick] (\xshift+4.5*\d,\yshift+1.5*\d)--(\xshift+3.5*\d,\yshift+\d);
\draw[thick] (\xshift+5.5*\d,\yshift+\d)--(\xshift+4.5*\d,\yshift+1.5*\d);
\draw[thick] (\xshift+4.5*\d,\yshift+1.5*\d)--(\xshift+4.5*\d,\yshift+2*\d);

\draw[thick] (\xshift+3.5*\d,\yshift+\d)--(\xshift+4.5*\d,\yshift+0.5*\d);
\draw[thick] (\xshift+5.5*\d,\yshift+\d)--(\xshift+4.5*\d,\yshift+0.5*\d);
\draw[thick] (\xshift+4.5*\d,\yshift+0.5*\d)--(\xshift+4.5*\d,\yshift);

\draw[whiteCircle] (\xshift+4.5*\d,\yshift+0.5*\d) circle;
\draw[whiteCircle] (\xshift+4.5*\d,\yshift+1.5*\d) circle;
\draw[blackCircle] (\xshift+3.5*\d,\yshift+\d) circle;
\draw[blackCircle] (\xshift+5.5*\d,\yshift+\d) circle;
\draw[whiteCircle] (\xshift+4.5*\d,\yshift+2*\d) circle;
\draw[blackCircle] (\xshift+4.5*\d,\yshift+0) circle;

\draw[\colMut, line width=\w mm] (\xshift+0.5*\d, \yshift+1*\d) -- (\xshift+1.5*\d, \yshift+0.5*\d) -- (\xshift+2.5*\d, \yshift+1*\d) -- (\xshift+1.5*\d, \yshift+1.5*\d) -- (\xshift+0.5*\d, \yshift+1*\d);

\draw[\colMut, line width=\w mm] (\xshift+3.5*\d, \yshift+1*\d) -- (\xshift+4.5*\d, \yshift+0.5*\d) -- (\xshift+5.5*\d, \yshift+1*\d) -- (\xshift+4.5*\d, \yshift+1.5*\d) -- (\xshift+3.5*\d, \yshift+1*\d);

\draw[->, very thick] (\xshift+6.25*\d,\yshift + 1*\d) -- (\xshift+6.75*\d,\yshift+1*\d);


\tikzmath{\xshift=7*\d;\yshift=-15;};

\draw[thick] (\xshift+0*\d,2*\d+\yshift)--(\xshift+6*\d,2*\d+\yshift);
\draw[thick] (\xshift+0*\d,\d+\yshift)--(\xshift+0.5*\d,\d+\yshift);
\draw[thick] (\xshift+2.5*\d,\d+\yshift)--(\xshift+3.5*\d,\d+\yshift);
\draw[thick] (\xshift+5.5*\d,\d+\yshift)--(\xshift+6*\d,\d+\yshift);
\draw[thick] (\xshift+0*\d,\yshift)--(\xshift+6*\d,\yshift);


\draw[thick] (\xshift+1.5*\d,\yshift+1.5*\d)--(\xshift+0.5*\d,\yshift+\d);
\draw[thick] (\xshift+2.5*\d,\yshift+\d)--(\xshift+1.5*\d,\yshift+1.5*\d);
\draw[thick] (\xshift+1.5*\d,\yshift+1.5*\d)--(\xshift+1.5*\d,\yshift+2*\d);

\draw[thick] (\xshift+0.5*\d,\yshift+\d)--(\xshift+1.5*\d,\yshift+0.5*\d);
\draw[thick] (\xshift+2.5*\d,\yshift+\d)--(\xshift+1.5*\d,\yshift+0.5*\d);
\draw[thick] (\xshift+1.5*\d,\yshift+0.5*\d)--(\xshift+1.5*\d,\yshift);

\draw[whiteCircle] (\xshift+1.5*\d,\yshift+0.5*\d) circle;
\draw[whiteCircle] (\xshift+1.5*\d,\yshift+1.5*\d) circle;
\draw[blackCircle] (\xshift+0.5*\d,\yshift+\d) circle;
\draw[blackCircle] (\xshift+2.5*\d,\yshift+\d) circle;
\draw[blackCircle] (\xshift+1.5*\d,\yshift+2*\d) circle;
\draw[whiteCircle] (\xshift+1.5*\d,\yshift+0) circle;


\draw[thick] (\xshift+4.5*\d,\yshift+1.5*\d)--(\xshift+3.5*\d,\yshift+\d);
\draw[thick] (\xshift+5.5*\d,\yshift+\d)--(\xshift+4.5*\d,\yshift+1.5*\d);
\draw[thick] (\xshift+4.5*\d,\yshift+1.5*\d)--(\xshift+4.5*\d,\yshift+2*\d);

\draw[thick] (\xshift+3.5*\d,\yshift+\d)--(\xshift+4.5*\d,\yshift+0.5*\d);
\draw[thick] (\xshift+5.5*\d,\yshift+\d)--(\xshift+4.5*\d,\yshift+0.5*\d);
\draw[thick] (\xshift+4.5*\d,\yshift+0.5*\d)--(\xshift+4.5*\d,\yshift);

\draw[blackCircle] (\xshift+4.5*\d,\yshift+0.5*\d) circle;
\draw[blackCircle] (\xshift+4.5*\d,\yshift+1.5*\d) circle;
\draw[whiteCircle] (\xshift+3.5*\d,\yshift+\d) circle;
\draw[whiteCircle] (\xshift+5.5*\d,\yshift+\d) circle;

\draw[whiteCircle] (\xshift+4.5*\d,\yshift+2*\d) circle;
\draw[blackCircle] (\xshift+4.5*\d,\yshift+0) circle;

\draw[\colMut, line width=\w mm] (\xshift+1.25*\d, \yshift+0*\d) -- (\xshift+1.5*\d, \yshift+0*\d) -- (\xshift+1.5*\d, \yshift+0.5*\d) -- (\xshift+1.2*\d, \yshift+0.65*\d);
\draw[\colMut, line width=\w mm] (\xshift+1.75*\d, \yshift+0*\d) -- (\xshift+1.5*\d, \yshift+0*\d) -- (\xshift+1.5*\d, \yshift+0.5*\d) -- (\xshift+1.8*\d, \yshift+0.65*\d);

\draw[\colMut, line width=\w mm] (\xshift+4.25*\d, \yshift+0*\d) -- (\xshift+4.5*\d, \yshift+0*\d) -- (\xshift+4.5*\d, \yshift+0.5*\d) -- (\xshift+4.2*\d, \yshift+0.65*\d);
\draw[\colMut, line width=\w mm] (\xshift+4.75*\d, \yshift+0*\d) -- (\xshift+4.5*\d, \yshift+0*\d) -- (\xshift+4.5*\d, \yshift+0.5*\d) -- (\xshift+4.8*\d, \yshift+0.65*\d);


\tikzmath{\xshift=7*\d;\yshift=-10;};

\draw[thick] (\xshift+0*\d,0*\d+\yshift)--(\xshift+6*\d,0*\d+\yshift);
\draw[thick] (\xshift+0*\d,\d+\yshift)--(\xshift+6*\d,\d+\yshift);
\draw[thick] (\xshift+0*\d,2*\d+\yshift)--(\xshift+6*\d,2*\d+\yshift);


\draw[thick] (\xshift+0.5*\d,\yshift+0*\d)--(\xshift+0.5*\d,\yshift+\d);
\draw[thick] (\xshift+2.5*\d,\yshift+\d)--(\xshift+2.5*\d,\yshift+0*\d);

\draw[whiteCircle] (\xshift+0.5*\d,\yshift+0*\d) circle;
\draw[blackCircle] (\xshift+0.5*\d,\yshift+\d) circle;
\draw[blackCircle] (\xshift+2.5*\d,\yshift+\d) circle;
\draw[whiteCircle] (\xshift+2.5*\d,\yshift+0*\d) circle;


\draw[thick] (\xshift+1.5*\d,\yshift+\d)--(\xshift+1.5*\d,\yshift+2*\d);
\draw[thick] (\xshift+4.5*\d,\yshift+2*\d)--(\xshift+4.5*\d,\yshift+\d);

\draw[whiteCircle] (\xshift+1.5*\d,\yshift+\d) circle;
\draw[blackCircle] (\xshift+1.5*\d,\yshift+2*\d) circle;
\draw[whiteCircle] (\xshift+4.5*\d,\yshift+2*\d) circle;
\draw[blackCircle] (\xshift+4.5*\d,\yshift+\d) circle;


\draw[thick] (\xshift+3.5*\d,\yshift+0*\d)--(\xshift+3.5*\d,\yshift+1*\d);
\draw[thick] (\xshift+5.5*\d,\yshift+1*\d)--(\xshift+5.5*\d,\yshift+0*\d);

\draw[blackCircle] (\xshift+3.5*\d,\yshift+0*\d) circle;
\draw[whiteCircle] (\xshift+3.5*\d,\yshift+1*\d) circle;
\draw[whiteCircle] (\xshift+5.5*\d,\yshift+1*\d) circle;
\draw[blackCircle] (\xshift+5.5*\d,\yshift+0*\d) circle;

\draw[\colMut, line width=\w mm] (\xshift+2.5*\d, \yshift+0*\d) -- (\xshift+3.5*\d, \yshift+0*\d) -- (\xshift+3.5*\d, \yshift+1*\d) -- (\xshift+2.5*\d, \yshift+1*\d) -- (\xshift+2.5*\d, \yshift+0*\d);

\draw[<-, very thick] (\xshift+4.5,\yshift - 0.75) -- (\xshift+4.5,\yshift-1.25);


\tikzmath{\xshift=7*\d;\yshift=-5;};

\draw[thick] (\xshift+0*\d,0*\d+\yshift)--(\xshift+6*\d,0*\d+\yshift);
\draw[thick] (\xshift+0*\d,\d+\yshift)--(\xshift+6*\d,\d+\yshift);
\draw[thick] (\xshift+0*\d,2*\d+\yshift)--(\xshift+6*\d,2*\d+\yshift);


\draw[thick] (\xshift+0.5*\d,\yshift+0*\d)--(\xshift+0.5*\d,\yshift+\d);
\draw[thick] (\xshift+2.5*\d,\yshift+\d)--(\xshift+2.5*\d,\yshift+0*\d);

\draw[whiteCircle] (\xshift+0.5*\d,\yshift+0*\d) circle;
\draw[blackCircle] (\xshift+0.5*\d,\yshift+\d) circle;
\draw[whiteCircle] (\xshift+2.5*\d,\yshift+\d) circle;
\draw[blackCircle] (\xshift+2.5*\d,\yshift+0*\d) circle;


\draw[thick] (\xshift+1.5*\d,\yshift+\d)--(\xshift+1.5*\d,\yshift+2*\d);
\draw[thick] (\xshift+4.5*\d,\yshift+2*\d)--(\xshift+4.5*\d,\yshift+\d);

\draw[whiteCircle] (\xshift+1.5*\d,\yshift+\d) circle;
\draw[blackCircle] (\xshift+1.5*\d,\yshift+2*\d) circle;
\draw[whiteCircle] (\xshift+4.5*\d,\yshift+2*\d) circle;
\draw[blackCircle] (\xshift+4.5*\d,\yshift+\d) circle;


\draw[thick] (\xshift+3.5*\d,\yshift+0*\d)--(\xshift+3.5*\d,\yshift+1*\d);
\draw[thick] (\xshift+5.5*\d,\yshift+1*\d)--(\xshift+5.5*\d,\yshift+0*\d);

\draw[whiteCircle] (\xshift+3.5*\d,\yshift+0*\d) circle;
\draw[blackCircle] (\xshift+3.5*\d,\yshift+1*\d) circle;
\draw[whiteCircle] (\xshift+5.5*\d,\yshift+1*\d) circle;
\draw[blackCircle] (\xshift+5.5*\d,\yshift+0*\d) circle;

\draw[\colMut, line width=\w mm] (\xshift+1.25*\d, \yshift+1*\d) -- (\xshift+2.75*\d, \yshift+1*\d);
\draw[\colMut, line width=\w mm] (\xshift+1.5*\d, \yshift+1*\d) -- (\xshift+1.5*\d, \yshift+1.25*\d);
\draw[\colMut, line width=\w mm] (\xshift+2.5*\d, \yshift+1*\d) -- (\xshift+2.5*\d, \yshift+0.75*\d);

\draw[\colMut, line width=\w mm] (\xshift+3.25*\d, \yshift+1*\d) -- (\xshift+4.75*\d, \yshift+1*\d);
\draw[\colMut, line width=\w mm] (\xshift+3.5*\d, \yshift+1*\d) -- (\xshift+3.5*\d, \yshift+0.75*\d);
\draw[\colMut, line width=\w mm] (\xshift+4.5*\d, \yshift+1*\d) -- (\xshift+4.5*\d, \yshift+1.25*\d);

\draw[<-, very thick] (\xshift+4.5,\yshift - 0.75) -- (\xshift+4.5,\yshift-1.25);


\tikzmath{\xshift=7*\d;\yshift=0;};

\draw[thick] (\xshift+0*\d,0*\d+\yshift)--(\xshift+6*\d,0*\d+\yshift);
\draw[thick] (\xshift+0*\d,\d+\yshift)--(\xshift+6*\d,\d+\yshift);
\draw[thick] (\xshift+0*\d,2*\d+\yshift)--(\xshift+6*\d,2*\d+\yshift);


\draw[thick] (\xshift+0.5*\d,\yshift+0*\d)--(\xshift+0.5*\d,\yshift+\d);
\draw[thick] (\xshift+2.5*\d,\yshift+2*\d)--(\xshift+2.5*\d,\yshift+\d);

\draw[whiteCircle] (\xshift+0.5*\d,\yshift+0*\d) circle;
\draw[blackCircle] (\xshift+0.5*\d,\yshift+\d) circle;
\draw[whiteCircle] (\xshift+2.5*\d,\yshift+\d) circle;
\draw[blackCircle] (\xshift+2.5*\d,\yshift+2*\d) circle;


\draw[thick] (\xshift+1.5*\d,\yshift+\d)--(\xshift+1.5*\d,\yshift+0*\d);
\draw[thick] (\xshift+4.5*\d,\yshift+0*\d)--(\xshift+4.5*\d,\yshift+\d);

\draw[whiteCircle] (\xshift+1.5*\d,\yshift+\d) circle;
\draw[blackCircle] (\xshift+1.5*\d,\yshift) circle;
\draw[whiteCircle] (\xshift+4.5*\d,\yshift) circle;
\draw[blackCircle] (\xshift+4.5*\d,\yshift+\d) circle;


\draw[thick] (\xshift+3.5*\d,\yshift+2*\d)--(\xshift+3.5*\d,\yshift+1*\d);
\draw[thick] (\xshift+5.5*\d,\yshift+1*\d)--(\xshift+5.5*\d,\yshift+0*\d);

\draw[whiteCircle] (\xshift+3.5*\d,\yshift+2*\d) circle;
\draw[blackCircle] (\xshift+3.5*\d,\yshift+1*\d) circle;
\draw[whiteCircle] (\xshift+5.5*\d,\yshift+1*\d) circle;
\draw[blackCircle] (\xshift+5.5*\d,\yshift+0*\d) circle;

\draw[<-, very thick] (\xshift+4.5,\yshift - 0.75) -- (\xshift+4.5,\yshift-1.25);

\end{tikzpicture}
}
\end{center}
\caption{Tetrahedron transformation as sequence of eight two-moves and four spider-moves. Red colour highlights those parts of graph which being transformed by two- or four- moves.}
\label{fig:tetraMovesDetailed}
\end{figure}
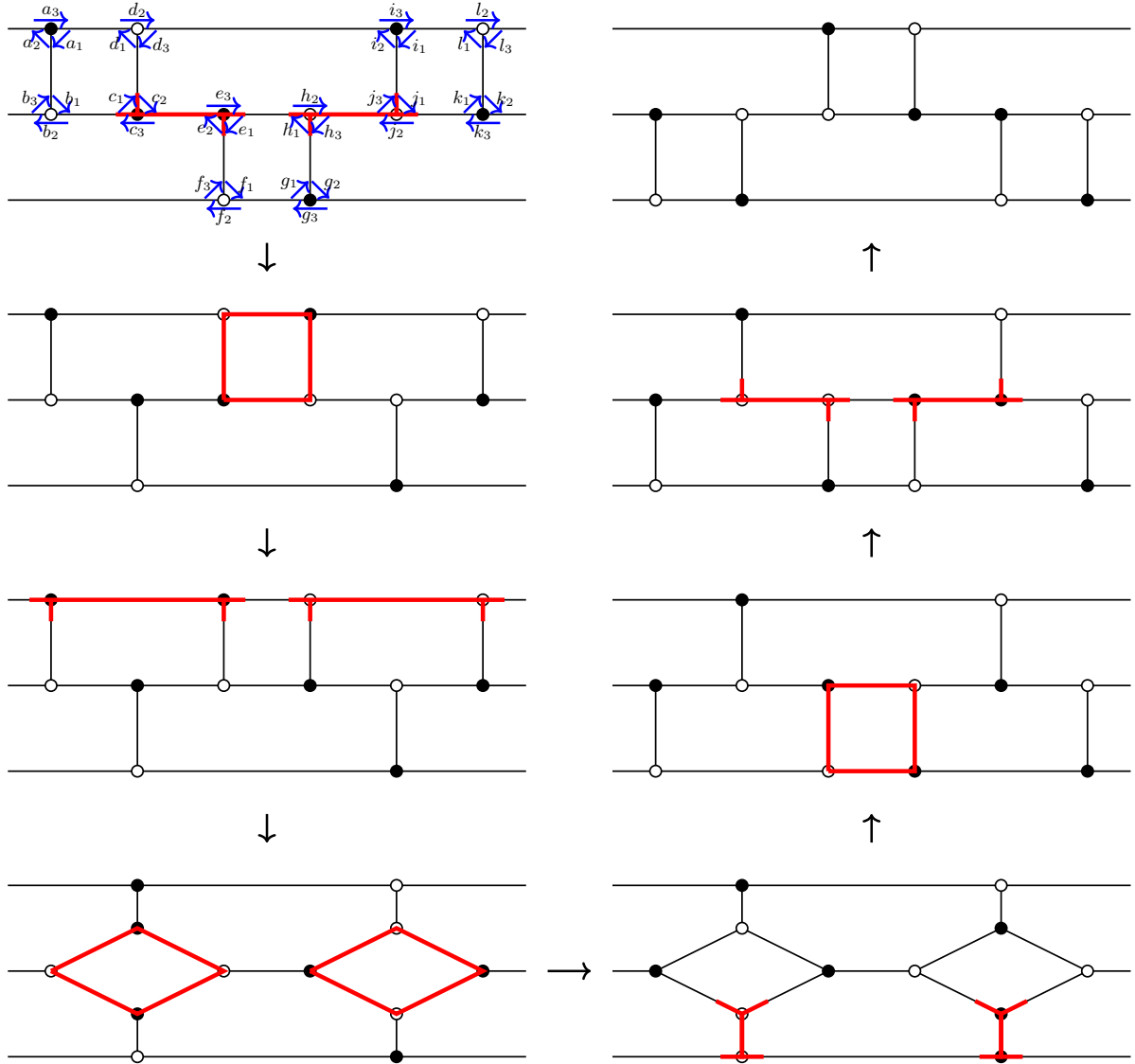

As it was said in Section \ref{ss:tetra}, it is easy to check that transformation of cluster variables (\ref{eq:tetraXtrans}) is agreed with tetrahedron transformation (\ref{eq:BSvarTrans}) via (\ref{eq:BStoClustVar}). However, it is not that easy to derive transformation rules for $\gamma$ variables (\ref{eq:tetraClustGamma}) directly from sequence of two- and four- moves. The major difficulty is that after sequence of moves shown in Fig.~\ref{fig:tetraMovesDetailed} new variables $\gamma'$ defined in Fig.~\ref{fig:tetrahedron} can not be expressed using $\gamma_{x,i}$ with $x=a,b,c;~i=1,2,3,4$ variables only, but more refined corner variables, $a_{1},a_{2},a_{3},...,l_{1},l_{2},l_{3}$, as indicated in Fig.~\ref{fig:tetraMovesDetailed}, should be involved.

It turns out that this problem might be treated by choosing of appropriate gauge. After application of two- and four- moves one can still apply gauge transformations at points shown by grey crosses in Fig.~\ref{fig:tetrahedron}, left, bottom, which transform $\gamma'$ variables by
\begin{equation}
\begin{array}{lll}
\gamma'_{a,1} \to X\gamma'_{a,1},
&
\gamma'_{a,2} \to X^{-1} Y\gamma'_{a,2},
&
\gamma'_{a,3} \to Y^{-1}\gamma'_{a,3},
\\
\gamma'_{b,2} \to Z \gamma'_{b,2},
&
\gamma'_{b,3} \to X Z^{-1} \gamma'_{b,3},
&
\gamma'_{b,4} \to X^{-1}\gamma'_{b,4},
\\
\gamma'_{c,1} \to Z^{-1}\gamma'_{c,1},
&
\gamma'_{c,4} \to Y^{-1} Z\gamma'_{c,4},
&
\gamma'_{c,3} \to Y \gamma'_{c,3},
\end{array}
\end{equation}
and change transfer matrix of each four-gonal block, but do not affect transfer matrix of whole network. Direct check shows\footnote{With four-move parameters chosen to be $\spA=\spB=0,~\spC=-\frac{1}{2}$ in (\ref{eq:fourMove})} that once $X,Y,Z$ are chosen to be
\begin{equation}
\begin{array}{l}
X =
\sqrt{\dfrac{f_2 \gamma_{a,4}}{e_3 \gamma_{a,2}}}
\left(\dfrac{
\gamma_{a,1} \gamma_{a,4} \gamma_{b,3}^{2} \gamma_{c,1} \gamma_{c,2}
}{
\gamma_{a,2} \gamma_{a,3} \gamma_{b,1}^2 \gamma_{c,3} \gamma_{c,4}
}\right)^{1/8},
\\
Y = \sqrt{\dfrac{l_2 b_2}{a_3 k_3}}
\left( \dfrac{d_3 l_3}{i_2 a_2} \right)^{3/8}
\left(\dfrac{\gamma_{a,2} \gamma_{b,4} \gamma_{c,2}}{\gamma_{a,4}\gamma_{b,2} \gamma_{c,4}}\right)^{1/4},
\\
Z = \sqrt{\dfrac{h_2 \gamma_{c,4}}{g_3 \gamma_{c,2}}}
\left( \dfrac{ \gamma_{a,2} \gamma_{a,3} \gamma_{b,1}^{2} \gamma_{c,3} \gamma_{c,4} }{\gamma_{a,1} \gamma_{a,4} \gamma_{b,3}^2 \gamma_{c,1} \gamma_{c,2}}\right)^{1/8},
\end{array}
\end{equation}
transformed $\gamma'$ variables match (\ref{eq:tetraClustGamma}) obtained directly from (\ref{eq:BSvarTrans}) via (\ref{eq:BStoClustVar}).


\begin{thebibliography}{99}

\bibitem{AGPR}
N.~Affolter,  M.~Glick, P.~Pylyavskyy, S.~Ramassamy,
{\it Vector-relation configurations and plabic graphs},
[\href{https://arxiv.org/abs/1908.06959}{{\tt arXiv:1908.06959}}].

\bibitem{BB1}
V.~V.~Bazhanov, R.~J.~Baxter,
{\it New solvable lattice models in three dimensions},
J Stat Phys 69, 453–485 (1992);

\bibitem{BB2}
V.~V.~Bazhanov, R.~J.~Baxter,
{\it Star-triangle relation for a three-dimensional model},
J Stat Phys 71, 839–864  (1993),
[\href{https://arxiv.org/abs/hep-th/9212050}{{\tt arXiv:hep-th/9212050}}].

\bibitem{BS}
V.~V.~Bazhanov, S.~Sergeev,~
{\it Zamolodchikov's tetrahedron equation and hidden structure of quantum groups}, Journal of Physics A 39 (2006) 13,
[\href{https://arxiv.org/abs/hep-th/0509181}{{\tt arXiv:hep-th/0509181}}].

\bibitem{BGM}
M.~Bershtein, P.~Gavrylenko, A.~Marshakov,~
{\it Cluster integrable systems, q-Painleve equations and their quantization}, JHEP 1802:077, 2018,
[\href{https://arxiv.org/abs/1711.02063}{{\tt arXiv:1711.02063}}].

\bibitem{BFM}
M.~Bershtein, B.~Feigin, G.~Merzon,~
{\it Plane partitions with a "pit": generating functions and representation theory },
Sel. Math. New Ser. 24(1) (2018) 21-62,
[\href{https://arxiv.org/abs/1512.08779}{{\tt arXiv:1512.08779}}].

\bibitem{BM}
G.~Bosnjak, V.~Mangazeev,~
{\it Construction of R-matrices for symmetric tensor representations related to $U_q(\widehat{sl}_n)$ },
J. Phys. A: Math. Theor. 49 (2016) 495204,
[\href{https://arxiv.org/abs/1607.07968v1}{{\tt arXiv:1607.07968}}].

\bibitem{BMS08}
V.~V.~Bazhanov, V.~V.~Mangazeev, S.~M.~Sergeev,
{\it Quantum geometry of 3-dimensional lattices},
J. Stat. Mech. (2008), P07006,
[\href{https://arxiv.org/abs/0801.0129}{\tt arXiv:0801.0129}].

\bibitem{BMS09}
V.~V.~Bazhanov, V.~V.~Mangazeev, S.~M.~Sergeev,
{\it Quantum Geometry of 3-Dimensional Lattices and Tetrahedron Equation},
in Proc. XVIth International Congress on Mathematical Physics, pp. 23-44 (2010),
Prague, Czech Republic, 3 - 8 August 2009, World Scientific:2010,
[\href{https://arxiv.org/abs/0911.3693}{\tt arXiv:0911.3693[math-ph]}].

\bibitem{BPZ}
F.~Benini, D.~S.~Park, P.~Zhao,~
{\it Cluster algebras from dualities of 2d $\mathcal{N}=(2,2)$ quiver gauge theories},
Commun. Math.Phys. 340 (2015) 1, 47-104,
[\href{https://arxiv.org/abs/1406.2699}{{\tt arXiv:1406.2699}}].

\bibitem{CW}
S.~A.~Cherkis, R.~S.~Ward,~
{\it Moduli of Monopole Walls and Amoebas},
J. High Energ. Phys. 2012, 90 (2012),
[\href{https://arxiv.org/abs/1202.1294}{{\tt arXiv:1202.1294}}].

\bibitem{EFS}
R.~Eager, S.~Franco, K.~Schaeffer,~
{\it Dimer Models and Integrable Systems},
J. High Energ. Phys. 2012, 106 (2012),
[\href{https://arxiv.org/abs/1107.1244}{{\tt arXiv:1107.1244}}].

\bibitem{FG:2003}
V.~V.~Fock, A.~B.~Goncharov,~
{\it Cluster ensembles, quantization and the dilogarithm},
Ann. Sci. Ec. Norm. Super. \textbf{(4) 42} (2009), 865--930,
[\href{http://arxiv.org/abs/math.AG/0311245}{{\tt arXiv:math.AG/0311245}}].
	
\bibitem{FG:2005}
V.V.~Fock, A.~B.~Goncharov,~
{\it Cluster X-varieties, amalgamation and Poisson-Lie groups},
In Algebraic Geometry Theory and Number Theory, pp. 27–68, Progr. Math., 253, Birkh\"auser	Boston, Boston, MA, 2006, [\href{http://arxiv.org/abs/math.RT/0508408}{{\tt arXiv:math.RT/0508408}}].

\bibitem{FM:2014}
V.~V.~Fock, A.~Marshakov,~
{\it Loop groups, Clusters, Dimers and Integrable systems},
in Geometry and Quantization of Moduli Spaces 1--65;
[\href{http://arxiv.org/abs/1401.1606}{{\tt arXiv:1401.1606}}].

\bibitem{FJMM}
B.~Feigin, M.~Jimbo, T.~Miwa, E.~Mukhin,~
{\it Branching rules for quantum toroidal $\mathfrak{gl}(n)$},
[\href{https://arxiv.org/abs/1309.2147v3}{{\tt arXiv:1309.2147}}].

\bibitem{FZ}
S.~Fomin, A.~Zelevinsky,~
{\it Total positivity: tests and parametrizations},
[\href{https://arxiv.org/abs/math/9912128}{{\tt arXiv:math/9912128}}].

\bibitem{GI}
T.~George, G.~Inchiostro,~
{\it Cluster modular groups of dimer models and networks},
[\href{https://arxiv.org/abs/1909.12896}{{\tt arXiv:1909.12896}}].

\bibitem{GK:2011}
A.~B.~Goncharov, R.~Kenyon,~
{\it Dimers and cluster integrable systems},
Ann. Sci. Ec. Norm. Sup (2013) \textbf{46} 5, 747--813,
[\href{http://arxiv.org/abs/1107.5588}{{\tt arXiv:1107.5588}}].

\bibitem{GV}
I.~Gessel, G.~Viennot,~
{\it Binomial Determinants, Paths, and Hook Length Formulae}, 
Advances in Mathematics 58, 300-321 (1985) 

\bibitem{GSV:2008}
M.~Gekhtman, M.~Shapiro, A.~Vainshtein,~
{\it Poisson Geometry of Directed Networks in a Disk}, 
Selecta Math., (2009) 15, 61-103,
[\href{https://arxiv.org/abs/0805.3541}{{\tt arXiv:0805.3541}}].

\bibitem{GSV:2009-1}
M.~Gekhtman, M.~Shapiro, A.~Vainshtein,~
{\it Poisson Geometry of Directed Networks in an Annulus}, 
Journal of the European Mathematical Society, (2012) 541–570,
[\href{https://arxiv.org/abs/0901.0020}{{\tt arXiv:0901.0020}}].

\bibitem{GSV:2009}
M.~Gekhtman, M.~Shapiro, A.~Vainshtein,~
{\it Generalized B\"acklund-Darboux transformations for Coxeter-Toda flows from a cluster algebra perspective}, 
Acta Math., 206 (2011), 245-310,
[\href{https://arxiv.org/abs/0906.1364}{{\tt arXiv:0906.1364}}].

\bibitem{GSTV:2014}
M.~Gekhtman, M.~Shapiro, S.~Tabachnikov, A.~Vainshtein,~
{\it Integrable cluster dynamics of directed networks and pentagram maps},
Adv. Math. 300 (2016), 390-450,
[\href{https://arxiv.org/abs/1406.1883}{{\tt arXiv:1406.1883}}].

\bibitem{HI13}
K.~Hikami, R.~Inoue,~
{\it Braids, Complex Volume, and Cluster Algebra},
Algebr. Geom. Topol. 15 (2015) 2175-2194,
[\href{https://arxiv.org/abs/1304.4776v6}{{\tt arXiv:1304.4776}}].

\bibitem{ILP}
R.~Inoue, T.~Lam, P.~Pylyavskyy,~
{\it On the cluster nature and quantization of geometric $R$-matrices},
[\href{https://arxiv.org/abs/1607.00722}{{\tt arXiv:1607.00722}}].

\bibitem{K93}
I.~G.~Korepanov,~
{\it Tetrahedral  Zamolodchikov  Algebras Corresponding to Baxter's L-Operators},
Commun. Math. Phys. 154, 85–97 (1993).

\bibitem{K94}
I.~G.~Korepanov,~
{\it A Dynamical System Connected with inhomogeneous 6-Vertex Model},
Zapiski Nauchn. Semin. POMI (S-Petersburg) 215 (1994) 178-196,
[\href{https://arxiv.org/abs/hep-th/9402043}{{\tt arXiv:hep-th/9402043}}].

\bibitem{K95}
I.~G.~Korepanov,~
{\it Algebraic integrable dynamical systems, 2+1-dimensional models in wholly discrete space-time, and inhomogeneous models in 2-dimensional statistical physics},
[\href{https://arxiv.org/abs/solv-int/9506003v1}{{\tt solv-int/9506003}}].

\bibitem{KKS}
R.~M.~Kashaev, I.~G.~Korepanov, S.~M.~Sergeev,~
{\it Functional tetrahedron equation},
Theor. Math. Phys. 117:3 (1998) 1402 - 1413,
[\href{https://arxiv.org/abs/solv-int/9801015}{{\tt arXiv:solv-int/9801015}}].

\bibitem{KMS}
R.~M.~Kashaev, V.~V.~Mangazeev, Yu.~G.~Stroganov,~
{\it Spatial symmetry, local integrability and tetrahedron equations in the Baxter-Bazhanov model},
International Journal of Modern Physics A, 8 (3) (1993) 587-601.

\bibitem{KOS}
A.~Kuniba, M.~Okado, S.~Sergeev,~
{\it Tetrahedron equation and generalized quantum groups},
J. Phys. A: Math. Theor. 48 (2015) 304001 (38pp),
[\href{https://arxiv.org/abs/1503.08536v2}{{\tt arXiv:1503.08536}}].

\bibitem{KP}
R.~Kenyon, R.~Pemantle,~
{\it Double-dimers, the Ising model and the hexahedron recurrence},
[\href{https://arxiv.org/pdf/1308.2998.pdf}{{\tt arXiv:1308.2998}}].

\bibitem{KS}
A.~Kuniba, S.~Sergeev,~
{\it Tetrahedron Equation and Quantum R Matrices for Spin Representations of $B^{(1)}_n$, $D^{(1)}_n$ and $D^{(2)}_{n+1}$},
Commun. Math. Phys. 324, 695--713 (2013),
[\href{https://arxiv.org/abs/1203.6436v3}{{\tt arXiv:1203.6436}}].

\bibitem{KV}
M.~Kapranov, V.~Voevodsky,~
{\it 2-categories and Zamolodchikov tetrahedra equations},
Proc. Sympos. Pure Math., 56 (2) 1994.

\bibitem{L}
B.~Lindstr\"om,~
{\it On the Vector Representations of Induced Matroids},
Bulletin of the London Mathematical Society, 5: 85-90 (1973).

\bibitem{LP}
T.~Lam, P.~Pylyavskyy,~
{\it Total positivity in loop groups, I: Whirls and curls},
Advances in Mathematics, 230 (2012) 3: 1222 - 1271,
[\href{https://arxiv.org/abs/0812.0840}{{\tt arXiv:0812.0840}}].

\bibitem{LS16}
A.~Litvinov, L.~Spodyneiko,~
{\it On $W$ algebras commuting with a set of screenings},
J. High Energ. Phys. 2016, 138 (2016),
[\href{https://arxiv.org/abs/1609.06271}{{\tt arXiv:1609.06271}}].

\bibitem{LS18}
A.~Litvinov, L.~Spodyneiko,~
{\it On dual description of the deformed $O(N)$ sigma model},
J. High Energ. Phys. 2018, 139 (2018),
[\href{https://arxiv.org/abs/1804.07084}{{\tt arXiv:1804.07084}}].

\bibitem{M}
A.~Marshakov,~
{\it Lie groups, cluster variables and integrable systems},
Journal of Geometry and Physics (2013) 67: 16-36,
[\href{https://arxiv.org/abs/1207.1869}{{\tt arXiv:1207.1869}}].

\bibitem{MBS}
V.~Mangazeev, V.~Bazhanov, S.~Sergeev,~
{\it An integrable 3D lattice model with positive Boltzmann weights},
J. Phys. A: Math. Theor., v. 46, 465206 (2013),
[\href{https://arxiv.org/abs/1308.4773v2}{{\tt arXiv:1308.4773}}].

\bibitem{MS}
A.~Marshakov, M.~Semenyakin,~
{\it Cluster integrable systems and spin chains},
J. High Energ. Phys. (2019) 2019: 100,
[\href{https://arxiv.org/abs/1905.09921}{{\tt arXiv:1905.09921}}].

\bibitem{OS}
V.~Ovsienko, M.~Shapiro,~
{\it Cluster algebras with Grassmann variables},
to appear in Electron. Res. Announc. Math. Sci.,
[\href{https://arxiv.org/abs/1809.01860}{{\tt arXiv:1809.01860}}].

\bibitem{P}
A.~Postnikov,~
{\it Total positivity, Grassmannians, and networks},
[\href{https://arxiv.org/abs/math/0609764}{{\tt arXiv:math/0609764}}].

\bibitem{S}
S.~M.~Sergeev,~
{\it Quantum 2 + 1 evolution model},
Journal of Physics A: Mathematical and General, 32 (30),
[\href{https://arxiv.org/abs/solv-int/9811003}{{\tt arXiv:solv-int/9811003}}].

\bibitem{S97}
S.~M.~Sergeev,~
{\it Solutions of the functional tetrahedron equation connected with the local Yang -- Baxter equation for the ferro-electric},
[\href{https://arxiv.org/abs/solv-int/9709006}{{\tt arXiv:solv-int/9709006}}].

\bibitem{S09}
S.~M.~Sergeev,
{\it Supertetrahedra and superalgebras},
J. Math. Phys. 50, 083519 (2009),
[\href{https://arxiv.org/abs/0805.4653}{\tt arXiv:0805.4653}].

\bibitem{SMS}
S.~Sergeev, V.~V.~Mangazeev, Yu.~G.~Stroganov,~
{\it The vertex formulation of the Bazhanov-Baxter Model},
J. Stat Phys 82, 31–49 (1996),
[\href{https://arxiv.org/abs/hep-th/9504035v1}{{\tt arXiv:hep-th/9504035}}].

\bibitem{SS}
G.~Schrader, A.~Shapiro,~
{\it A cluster realization of $U_q(\mathfrak{sl}_n)$ from quantum character varieties},
[\href{https://arxiv.org/abs/1607.00271}{{\tt arXiv:1607.00271}}].

\bibitem{Talaska}
K.~Talaska,~
{\it A formula for Plucker coordinates associated with a planar network},
Int Math Res Notices (2008), ID: rnn081,
[\href{https://arxiv.org/abs/0801.4822v2}{{\tt arXiv:0801.4822}}].

\bibitem{Thurston}
D.~Thurston,~
{\it From Dominoes to Hexagons},
[\href{https://arxiv.org/abs/math/0405482}{{\tt arXiv:math/0405482}}].

\bibitem{Y16}
M.~Yamazaki,~
{\it Cluster-Enriched Yang-Baxter Equation from SUSY Gauge Theories},
Lett Math Phys 108, 1137–1146 (2018),
[\href{https://arxiv.org/abs/1611.07522v1}{{\tt arXiv:1611.07522}}].

\bibitem{Za1}
A.~B.~Zamolodchikov,~
{\it Tetrahedra equations and integrable systems in three-dimensional space},
JETP, Vol. 52, No 2, p. 325.

\bibitem{Za2}
A.~B.~Zamolodchikov,~
{\it Tetrahedron equations and the relativistic S-matrix of straight-strings in 2+1-Dimensions},
Commun. Math. Phys. 79, 489–505 (1981).

\bibitem{Ze}
Y.~Zenkevich,
{\it Higgsed network calculus},~
[\href{https://arxiv.org/abs/1812.11961}{{\tt arXiv:1812.11961}}].

\end{thebibliography}
\end{document}